\documentclass[12pt]{article}
\usepackage{latexsym}
\usepackage{amsmath}
\usepackage{amssymb}
\usepackage{epsfig,graphics}
\usepackage{booktabs}
\usepackage{multirow}

\newcommand{\be}{\begin{equation}}
\newcommand{\ee}{\end{equation}}
\newcommand{\otoprule}{\midrule[\heavyrulewidth]}

\begin{document}

\begin{titlepage}
\begin{flushright} DESY 10-116\end{flushright}

\vspace*{0.25in}
 
\begin{center}
{\large\bf Closed flux tubes and their string description\\
in D=3+1 SU(N) gauge theories\\}
\vspace*{0.85in}
{Andreas Athenodorou$^{a}$, Barak Bringoltz$^{b}$  and Michael Teper$^{c}$\\
\vspace*{.2in}
$^{a}$DESY, Platanenallee 6, 15738 Zeuthen, Germany \\
\vspace*{.1in}
$^{b}$Department of Physics, University of Washington, Seattle,
WA 98195-1560, USA\\
\vspace*{.1in}
$^{c}$Rudolf Peierls Centre for Theoretical Physics, University of Oxford,\\
1 Keble Road, Oxford OX1 3NP, UK
}
\end{center}

\vspace*{0.4in}

\begin{center}
{\bf Abstract}
\end{center}

We calculate the energy spectrum of a confining flux tube that is
closed around a spatial torus, as a function of its length $l$. We do 
so for various SU($N$) gauge theories in 3+1 dimensions, and
for various values of spin, parity and longitudinal momentum. 
We are able to present
usefully accurate results for about $20$ of the lightest such states,
for a range of $l$ that begins close to the (finite volume) 
deconfining phase transition at $l\surd\sigma \sim 1.6$, and extends 
up to $l\surd\sigma \sim 6$ (where $\sigma$ is the string tension). 
We find that most of these low-lying states are well described by the
spectrum of the Nambu-Goto free string theory in flat space-time. 
Remarkably, this is so not only at the larger values of $l$, 
where the gap between the ground state energy and the low-lying 
excitations becomes small compared to the mass gap, but also down to 
much shorter lengths where these excitation energies become large 
compared to $\surd\sigma$, the flux-tube no longer `looks' anything 
like a thin string, and an expansion of the effective string action
in powers of $1/l$ no longer converges. All this is for flux in the
fundamental representation. We also calculate the $k=2$ (anti)symmetric
ground states and these show larger corrections at small $l$.
So far all this closely resembles 
our earlier findings in 2+1 dimensions. However, and in contrast to the 
situation in $D=2+1$, we also find that there are some states, 
with $J^P=0^-$ quantum numbers, that show large deviations from the
Nambu-Goto spectrum. We investigate the possibility that (some of)
these states may encode the massive modes associated with the internal
structure of the flux tube, and we discuss how the precocious free string
behaviour of most states constrains the effective string action, 
on which much interesting theoretical progress has recently been made.

\end{titlepage}

\setcounter{page}{1}
\newpage
\pagestyle{plain}

\section{Introduction}
\label{section_intro}

The idea that the strong interactions may be described by a string
theory is even older than Quantum Chromodynamics, e.g.
\cite{strings_old}, 
and the idea that gauge theories (at least in the planar limit) may 
have such a description is only a little younger
\cite{tHooft_N}. 
The more recent and radical version of this idea is gauge-string duality
\cite{duality}. 
To learn something about 
this string theory it is natural to start by focusing upon any degrees of 
freedom that are manifestly string-like and, in linearly confining 
theories such as SU($N$) gauge theories in $D=2+1$ and $D=3+1$, these are 
long confining flux tubes. One can ask what effective string theory 
describes their dynamics. Recently there has been substantial 
analytic progress towards answering this (old) question which, roughly
speaking, tells us that the dynamics governing very long flux tubes is, 
to a certain approximation in powers of $1/l$, that of a Nambu-Goto
free bosonic string theory
\cite{LW04,JD,DM06,AHEK09,DM09}. 
(We shall review this in more detail below.) At the same time our 
numerical calculations of the  the low-lying excitations of closed 
flux tubes in $D=2+1$  
\cite{AABBMT_k1d3}
have shown that all the calculated energies are remarkably close to 
those of the free bosonic string theory even when the flux tube length
$l$ is not much greater than its width so that an expansion in powers 
of $1/\sigma l^2$ (where $\sigma$ is the string tension) no longer 
converges. Moreover the fact that we have not observed additional massive 
string modes associated with the non-trivial structure of the flux-tube, 
suggests that these interact very weakly with the usual massless stringy 
modes. These complementary numerical
and analytic results strongly suggest that the effective string action 
in $D=2+1$ SU($N$) gauge theories is some small perturbation about the 
Nambu-Goto action (in flat space-time), so that the latter, rather 
than the traditional classical string configuration, should
be the appropriate starting point for any analytic investigation.

In this paper we extend our calculations to $3+1$ dimensions. As in 
$2+1$ dimensions, we close our flux tubes around a spatial torus of 
length $l$ so as to stabilise them at a chosen length. Such flux
tubes have non-trivial rotational properties in 3 spatial dimensions 
and hence a richer spectrum than in 2 spatial dimensions. Thus, despite 
the fact that our calculations turn out to be less accurate than in 
$D=2+1$, we are able to obtain usefully accurate results on a substantial
number of low-lying eigenstates. What we shall show is that most
of these states are well-described by the Nambu-Goto model down
to very short flux tube lengths, just as in $D=2+1$. However we shall
also find that a few states, with specific quantum numbers, are very far 
from the Nambu-Goto prediction, leaving it unclear whether they approach 
the free bosonic string value as $l$ increases or not. We discuss in
some detail the possibility that this is a signal of the massive modes 
that had eluded us in our earlier  $D=2+1$ calculations.

We shall begin, in Section~\ref{section_EST}, by summarising the theoretical
background and briefly describing where the analytic calculations
of effective string theories now stand. In Section~\ref{section_lattice} 
we turn to the technicalities of our numerical calculation. We describe
the lattice Monte Carlo set-up, how to calculate the eigenspectrum
of closed flux tubes, the specific lattice operators that we use, and
the quantum numbers of the states. We finish Section~\ref{section_lattice}
with some brief remarks designed to place our calculations in the context
of earlier lattice work. We then turn to our results in 
Section~\ref{section_results}. We begin, in Section~\ref{subsection_groundq0}, 
with an analysis of how the (absolute) ground 
state energy of such a closed flux tube varies with its length $l$. 
We compare our values of the energy to the free string prediction and fit
the observed (small) deviation to some theoretically motivated correction 
terms. In Section~\ref{subsection_groundq12} we consider non-zero momentum 
along the flux tube and calculate the energies of the lightest states 
in each such sector. In Section~\ref{subsection_excited} we calculate 
a number of excited state energies. It is here that we shall identify 
and discuss a number of states with energies very far from the free 
string prediction. In Section~\ref{subsection_k2} we briefly deviate 
from the main theme of this paper and consider the properties
of the $k=2$ (anti)symmetric ground states. Finally, we discuss 
the implication of these results in  Section~\ref{subsection_discussion}
Section~\ref{section_conclusion} contains our conclusions.

An important aim of our work is to provide `data' that can be used
to inform theoretical analyses of the effective string action. While 
we do attempt to provide comparisons with those analyses available to
us at the time of writing, this is an area where there is rapid
current progress, and other comparisons may soon be of interest
\cite{OA_ECT}.
So to maximise the usefulness of our work, we have deliberatively 
provided in an Appendix a comprehensive tabulation  of our numerically 
determined eigenenergies, in a form that is intended to be readily usable.

We remark that a brief summary of some of this work has appeared 
previously in \cite{lat09}.

\section{Effective string theory}
\label{section_EST}

We are in the confining phase of an SU($N$) gauge theory on a 4-torus,
with partition function $Z$. We label the coordinates by (x,y,z,t).
Since space-time is Euclidean we may in fact choose any co-ordinate as 
`time' and write $Z$ as a sum over eigenstates of the Hamiltonian defined
on the orthogonal 3-dimensional `space'. The same is true for partition
functions including sources, which we define below, and the resulting 
`dual' descriptions play an important role both in constraining the 
effective string theory and in the set-up of our numerical calculations.

\subsection{General considerations}
\label{subsection_general_EST}

Consider a static fundamental source at $\vec{x} = (0,y,z)$, with a conjugate 
source at $\vec{x} = (r,y,z)$. Let $\tau$ be the the Euclidean time extent 
of our 4-torus (with all other tori large). The partition function 
for the system with these two sources is 
\begin{equation}
Z_{s\bar{s}}(r,\tau) = \int dA \  l^\dagger(x=r,y,z)
l(x=0,y,z) \, \exp\{-S[A]\}
\label{eqn_Zss}
\end{equation}
where $l(\vec{x})$ is a Polyakov loop. (The traced path-ordered exponential 
of the gauge potential along a path that encircles the t-torus
at $\vec{x}=(x,y,z)$, which is the phase factor arising from the minimal 
coupling, $j_\mu A_\mu =j_0 A_0$, of our static source.)

If $r \gg 1/\surd\sigma$ there will be a flux tube between the
sources which, as it evolves in time, sweeps out a sheet bounded
by the periodic sources. This sheet clearly has the topology 
of a cylinder. The partition function can be written as a sum over 
energy eigenstates   
\begin{equation}
\frac{1}{Z} Z_{s\bar{s}}(r,\tau) = \sum_n e^{-E_n(r)\tau}
\label{eqn_Zss_open}
\end{equation}
where $E_n(r)$ is an energy of two sources separated by $r$. The states
are (excited) flux tubes that begin and end on a source and which
evolves around the $t$-torus.

There is another way to interpret $Z_{s\bar{s}}(r,\tau)$. Take $x$ to label
the `time', so $l$ is now a Wilson line that winds around what is now 
a spatial torus of length $\tau$. What $Z_{s\bar{s}}(r,\tau)$
represents, in this point of view, is a correlation function whose 
intermediate states consist of flux tubes that wind around this same 
`spatial' torus of length $\tau$ and propagate the distance $r$ 
between the two Wilson lines. This partition function can 
therefore be written as a sum over these energy eigenstates
\begin{equation}
\frac{1}{Z} Z_{s\bar{s}}(r,\tau) = 
\sum_{n,p} c_n(p,\tau) e^{-\tilde{E}_n(p,\tau)r}
\label{eqn_Zss_closed}
\end{equation}
where $\tilde{E}_n$ is an energy of an (excited) flux tube that winds 
around a spatial torus of length $\tau$. $\tilde{E}_n$ and $E_n$ are 
different functions because the flux tubes have different boundary 
conditions (although later on we shall use $E_n$ for both). Since the 
operators $l$ are localised at $y,z$, the winding flux tubes are
delocalised in  transverse momentum, and we have made explicit in  
eqn(\ref{eqn_Zss_closed}) the sum over that momentum.
The $c_n$ are the wave-function factors for the overlap of a state 
$|n,p\rangle$ on the state obtained by applying the operator $l$
to the vacuum state $| vac \rangle$:
$c_n = |\langle vac | l^\dagger | n,p \rangle |^2$.

Suppose that we have an effective string theory, governed by an effective 
action $S_{eff}$, which reproduces the long distance physics of flux tubes. 
Then the string partition function, taken over the $r\times\tau$ cylinder 
considered above, should reproduce the flux tube contribution to the
field-theoretic partition function in eqn(~\ref{eqn_Zss}):
\begin{equation}
Z_{cyl}(r,\tau) = \int_{cyl = r\times\tau} dS e^{-S_{eff}[S]}
=
\frac{1}{Z} Z_{s\bar{s}}(r,\tau)
\label{eqn_Zcyl}
\end{equation}
where we integrate over all surfaces $S$ spanning the cylinder.
From eqn(\ref{eqn_Zcyl}) we see that $Z_{cyl}(r,\tau)$ can be written 
as a sum over open flux tube states as in eqn(\ref{eqn_Zss_open}).
This is nothing but a Laplace transform in $\tau$ of $Z_{cyl}(r,\tau)$. So
if we have a candidate string action, $S_{eff}[S]$, we can perform this 
Laplace transform and extract a prediction for the the open flux tube
energies $E_n(r)$. We can also express $Z_{cyl}(r,\tau)$ as a sum over 
closed flux tube states using eqn(\ref{eqn_Zss_closed}). This is a
Laplace transform in $r$ of $Z_{cyl}(r,\tau)$, and so a candidate 
$S_{eff}[S]$ will imply a prediction for the the closed flux tube
energies, $\tilde{E}_n(p,\tau)$. However in this case the energies
are not all independent: the relationship between energy eigenstates 
that differ only by their transverse momentum $p$ is determined by Lorentz 
invariance. (In fact one can show that doing  the integral over $p$ turns 
eqn(\ref{eqn_Zss_closed}) into a sum over Bessel functions 
\cite{LW04,HM06}.)
There is no reason that, for some arbitrary choice of $S_{eff}[S]$, the
Laplace transform of $Z_{cyl}(r,\tau)$ in $r$ should reproduce an energy 
spectrum that satisfies this relationship. As first realised in
\cite{LW04},
this provides a powerful constraint on the permitted form of the
effective string action.  

It was then observed in 
\cite{AHEK09}
that the  constraints arising from the above open-closed duality 
associated with a cylinder 
can be extended to other geometries that are less obvious from a 
field-theoretic perspective. In particular one can extend the analysis 
to an $r\times\tau$ torus 
\cite{AHEK09}.
A world sheet spanning such a torus corresponds to a closed flux tube 
of length $r$ propagating over a time $\tau$ or one of length $\tau$ 
propagating over a time $r$. Thus we have a closed-closed string duality: 
\begin{equation}
Z_{torus}(r,\tau) = \int_{T^2=r\times\tau} dS e^{-S_{eff}[S]}
=
\sum_{n,p} e^{-\tilde{E}_n(p,\tau)r}
=
\sum_{n,p} e^{-\tilde{E}_n(p,r)\tau}
\label{eqn_Ztorus}
\end{equation}
which turns out to provide useful new constraint on  $S_{eff}[S]$
\cite{AHEK09}.
It may perhaps be that we have not exhausted all the possibilities and 
that other boundary conditions and set-ups may provide further
useful constraints.

This general discussion needs some qualifications. Our effective string
theory is in $2+1$ or $3+1$ dimensions and far from the critical
dimension where we can consistently define a string theory. However the 
resulting anomalies, which show up in different ways depending on how one 
`gauge-fixes' the diffeomorphism invariance in one's calculation, typically 
die off at long distances
\cite{Olesen}
(although the implications of this are not unambiguous),
and when one considers smooth fluctuations around a long string
\cite{PS}.
Thus it can make sense, at least technically, to consider a string path 
integral over single large surfaces, in an effective string theory 
approach outside the critical dimension
\cite{PS}.
Since this represents the world sheet swept out by a single long fluctuating 
string, this means that an effective string theory is limited to describing 
the dynamics of a single long fluctuating flux tube. This is clearly an 
important physical limitation. In reality, a sufficiently excited flux tube 
can decay into a flux tube of lower energy and a glueball, and such states 
inevitably appear in the field-theoretic sum over states in 
eqn(\ref{eqn_Zss_open}) and eqn(\ref{eqn_Zss_closed}). In the string picture 
a  glueball is a contractible closed loop of string whose length is 
$O(1/\surd\sigma)$ (for light glueballs). There is no guarantee that an 
effective string theory can consistently describe such extra small surfaces. 
One can partially circumvent this by only considering low-lying
string states which are too light to decay:
\begin{equation}
E_n(r) - E_0(r) \ll m_G
\label{eqn_Elight}
\end{equation}
where $m_G$ is the energy of the lightest glueball. (Or of the lightest 
state composed of a flux loop and its conjugate, if that is lighter.)
However even such states will be affected by mixing through virtual
glueball emission, which corresponds to small handles on our large
surface - again something that would be problematic for the string
theory. 

There is of course a limit in which mixings and decays do vanish, and
that is the $N\to\infty$ limit. In the  SU($\infty$) theory one
can consistently discuss a partition function over states containing
a single flux-tube, as in eqns(\ref{eqn_Zss_open},\ref{eqn_Zss_closed}), 
and ask if it is given by an integration over single surfaces with
some effective string action. It is then also plausible 
that as we move continuously away from that limit, to finite $N$, the 
corrections will be small, so that it makes sense to analyse all SU($N$) 
gauge theories within such an effective string framework
\cite{AHEK09}.
The fact that there is very little dependence on $N$ in the low-lying
flux tube spectrum, both in  $D=2+1$
\cite{AABBMT_k1d3}
and in $D=3+1$ (this paper), lends credibility to this assumption.

Consider a flux tube that winds around a spatial torus of length $l$.
The excited states of this flux tube are obtained from the ground state, 
$E_0(l)$, by exciting some of the modes living on the flux tube. If the 
excited mode is massive we expect the energy to be
\begin{equation}
E(l)  = E_0(l) + O(\surd\sigma).
\label{eqn_massivemode}
\end{equation}
If the excited mode is massless, we would expect the extra energy 
to equal the (longitudinal) momentum which, for bosons, is quantised 
to be $p=2\pi k/l ; k=\pm 1, ...$ by periodicity. So we then expect
\begin{equation}
E(l)  = E_0(l) + O\left(\frac{\pi}{l}\right).
\label{eqn_masslessmode}
\end{equation}
So once our flux tube is long enough, with $\pi/l \ll\surd\sigma$, its
lightest excitations are given solely by the excitations of the massless modes.
One should therefore be able to construct an effective string theory solely
out of its massless modes if what we want is a description of the low-lying
spectrum of long flux tubes.

In general we expect modes to be massless for symmetry reasons.
In the case of a flux tube there are $D-2$ obvious massless modes.
These are the Goldstone modes that arise from the fact that once
we have specified the location of our flux tube, we have broken
spontaneously the translation invariance in the $D-2$ directions
transverse to the flux tube. So one usually starts with an effective
bosonic string theory involving just these Goldstone fields. By
comparing with the numerical spectrum at large-$l$ one can see
if the presence of other, less obvious, massless modes is indicated.

In practice most numerical work has involved the ground state energy,
$E_0(l)$. The massless modes, when quantised, contribute  $O(1/l)$ 
zero-point energies to $E_0(l)$. Summing over frequencies, gives 
the `universal L\"uscher correction'
\cite{LSW}
\begin{equation}
E_0(l)  = \sigma l  - \frac{c}{l}, \qquad 
c = \frac{\pi(D-2)}{6}  \quad : \quad  \mathrm{bosonic}
\label{eqn_zeromode}
\end{equation}
where $c$ is proportional to the central charge of the string theory 
and will differ from the bosonic value shown if other massless
modes are present. For example, one expects it to be zero in the case
of ${{\cal{N}}=1}$ SUSY \footnote{We thank Ofer Aharony for this 
observation.}.
(While this is for closed flux tubes, there is a similar
expression for open flux tubes.) There have been increasingly accurate 
lattice determinations of $c$ over the last 25 years that leave little
doubt that in pure gauge theories the only massless modes on the flux 
tube are indeed the transverse Goldstone modes.

To proceed with explicit computations, one needs to fix convenient 
coordinates to describe the surface in the path integral. This 
`gauge-fixing' of the fundamental diffeomorphism invariance typically 
makes the constraints that follow from this string symmetry 
less obvious. Here we follow the `static gauge' approach of
\cite{LSW,LW04,AHEK09}
and do not discuss the general effective string approach of
\cite{PS},
which has been used in 
\cite{JD,DM06,DM09}
to obtain comparable results (in `conformal gauge').

Suppose we are integrating over the surfaces of the cylinder
discussed above. There is a minimal surface which can be
parameterised by $x \in [0,r]$ and $t\in [0,\tau]$. Other
surfaces are specified by a transverse displacement vector
$h(x,t)$ that has two components in the $(y,z)$ directions
(but only one in $D=2+1$). 
This way of parameterising a surface is called `static gauge'.
We can now write the effective string action in terms of 
this field $h$ and the integral over surfaces becomes an integral
over $h(x,t)$ at each value of $(x,t)\in [0,r]\times[0,\tau]$. Since the 
field $h$ is an integration variable in $(0,\infty)$, we can take it
to be dimensionless. Moreover, since the action cannot depend on the 
position of the flux tube (translation invariance), it cannot depend
on $\langle h \rangle$ but can only depend on $\partial_\alpha h$ where
$\alpha = x,t$. That is to say, schematically, 
\begin{equation}
S_{eff}[S] \longrightarrow S_{eff}[h] \longrightarrow S_{eff}[\partial h]
\label{eqn_Seff_Stohtodh}
\end{equation}
and we can perform a derivative expansion of  $S_{eff}$ in powers 
of derivatives of $h$: (very) schematically
\begin{equation}
S_{eff} = \sigma r \tau
+ \int^\tau_0 dt \int^r_0 dx \frac{1}{2} \partial h\partial h
+ \sum_{n=2} c_n \int^\tau_0 dt \int^r_0 dx  (\partial h)^{2n}
\label{eqn_Seff_expansion}
\end{equation}
where the derivatives are with respect to $x$ and $t$ and indices
are appropriately contracted. 
The coefficients $c_n$ have dimensions [length]$^{(2n-2)}$
to keep the terms dimensionless. So we can expect that for the 
long wavelength fluctuations of a long string, such a higher
order term will make a contribution to the energy that is down
by $O(1/(\sigma l^2)^n)$ compared to the leading linear piece
and so the importance of these terms is naturally ordered by
the number of derivatives. All this is entirely analogous to the 
familiar way chiral Lagrangians depend on their Goldstone fields.
And just as the applicability of chiral Lagrangians is typically 
bounded by the energy scale of the lowest resonance, this
derivative expansion is designed to capture the physics on energy 
scales smaller than the $O(\surd\sigma)$ dynamical mass scale.

For a surface with a boundary, like a cylinder, there can also be 
operators localised on the boundary, which contribute odd powers of 
$1/r$, and these we have ignored. Since we are interested
in the spectrum of winding strings, that can be obtained from the 
torus geometry which has no boundaries, it is safe for us to do so. 

Note that our chosen `static-gauge' parameterisation does not work 
for general surfaces. To describe a string with an `overhang' or
any kind of `back-tracking', the field $h(x,t)$ would have to multivalued, 
which is something the standard treatments do not allow. That is to say, 
we arbitrarily exclude such rough surfaces from the path integral. For a 
flux tube, its finite width provides a physical lower distance cutoff on such 
fluctuations: any overhang that is within a distance $ \lesssim 1/\surd\sigma$
will in effect be a fluctuation in the intrinsic width of the flux tube 
i.e. a massive mode excitation. Any backtracking/overhang that is larger
will increase the length by $\Delta l >  1/\surd\sigma$ and hence
the energy by  $\Delta E \sim \sigma\Delta l > \surd\sigma$.
In both cases the associated excitation energies will be much
greater than the $O(1/l)$ gap to the stringy modes, once $l$
is large enough.  Thus this should not be a significant issue for the 
long wavelength massless oscillations we have discussed above. But it
needs to be addressed in any analytic treatment that aims to be more 
ambitious than that.

\subsection{Gaussian action}
\label{subsection_gaussian_EST}

The first non-trivial term in our effective string action
in eqn(\ref{eqn_Seff_expansion}) is the Gaussian piece:
\begin{equation}
S^{gauss}_{eff} = \sigma r \tau
+\int^\tau_0 dt \int^r_0 dx \frac{1}{2} 
\sum_{\alpha=x,t}
\partial_\alpha h\partial_\alpha h
\label{eqn_Sgaussian}
\end{equation}
It has the fewest derivatives and so will provide the leading large $l$
correction to the linear piece of the string energy. Since the action
is Gaussian one can evaluate $Z_{cyl}$ exactly, giving
\begin{equation}
Z_{cyl}(r,\tau) = e^{-\sigma r\tau} |\eta(q)|^{-(D-2)} 
\quad :\quad q=e^{-\pi\tau/r}
\label{eqn_Zgaussian}
\end{equation}
in terms of the Dedekind eta function
\begin{equation}
\eta(q) = q^\frac{1}{24} \prod^\infty_{n=1} (1-q^n).
\label{eqn_dedekind}
\end{equation}
(See e.g.
\cite{LW04}
whose notation we borrow.) If we expand the product in 
eqn(\ref{eqn_dedekind}) we have a sum of powers of $q$, which,
using  $q=e^{-\pi\tau/r}$, becomes a sum of exponentials in
$\tau$. This is precisely of the form given in eqn(\ref{eqn_Zss_open}).
If we match this sum to eqn(\ref{eqn_Zss_open}), we obtain the prediction
\begin{equation}
E_n(r) = \sigma r + \frac{\pi}{r} 
\left\{n-\frac{1}{24}(D-2)\right\} \qquad : \ {\mathrm{Gaussian}}
\label{eqn_Engaussian}
\end{equation}
for the energy levels (as well as a prediction for their degeneracies). 
This is the exact result for the energy levels of a string with ends 
fixed to static sources, a distance $r$ apart, for the case of a 
Gaussian $S_{eff}$. We note 
that the excitation energies display an $O(1/r)$ gap as expected from 
eqn(\ref{eqn_masslessmode}).

To obtain a corresponding prediction for closed strings, we use
the Dedekind eta function's modular invariance under 
$q=\exp\{-\pi\tau/r\} \to \tilde{q}=\exp\{-4\pi r/\tau\}$. This 
allows us to rewrite the expression for $Z_{cyl}$ in 
eqn(\ref{eqn_Zgaussian}) as a sum of exponentials in $r$ rather than 
in $\tau$. However this sum turns out not to be precisely of the form 
shown in eqn(\ref{eqn_Zss_closed}), where the eigenstates fall into
subsets that are related by Lorentz invariance (so that, for example, 
their energies satisfy the usual energy-momentum dispersion relation).  
Thus a Gaussian $S_{eff}$ does not encode 
the open-closed string duality exactly and cannot be considered
as a candidate for an exact description of strings on a cylinder. 
However if, instead, we view the Gaussian $S_{eff}$ as an approximation
for long closed strings, we can expand the momentum dependence
in eqn(\ref{eqn_Zss_closed}) in powers of $p/\sigma\tau$, and then 
match to this approximate form. We thus obtain a prediction for
the closed string energy levels,
\begin{equation}
\tilde{E}_n(\tau) = \sigma \tau + \frac{4\pi}{\tau} 
\left\{n-\frac{1}{24}(D-2)\right\} + O\left(\frac{1}{\tau^2}\right)
 \qquad : \quad {\mathrm{Gaussian}}
\label{eqn_tildeEngaussian}
\end{equation}
and also for the overlaps $c_n$
\cite{LW04}.

The $O(1/r)$ correction to the leading linear term in $E_0(r)$ in 
eqn(\ref{eqn_Engaussian}) is the famous L\"uscher correction
\cite{LSW} 
for a flux tube with ends fixed on static sources. Physically it
arises from the regularised sum of the zero-point energies of all
the quantised oscillators on the string. It depends only on
the long wavelength massless modes and so is universal: any bosonic 
string theory in which the only massless modes are the transverse
oscillations will have precisely this leading correction. 
The same applies to the $O(1/\tau)$ correction to the leading 
linear term in $\tilde{E}_0(\tau)$ in eqn(\ref{eqn_tildeEngaussian}).

Although the above results for $E_n(r)$ are obtained in the Gaussian 
approximation to $S_{eff}[h]$, this approximation becomes exact as 
$r\to\infty$, and these predictions for the leading $O(1/r)$ correction 
are exact and universal. And similarly for $\tilde{E}_n(\tau)$ as 
$\tau\to\infty$.

\subsection{Nambu-Goto action}
\label{subsection_NG_EST}

A string theory that is simple enough to have a calculable energy spectrum 
\cite{Arvis}
is the free-string Nambu-Goto model in flat space time: 
\begin{equation}
Z = \int dS e^{-\kappa A[S]}
\label{eqn_ZNG}
\end{equation}
where we integrate over all surfaces, with the action proportional
to the invariant area. A free string theory is not necessarily
unrealistic as an effective string theory: after all, we recall 
that at $N=\infty$ confining flux tubes do not interact. Moreover we
have found that the flux tube spectrum in $D=2+1$ is remarkably close
to the Nambu-Goto spectrum
\cite{AABBMT_k1d3}. 
For these reasons we shall use this model as our main point of comparison. 

Consider a string winding once around the $x$-torus. Working in static 
gauge, perform a Fourier decomposition of $h(x,t)$. Upon quantisation 
the coefficients become creation operators for massless `phonons' 
on the string with momenta $\pm 2\pi k/l$ 
along the string and energy $2\pi k/l$. (The $k=0$ mode is not included 
since it corresponds to a shift to a different transverse position 
of the whole string i.e. to another vacuum of the spontaneously
broken symmetry.) There are 2 transverse directions, so $h$ is a 
2-component vector and the phonons carry angular momentum of $\pm 1$. 
We call positive momenta left-moving (L) and the
negative ones right-moving (R). Let $n^{\pm}_{L(R)}(k)$ be the number 
of left(right) moving phonons of momentum $2\pi k/l$ and angular
momentum  $\pm 1$. Define the total energy and momentum of the 
left(right) moving phonons as $2\pi N_{L(R)}/l$, then:
\begin{equation}
N_L = \sum_k \sum_{n_L(k)} k(n^+_L(k)+ n^-_L(k)), \qquad
N_R = \sum_k \sum_{n_R(k)} k(n^+_R(k)+ n^-_R(k)),
\label{eqn_NLR}
\end{equation}
If $p=2\pi q/l$ is the total longitudinal momentum of the string
then, since the phonons provide that momentum, we must have
\begin{equation}
N_L - N_R = q.
\label{eqn_NLRmom}
\end{equation}
The angular momentum around the string is 
\begin{equation}
J = \sum_{k,n_L(k),n_R(k)} n^+_L(k) + n^+_R(k) - n^-_L(k) - n^-_R(k)
\label{eqn_NLRspin}
\end{equation}

One can now write down the expression for the energy levels of the 
Nambu-Goto string:
\begin{equation}
E^2_{N_L,N_R}(q,l)
=
(\sigma l)^2 
+
8\pi\sigma \left(\frac{N_L+N_R}{2}-\frac{D-2}{24}\right)
+
\left(\frac{2\pi q}{l}\right)^2
\label{eqn_EnNG}
\end{equation}
where the degeneracies corresponding to particular values of $N_L$ and $N_R$ 
will depend on the number of different states that can be formed from 
combinations of values of $n^\pm_L$ and $n^\pm_R$ that give the same
values of $N_L + N_R$ and $q$ in eqn(\ref{eqn_NLR}). In discussing the 
states, we shall often refer to the left and right moving phonon creation 
operators of (absolute) momentum $2\pi k/l$  and angular momentum 
$\pm 1$ as $a^\pm_k$ and $a^\pm_{-k}$ respectively, and the unexcited 
string ground state as $|0\rangle$.

Let us specialise, for simplicity, to $q=0$, i.e. $N_L=N_R=n$, and make some 
general comments.\\
$\bullet$ The energy $E_n(l)$  can be expanded for large $l$ in 
inverse powers of $1/\sigma l^2$:
\begin{eqnarray}
E_n(l)
& = &
\sigma l \left(1 + 
\frac{8\pi}{\sigma l^2} 
\left(n -\frac{D-2}{24}\right)\right)^\frac{1}{2} \nonumber \\
& = &
\sigma l + \frac{4\pi}{l} 
\left(n -\frac{D-2}{24}\right) + O\left( \frac{1}{\sigma l^3} \right)
\label{eqn_EnNGexpansion}
\end{eqnarray}
We note that the first correction to the linear piece is exactly as in 
eqn(\ref{eqn_tildeEngaussian}) - as it should be if it is `universal'.\\
$\bullet$ The ground state energy becomes tachyonic at small $l$:
\begin{equation}
E^2_0(l)
= (\sigma l)^2 - \frac{\pi\sigma (D-2)}{3}
< 0 \qquad : \quad \sigma l^2 < \frac{\pi (D-2)}{3}.
\label{eqn_E0NG}
\end{equation}
One can regard this as the Hagedorn/deconfining transition in the Nambu-Goto 
model. We note that the length at which this transition occurs, 
$l_H\surd\sigma = \sqrt{2\pi/3} \simeq 1.45$, is not accessible in 
pure gauge theories at large $N$, because of a first-order `deconfining' 
transition that occurs at $l_c\surd\sigma = \surd\sigma/T_c \simeq 1.65$
\cite{BLMTUW_Tc}. \\ 
$\bullet$ The expansion of the square root expression for the energy $E_n$ 
in eqn(\ref{eqn_EnNGexpansion}) only converges for $\sigma l^2 > 8\pi n$ 
(ignoring the negligible $D-2$ term). So the more excited is the state,
the larger is the value of $l$ to which we must go to be able to use 
such an expansion. The derivative expansion of the action is 
not uniform in frequency: it is formal. \\
$\bullet$ One can show (see Appendix C of
\cite{LW04})
that the Nambu-Goto model satisfies open-closed duality exactly. This is 
in contrast to the Gaussian string action. We assume (although we have 
been unable to locate the proof in the literature) that it also satisfies 
the closed-closed duality associated with a world sheet on a 2-torus.
Thus, if these dualities are our only 
constraints, the Nambu-Goto model is a viable candidate for providing a 
consistent effective string action. An important consequence of this 
is that where imposing these constraints allows us to completely 
fix the expansion coefficients of $E_n(l)$ up to some order in $1/l$, 
these coefficients will have to be precisely the same as those obtained 
by expanding the Nambu-Goto expression in eqn(\ref{eqn_EnNGexpansion})
to that order. (Or the corresponding expression for strings with fixed ends.)

Finally we note that a comparison between our lattice results for closed 
flux-tubes and the Nambu-Goto spectrum of eqn(\ref{eqn_EnNG}) can be 
seen as analogous to comparing glueball spectra to Regge trajectories 
\cite{hmmt_regge}.
Specifically, if we focus on excited states for which the second term 
in eqn(\ref{eqn_EnNG}) is dominant, and if for simplicity we take $q=0$, 
or $N_R=N_L\equiv N$, then we can write
\begin{equation}
E^2 \simeq 8\pi\sigma N.
\label{eqn_NGregge}
\end{equation}
This linear relation between the energy squared and the quantum number of the 
state is very reminiscent of the $M^2 \sim J+n $ Regge relation that
one might expect for glueballs (the `Pomeron' trajectory) and 
indeed one immediately sees from eqns(\ref{eqn_NLR}-\ref{eqn_NLRspin}) that 
the maximum value of $J$ increases linearly with the value of $N$.

\subsection{Recent  theoretical progress}
\label{subsection_recent_EST}

The seminal work in analysing flux tubes in a string
description in static gauge 
\cite{LSW}
(reviewed above) and the later more general 
Polchinski-Strominger approach using conformal gauge
\cite{PS}
(not reviewed here) led to an understanding of the universality
of the leading $O(1/l)$ L\"uscher correction to the linear growth
of the flux tube energy. Until recently there was, however, very
little further analytic progress along these lines. 

The situation changed in 2004 when major progress took place
within both approaches.\\ 
1) In
\cite{LW04}
it was shown that the open-closed duality (discussed above) could 
be used to provide useful constraints on the  higher order terms 
in the expansion of the effective string action. In particular
it was shown that in $D=2+1$ the next, $O(1/l^3)$, term is also
universal and that the coefficient is precisely what you get by 
expanding the Nambu-Goto square-root expression to that order
(as must be the case). In $D=3+1$
the coefficient is not fixed but there is a relationship predicted
between the coefficients of the two terms in the effective action
that contribute at that order. \\
2) Simultaneously, the next $O(1/l^3)$ term was calculated within the 
Polchinski-Strominger framework in 
\cite{JD}
(and also, later, in
\cite{DM06}).
The same conclusion was reached as in 
\cite{LW04}
for $D=2+1$, but a stronger conclusion was obtained in $D=3+1$, where 
the $O(1/l^3)$ term in the action was shown to be universal (and equal 
to the value in the Nambu-Goto expansion).

In 2009 further substantial progress was achieved 
\cite{AHEK09}
using the static gauge approach and including the new constraints 
that arise from the less obvious closed-closed (torus) duality.
In $D=2+1$ the terms up to and including $O(1/l^5)$ have been shown to 
be universal (and equal to the Nambu-Goto values)
\cite{AHEK09}. 
In $D=3+1$ the  $O(1/l^3)$ contribution has been shown to be universal
\cite{AHEK09},
in agreement with 
\cite{JD},
and the $O(1/l^5)$ contribution to the string ground state (and partition 
function) has also been shown to be universal. 
Moreover, during the writing of this paper, some papers have appeared
\cite{DM09}
extending the Drummond-Polchinski-Strominger approach
\cite{PS,JD}
and claiming that the terms up to  $O(1/l^5)$ are universal in all the 
eigenstates.

It is now clear that the effective string theory of long flux tubes
is that of the Nambu-Goto free string theory to quite high order
in the derivative expansion.

\section{Lattice methodology}
\label{section_lattice}

In Section~\ref{subsection_setup_lat} we outline the lattice framework
of our calculations. In Section~\ref{subsection_energies_lat} we discuss
in some detail how we extract the energies of (excited) flux tubes. For
the reader who wishes to skip this sub-section, 
the important message for interpreting our subsequent results is 
that the larger the energy of the flux tube, whether because it is longer 
or because it is more highly excited, the larger are the errors.
(Particularly the systematic errors that can be hard to estimate.)
In  Section~\ref{subsection_quanta} we describe our labelling of the 
quantum numbers of the flux tubes and tabulate the corresponding Nambu-Goto 
description -- all of which is important for understanding our later results. 
We present the details of our lattice operators in 
Section~\ref{subsection_op_lat}, which the general reader may again
wish to omit. Finally in Section~\ref{subsection_previouslat} we provide
a very brief and incomplete sketch of relevant earlier work, to provide some
context for our calculations.

\subsection{Setup}
\label{subsection_setup_lat}

Our numerical calculations are entirely conventional. In order
to remove the troublesome phase factor in the Minkowski path
integral, we perform our calculations in the Euclidean theory,
and to make the problem finite we work on a hypercubic lattice
on a finite hypertorus. Schematically,
\begin{equation}
\int \prod_{x\in M_4} d\phi(x) e^{iS[\phi]}
\to
\int \prod_{x\in R^4} d\phi(x) e^{-S_E[\phi]}
\to
\int \prod_{n\in T^4} d\phi_L(n) e^{-S_L[\phi_L]}
\end{equation}
where we express the discrete space-time points as $x=na$ or
$x_\mu=n_\mu a$, where $\{n_\mu\}$ is a 
$D$-tuple of integers, $a$ is the lattice spacing, and $\phi_L$ is a 
dimensionless lattice field variable, with action $S_L$, chosen so that
\begin{eqnarray}
\phi_L(x) & \stackrel{a\to 0}{\longrightarrow} &  a^{-dim(\phi)}\phi(x) 
\nonumber \\
S_L  & \stackrel{a\to 0}{\longrightarrow} & S_E 
\end{eqnarray}
where $dim(\phi)$ is the length dimension of the field $\phi$.
Because the theory has a finite mass gap, the leading finite size 
corrections are exponentially small, and because it is asymptotically
free one can show that the leading finite lattice spacing corrections
for the action we shall use are $O(a^2)$.

The gauge degrees of freedom are  SU($N$) group elements, $U_l$, 
that are assigned to the forward going links, $l$, of the lattice. 
(With $U^\dagger_l$ for backward-going links.) For a link
that joins the site $x$ to the site $x+a \hat{\mu}$, the link
matrix $U_\mu(x)$ will transform as
$U_\mu(x) \longrightarrow V(x)U_\mu(x)V^\dagger(x+a \hat{\mu})$
under a gauge transformation $V(x)$. To construct a gauge-invariant
action we note that the trace of the product of link-matrices around 
any closed path $c$, $ \mathrm{Tr} \prod_{l\in c} U_l$, is gauge 
invariant. So the simplest choice for the lattice gauge action 
is to use the path that is an elementary square on the lattice, called
a plaquette:
\begin{equation}
S = \sum_p \left\{1 - \frac{1}{N} \mathrm{ReTr}U_p\right\}
\label{eqn_Splaq}
\end{equation}
where $U_p$ is the path-ordered product of link matrices around  
the plaquette $p$.  The $\sum_p$ ensures that the action is
translation and rotation invariant. Taking the real part of
the trace ensures that it has $C=P=+$. So our lattice path
integral is  
\begin{equation}
Z = \int \prod_{l} dU_l e^{-\beta S}
\end{equation}
where $\beta$ is a constant whose value will determine the
lattice spacing $a$. Since
\begin{equation}
\int \prod_{l} dU_l e^{-\beta S} 
\stackrel{a\to 0}{\propto}
\int \prod_{x,\mu} dA_\mu(x) 
e^{-\frac{4}{g^2} \int d^4 x Tr F_{\mu\nu} F_{\mu\nu} } 
\end{equation}
we see that 
$\beta = 2N/g^2 \stackrel{a\to 0}{\longrightarrow} \infty$.
In  $D=2+1$ where $g^2$ has dimensions of mass, the dimensionless 
bare lattice coupling is $ag^2$ and $\beta = 2N/ag^2$. Since
a smooth large $N$ limit requires keeping $g^2N$ fixed, this
means that we must vary $\beta \propto N^2$ in order to keep
$a$ approximately fixed as we vary $N$. 

To calculate the expectation value of some functional $\Phi[U]$ 
of the gauge fields, 
we generate a set of $n_g$ gauge fields $\{U^I\}; I=1,...,n_g$ distributed 
with the Boltzmann-like action factor included in the probability 
measure, i.e. $ dP \propto \prod_l dU_l \exp\{-\beta S[U]\}$.
We use a Cabbibo-Marinari algorithm applied to the $N(N-1)/2$
SU(2) subgroups of the SU($N$) matrix, with a mix of standard heat-bath 
and over-relaxation steps. We can now calculate the expectation value of 
some functional $\Phi[U]$ of the gauge fields as follows: 
\begin{equation}
\langle \Phi \rangle
=
\frac{1}{Z}
\int \prod_{l} dU_l  \Phi[U] e^{-\beta S}
=
\frac{1}{n_g} \sum^{n_g}_{I=1} \Phi[U^I] + 
O\left( \frac{1}{\surd n_g}\right)
\label{eqn_avPhi}
\end{equation}
where the last term is the statistical error.

\subsection{Calculating energies of closed flux tubes}
\label{subsection_energies_lat}

We begin by recalling the standard decomposition of a Euclidean correlator 
of some operator $\phi(t)$ in terms of the energy eigenstates of the
Hamiltonian $H$:
\begin{eqnarray}
\langle \phi^\dagger(t=an_t)\phi(0) \rangle
& = &
\langle \phi^\dagger e^{-Han_t} \phi \rangle
=
\sum_i |c_i|^2 e^{-aE_in_t} \nonumber \\
& \stackrel{t\to \infty}{=} & 
|c_0|^2 e^{-aE_0n_t}
\label{eqn_cortomass}
\end{eqnarray}
where the energies are ordered, $E_{i+1}\geq E_i$, with  $E_0$ that of 
the ground state. The only states that contribute are those that have
overlaps
$c_j = \langle vac | \phi^\dagger | j \rangle \neq 0$, so we need to
match the quantum numbers of the operator $\phi$ to those of the state 
we are interested in, which here is a flux tube winding around the
$x$-torus, of length  $l=l_x=La$.

The simplest such operator is the elementary Polyakov loop:
\begin{equation}
l_p(n_t) =  \sum_{n_y,n_z} \mathrm{Tr} 
\left\{\prod^L_{n_x=1} U_x(n_x,n_y,n_z,n_t)\right\} 
\label{eqn_poly}
\end{equation}
where we take a product of the link matrices in the 
$x$-direction once around the $x$-torus, and the sum over translations
projects onto zero transverse momentum $(p_y,p_z) = (0,0)$. The
operator is invariant under translations in $x$ and therefore
has zero longitudinal momentum $p_x =0$. It is invariant under rotations 
about its axis and so has $J=0$. It is clearly invariant under parity
(accompanied where necessary by charge conjugation $C$ to reverse
the flux). Clearly we will need to use deformed versions of the Polyakov
loop to access non-trivial quantum numbers for the flux tube (as 
discussed in detail later on). 

An operator like the Polyakov loop which winds once around a torus
has zero overlap onto states such as glueballs that are localised 
and which are described by operators constructed around closed 
loops that can be continuously contracted to a point. This orthogonality 
is enforced by a center symmetry that corresponds to SU($N$)
gauge transformations that are periodic - on the torus - up to an
element $z\in Z_N$. These leave any contractible loop, $l_c$, invariant
but not ones with unit winding, $l_p\to zl_p$. Thus
$\langle l^\dagger_c l_p\rangle = z \langle l^\dagger_c l_p\rangle = 0$
but only as long as this center symmetry is not spontaneously broken.
As we decrease $l$ there is a critical length $l_c=1/T_c$, where $T_c$ 
is the usual deconfining temperature, below which this center symmetry 
is spontaneously broken and we are in a `deconfining' phase in which we 
no longer have confining flux tubes around the $x$-torus. (Although 
large Wilson loops orthogonal to $x$ continue to display a
confining area law.) In this paper we always restrict $l$ to be $>l_c$.     

So, having chosen a suitable operator $\phi$, we calculate its correlator
using eqn(\ref{eqn_avPhi}) with $\Phi=\phi^\dagger(t)\phi(0)$. This 
will produce an estimate of $\langle \phi^\dagger(t)\phi(0) \rangle$ 
with a finite statistical error. To extract a value of $aE_0$ using 
eqn(\ref{eqn_cortomass}), we need to have significant evidence for 
the exponential behaviour $\propto e^{-a E_0 n_t}$, over some range of 
$t=an_t$ and this range needs to begin at small enough $n_t$ that the 
decreasing exponential is still clearly visible above the statistical
errors. (In a pure gauge theory one can show that the error tends to 
a non-zero value at large $t$, so that the error/signal ratio increases
exponentially with $t$.) This is obviously harder to achieve for larger 
$E_0$, so the corresponding systematic error will be larger for heavier 
states. However, even for the lightest state we need the normalised
overlap to be close to unity, $|c_0|^2 \simeq 1$, to obtain an accurate 
energy estimate: our operator needs to be a good wavefunctional 
for the state whose energy we are interested in, so that its
correlator is dominated by this state even at small $n_t$. For the
ground state of the flux tube this can be achieved by using the
Polyakov loop operators in eqn(\ref{eqn_poly}), but with `smeared' 
\cite{smear}
and `blocked'
\cite{block}
link matrices replacing the elementary ones. These produce winding 
operators that are smooth on various transverse size scales including
ones that are comparable to that of the physical flux tube, and, not
surprisingly, this works well for the ground state which presumably 
has a smooth wavefunctional. Starting with this basis of operators,
$\{\phi_i; i=1,...,n\}$, we consider the vector space $V_\phi$ 
generated by them, calculating all the correlation functions 
$\langle \phi^\dagger_i(t)\phi_j(0)\rangle$, and determine 
the linear combination that is the `best' operator $\psi_0$ by a 
variational criterion
\begin{equation}
\langle {\psi_0}^\dagger(t_0)\psi_0(0)\rangle 
=
\max_{\phi\in V_\phi}
\langle \phi^\dagger(t_0) \phi(0)\rangle 
=
\max_{\phi\in V_\phi}
\langle \phi^\dagger e^{-Ht_0} \phi\rangle 
\end{equation}
where $t_0$ is some convenient value of $t$. Then $\psi_0$ is our best 
variational estimate for the true eigenfunctional of the ground state. 
We now use the correlator $\langle {\psi_0}^\dagger(t)\psi_0(0)\rangle$ 
in eqn(\ref{eqn_cortomass}) to obtain our best estimate of the ground 
state energy. This generalises in an obvious 
way to calculating excited state energies. One constructs from $V_\phi$
the vector space orthogonal to $\psi_0$, repeats the above within this
reduced vector space, and obtains  $\psi_1$ which  is our best variational 
estimate for the true eigenfunctional of the first excited state. 
And so on. 

For the ground state the main systematic error is to overestimate $aE_0$
by extracting it at a value of $t$ where there is still a significant
contribution from excited states. This becomes an increasing problem
as we increase $l$ and hence $aE_0(l)$, and so decrease the range of 
$t$ where the statistical errors are small compared to the `signal'. 
At the same time the gap to the excited states decreases $\propto 1/l$, 
increasing
their relative contribution to the correlator. This is the main obstacle 
to accurate calculations of flux tube energies for very large $l$
and can obviously undermine an attempt to calculate corrections to
the large-$l$ behaviour of $E_0(l)$. We therefore restrict our 
calculations of the ground state energy to values of $l$ where we estimate 
this systematic error to be smaller than our quoted statistical errors.

For excited states this systematic error is less easily avoided
because  $a E_i(l)$ is often quite large for all values of $l$.
One therefore needs to be aware of this in any comparison between
our results and the predictions of an effective string model in
this paper. In addition there is now an additional error that arises
from the fact that our best variational wavefunctional, $\psi_i$,
may contain some admixture of lighter eigenstates. At sufficiently
large $t$ these will dominate and give an energy that is lower
than $E_i$. Note, however, that if the state is the lightest one for some
given quantum numbers, then this will not occur and we will often
refer to these states as `ground states', at least when there is no risk
of confusion with the absolute ground state. Such ground states will
be of particular interest to us because their calculation is typically 
much more reliable. 

Both the systematic errors described above can be controlled if
$\psi_i$ is a very good wave-functional for the corresponding excited 
state. In that case there will be a useful range $t\in[t_1,t_2]$
where the desired state dominates because the small admixture of  
higher excited states has died away by a small value of $t_1$ and 
any small admixture
of lower states only becomes significant beyond $t_2$. That we are
able to obtain usefully large overlaps onto quite a large number of 
states with different quantum numbers, is the main technical
feature of the present work. It requires the construction of a large
basis of winding operators, a strategy that proved successful in our 
earlier calculations of the closed flux tube spectrum in $D=2+1$
\cite{AABBMT_k1d3}
and which we describe in detail in Section~\ref{subsection_op_lat}. 
The detailed construction is partly motivated by the range of quantum 
numbers we wish to encode, and so we turn to this now.

\subsection{Quantum numbers}
\label{subsection_quanta}

Consider a confining flux tube, with the flux in the fundamental 
representation and let it wind around the $x$-torus. Its quantum numbers 
can be chosen as follows. \\
$\bullet$ The length $l=aL$ of the $x$-torus around which
the flux tube winds. \\
$\bullet$ The number $w$ of times (modulo $N$) that the flux tube winds 
around this 
torus. In this paper we shall only consider $w=1$, except for
the discussion of $k$-strings in Section~\ref{subsection_k2}. \\
$\bullet$ The momentum along the flux tube which is quantised by
periodicity to be  $p=2\pi q/l$ where $q$ is an integer. Since the
string ground state is (presumably) invariant under longitudinal
translations,  $p\neq 0$ will require some non-trivial excitation
along the string with an additional excitation energy. By contrast, 
transverse momentum $p_\perp$ should simply lead to the usual 
dispersion relation and so is uninteresting to us, and we only 
consider $p_\perp=0$. Thus $p$ will always refer to the longitudinal
momentum although we shall sometimes label it by $p_\shortparallel$, 
where this increases clarity.  Since the energy does not depend 
on the sign of $q$, we will restrict ourselves to $q\ge 0$.\\
$\bullet$ The flux tube can rotate about its symmetry axis and can
have a corresponding angular momentum of $J=0, \pm 1, ...$ along that axis.\\
$\bullet$ There is a `transverse parity', $P_t$, in the plane 
that is transverse to the symmetry axis and which is just like parity 
in $D=2+1$: $P_t : (y,z) \to (y,-z)$. It is plausible that the 
absolute ground state, with energy $E_0(l)$, is invariant under 
reflections, and so will have $P_t=+$. Thus the lightest $P_t=-$ state 
requires a non-trivial excitation of the flux tube and the difference
between the  $P_t=\pm$ ground state energies will measure the
energy of that excitation. This parity does not commute with the
angular momentum projection since under $P_t : J\to -J$. So when we 
choose to use this parity to label our states, we have to use $|J|$ 
rather than $J$ as the spin label (although we shall continue to 
refer to  it as $J$ where there is no ambiguity). States with
$|J|\neq 0$ come in degenerate $P_t=\pm$ pairs.
\\ 
$\bullet$ There is also a reflection symmetry across the string midpoint, 
which defines a corresponding `longitudinal parity' that we call $P_l$; 
it reverses the momenta of the individual phonons. (The reversal of the 
direction of flux is compensated for by the simultaneous application of 
charge conjugation, $C$, which is to be understood from now on.) Since 
$P_l$ reverses the momentum it is only a useful quantum number if $p=0$.
So when $p\neq 0$ we label states by $P_t$, $|J|$  and $|p|$.\\
$\bullet$ Under charge conjugation, $C$, the direction of the flux is 
reversed. Thus a flux tube has zero overlap onto its charge-conjugated 
homologue (the center symmetry argument again) and so linear combinations 
of definite $C$ are trivially degenerate. (Except for SU(2), where $C$ is 
trivial, and for $k$-strings when $k=N/2$.) Thus $C$ is, by itself, 
not interesting in general.

Our comparison to the effective string theory will be in static gauge
(as described earlier) and so it is useful to discuss these
symmetries and quantum numbers in terms of the transverse deformation
of the string, $\vec{h}(x)$, and the associated phonon creation
operators, $a^\pm_k$, that arise from quantising the coefficients
in the Fourier decomposition of  $\vec{h}(x)$, and which carry momentum
$2\pi k/l$ and spin (around the flux-tube axis) of $\pm 1$. 
The longitudinal and angular momenta of the flux tube are simply the 
sum of all the phonon momenta and spins. Under $P_t$ we have 
$a^+_k \to a^-_k$ \footnote{In contrast to $D=2+1$ where the $P_t$
parity transforms the single $a_k$ operator as  $a_k \to -a_k$.} 
while under $P_l$ we have $a^\pm_k \to a^\pm_{-k}$. For later
reference we give in Table~\ref{table_NGstates} the explicit phonon 
content of the lowest-lying states in the free string Nambu-Goto model. 
\begin{table}[htp] 
\begin{center}
\centering{\scalebox{0.85}{
\begin{tabular}{c|c|c|c|c} \otoprule %\otoprule
$N_L,N_R$ & $|J|$ & $P_t$ & $P_l$  & \ \ String State \ \  \\ \otoprule %\otoprule
 $N_L=N_R=0$ & $0$ &  $+$ & $+$ & $| 0 \rangle$ \\ \midrule %\midrule 
 $N_L=1,N_R=0$ & $1$ &  $\pm$ & $$ & $\left( a^{+}_{1} \pm a^{-}_{1} \right) | 0 \rangle$ \\ \midrule %\midrule 
\multirow{4}*{$N_L=N_R=1$} & $0$ &  $+$ & $+$ & $\left( a^{+}_{1} a^{-}_{-1}+a^{-}_{1} a^{+}_{-1} \right) | 0 \rangle$ \\
& $0$ &  $-$ & $-$ & $\left( a^{+}_{1} a^{-}_{-1}-a^{-}_{1} a^{+}_{-1} \right) | 0 \rangle$ \\
& $2$ &  $+$ & $+$ & $\left( a^{+}_{1} a^{+}_{-1}+a^{-}_{1} a^{-}_{-1} \right) | 0 \rangle$ \\
& $2$ &  $-$ & $+$ & $\left( a^{+}_{1} a^{+}_{-1}-a^{-}_{1} a^{-}_{-1} \right) | 0 \rangle$ \\\midrule %\midrule
\multirow{4}*{$N_L=2,N_R=0$} & $0$ &  $+$ & $$ & $a^{+}_{1} {a}^{-}_{1} | 0 \rangle$ \\
& $1$ &  $\pm$ & $$ & $\left( a^{+}_{2} \pm  a^{-}_{2}  \right) | 0 \rangle$ \\
& $2$ &  $+$ & $$ & $\left( a^{+}_{1} {a}^{+}_{1}+a^{-}_{1}{a}^{-}_{1} \right) | 0 \rangle$ \\
& $2$ &  $-$ & $$ & $\left( a^{+}_{1} {a}^{+}_{1}-a^{-}_{1}{a}^{-}_{1} \right) | 0 \rangle$ \\\midrule %\midrule
\multirow{7}*{$N_L=2,N_R=1$} & $0$ &  $+$ & $$ & $\left( a^{+}_{2} a^{-}_{-1}+a^{-}_{2} a^{+}_{-1} \right) | 0 \rangle$ \\
& $0$ &  $-$ & $$ & $\left( a^{+}_{2} a^{-}_{-1}-a^{-}_{2} a^{+}_{-1} \right) | 0 \rangle$ \\
& $1$ &  $\pm$ & $$ & $\left( a^{+}_{1}a^{+}_{1} a^{-}_{-1} \pm  a^{-}_{1}a^{-}_{1}a^{+}_{-1} \right) | 0 \rangle$ \\
& $1$ &  $\pm$ & $$ & $\left( a^{+}_{1}a^{-}_{1} a^{-}_{-1} \pm a^{-}_{1}a^{+}_{1} a^{+}_{-1}  \right) | 0 \rangle$ \\
& $2$ &  $+$ & $$ & $\left( a^{+}_{2} a^{+}_{-1}+a^{-}_{2} a^{-}_{-1} \right) | 0 \rangle$ \\
& $2$ &  $-$ & $$ & $\left( a^{+}_{2} a^{+}_{-1}-a^{-}_{2} a^{-}_{-1} \right) | 0 \rangle$ \\
& $3$ &  $\pm$ & $$ & $\left( a^{+}_{1}a^{+}_{1} a^{+}_{-1} \pm a^{-}_{1}a^{-}_{1} a^{-}_{-1}  \right) | 0 \rangle$ \\\midrule %\midrule
\multirow{17}*{$N_L=2,N_R=2$} & $0$ &  $+$ & $+$ & $\left( a^{+}_{2} a^{-}_{-2}+a^{-}_{2} a^{+}_{-2} \right) | 0 \rangle$ \\
& $0$ &  $-$ & $-$ & $\left( a^{+}_{2} a^{-}_{-2}-a^{-}_{2} a^{+}_{-2} \right) | 0 \rangle$ \\
& $0$ &  $+$ & $+$ & $\left( a^{+}_{1}a^{+}_{1} a^{-}_{-1} a^{-}_{-1} +a^{-}_{1}  a^{-}_{1} a^{+}_{-1} a^{+}_{-1} \right) | 0 \rangle$ \\
& $0$ &  $-$ & $-$ & $\left( a^{+}_{1}a^{+}_{1} a^{-}_{-1} a^{-}_{-1} -a^{-}_{1}  a^{-}_{1} a^{+}_{-1} a^{+}_{-1} \right) | 0 \rangle$ \\
& $0$ &  $+$ & $+$ & $ a^{+}_{1}a^{-}_{1} a^{+}_{-1} a^{-}_{-1} | 0 \rangle$ \\
& $1$ &  $\pm$ & $+$ & $\big[ ( a^{+}_{1}a^{+}_{1} a^{-}_{-2}+ a^{-}_{2}a^{+}_{-1} a^{+}_{-1}) \pm  ( a^{-}_{1}a^{-}_{1} a^{+}_{-2}+ a^{-}_{2}a^{+}_{-1} a^{+}_{-1}) \big] | 0 \rangle$ \\
& $1$ &  $\pm$ & $-$ & $\big[ ( a^{+}_{1}a^{+}_{1} a^{-}_{-2}- a^{-}_{2}a^{+}_{-1} a^{+}_{-1}) \pm  ( a^{-}_{1}a^{-}_{1} a^{+}_{-2}- a^{-}_{2}a^{+}_{-1} a^{+}_{-1}) \big] | 0 \rangle$ \\
& $1$ &  $\pm$ & $+$ & $\big[ ( a^{+}_{1}a^{-}_{1} a^{+}_{-2}+ a^{+}_{2}a^{-}_{-1} a^{+}_{-1}) \pm  ( a^{-}_{1}a^{+}_{1} a^{-}_{-2}+ a^{-}_{2}a^{+}_{-1} a^{-}_{-1}) \big] | 0 \rangle$ \\ 
& $1$ &  $\pm$ & $-$ & $\big[ ( a^{+}_{1}a^{-}_{1} a^{+}_{-2}- a^{+}_{2}a^{-}_{-1} a^{+}_{-1}) \pm  ( a^{-}_{1}a^{+}_{1} a^{-}_{-2}- a^{-}_{2}a^{+}_{-1} a^{-}_{-1}) \big] | 0 \rangle$ \\ 
& $2$ &  $+$ & $+$ & $\left( a^{+}_{2} a^{+}_{-2}+a^{-}_{2} a^{-}_{-2} \right) | 0 \rangle$ \\
& $2$ &  $-$ & $+$ & $\left( a^{+}_{2} a^{+}_{-2}-a^{-}_{2} a^{-}_{-2} \right) | 0 \rangle$ \\
& $2$ &  $+$ & $+$ &  $\left[   \left( a^{+}_{1}a^{+}_{1} a^{+}_{-1} a^{-}_{-1} + a^{-}_{1}a^{-}_{1} a^{-}_{-1} a^{+}_{-1} \right)+\left( a^{+}_{1}a^{-}_{1} a^{-}_{-1} a^{-}_{-1} + a^{-}_{1}a^{+}_{1} a^{+}_{-1} a^{+}_{-1} \right) \right] | 0 \rangle$ \\
& $2$ &  $+$ & $-$ &  $\left[   \left( a^{+}_{1}a^{+}_{1} a^{+}_{-1} a^{-}_{-1} + a^{-}_{1}a^{-}_{1} a^{-}_{-1} a^{+}_{-1} \right)-\left( a^{+}_{1}a^{-}_{1} a^{-}_{-1} a^{-}_{-1} + a^{-}_{1}a^{+}_{1} a^{+}_{-1} a^{+}_{-1} \right) \right] | 0 \rangle$ \\
& $2$ &  $-$ & $+$ &  $\left[   \left( a^{+}_{1}a^{+}_{1} a^{+}_{-1} a^{-}_{-1} - a^{-}_{1}a^{-}_{1} a^{-}_{-1} a^{+}_{-1} \right)+\left( a^{+}_{1}a^{-}_{1} a^{-}_{-1} a^{-}_{-1} - a^{-}_{1}a^{+}_{1} a^{+}_{-1} a^{+}_{-1} \right) \right] | 0 \rangle$ \\
& $2$ &  $-$ & $-$ &  $\left[   \left( a^{+}_{1}a^{+}_{1} a^{+}_{-1} a^{-}_{-1} - a^{-}_{1}a^{-}_{1} a^{-}_{-1} a^{+}_{-1} \right)-\left( a^{+}_{1}a^{-}_{1} a^{-}_{-1} a^{-}_{-1} - a^{-}_{1}a^{+}_{1} a^{+}_{-1} a^{+}_{-1} \right) \right] | 0 \rangle$ \\
%& $3$ &  $\varnothing$ & $\varnothing$ & $\left( a^{+}_{1}a^{+}_{1} a^{+}_{-1}+ a^{-}_{1}a^{-}_{1} a^{-}_{-1}, a^{+}_{1}a^{+}_{1} a^{+}_{-1} - a^{-}_{1}a^{-}_{1} a^{-}_{-1}  \right) | 0 \rangle$ \\
& $3$ &  $\pm$ & $+$ & $\big[ ( a^{+}_{1}a^{+}_{1} a^{+}_{-2}+ a^{+}_{2}a^{+}_{-1} a^{+}_{-1}) \pm  ( a^{-}_{1}a^{-}_{1} a^{-}_{-2}+ a^{-}_{2}a^{-}_{-1} a^{-}_{-1}) \big] | 0 \rangle$ \\
& $3$ &  $\pm$ & $-$ & $\big[ ( a^{+}_{1}a^{+}_{1} a^{+}_{-2}- a^{+}_{2}a^{+}_{-1} a^{+}_{-1}) \pm  ( a^{-}_{1}a^{-}_{1} a^{-}_{-2}- a^{-}_{2}a^{-}_{-1} a^{-}_{-1}) \big] | 0 \rangle$ \\
& $4$ &  $+$ & $+$ & $\left( a^{+}_{1}a^{+}_{1} a^{+}_{-1} a^{+}_{-1} +a^{-}_{1}  a^{-}_{1} a^{-}_{-1} a^{-}_{-1} \right) | 0 \rangle$ \\
& $4$ &  $-$ & $-$ & $\left( a^{+}_{1}a^{+}_{1} a^{+}_{-1} a^{+}_{-1} -a^{-}_{1}  a^{-}_{1} a^{-}_{-1} a^{-}_{-1} \right) | 0 \rangle$ \\\midrule %\midrule
\end{tabular} }}
\end{center}
\caption{\label{table_NGstates}
Table of the states of the lowest six Nambu-Goto energy levels. These
have (rescaled) momenta $q=0,1,2$ where $q=N_R-N_L$.}
\end{table}

The above discussion has been for a flux tube in continuous space.
The cubic lattice has implications for angular momentum and the 
two-dimensional parity, $P_t$. The relevant irreducible representations 
are those of the two-dimensional lattice cubic symmetry 
\cite{HM_thesis}
which we denote by $C_4$.
It is a subgroup of $O(2)$ that corresponds to rotations by integer multiples 
of $\pi/2$ around the tube axis. This makes values of angular momenta that 
differ by an integer multiple of four indistinguishable on the lattice, and 
factorizes the Hilbert space into four sectors: $J_{{\rm mod}\, 4}=0$, 
$J_{{\rm mod}\, 4}=\pm 1$,  $J_{{\rm mod}\, 4}=2$. In the continuum states 
of nonzero $J$ are $P_t$ parity degenerate, but on the lattice this is 
true only for the odd $J$ sector. In the even $J\neq 0$ sector the
$P_t=\pm$ partners will in general experience $O(a^2)$ splittings.
All this means we can denote our states by the $5$ 
irreducible representations $A_{1,2}, E, B_{1,2}$ of $C_4$ whose $J$ and 
$P_t$ assignments are: $\left\{ A_1: \, |J_{{\rm mod}4}|=0, 
P_t=+\right\}, \left\{A_2:  \, |J_{{\rm mod}4}|=0, P_t=-\right\}$, 
$\left\{ E: |J_{{\rm mod}4}|=1, P_t=\pm\right\},$ 
$\left\{ B_1: \, |J_{{\rm mod}4}|=2, P_t=+\right\}$, and 
$\left\{B_2:  \, |J_{{\rm mod}4}|=2, P_t=-\right\}$.
All the representations of $C_4$  are one-dimensional except for $E$ which 
is two-dimensional.

\subsection{Operators}
\label{subsection_op_lat}

To calculate excited states efficiently we need a large basis of operators
that explicitly includes ones that are less symmetric than the
simple Polyakov loops. In our $D=2+1$ calculations
\cite{AABBMT_k1d3}
we contructed a basis of $\sim 200$ operators which provided us with
good overlaps onto the $\sim 10$ flux tube states that were light enough
to be accessible within our statistics. In $D=3+1$ we have two rather
than one transverse direction so, very naively, the corresponding number 
of operators would be $\sim 200^2 = 40000$. This is much too large. Instead 
we limit ourselves to $\sim 1000$ operators which are partly chosen by
looking at which operators turned out to be most important in the $D=2+1$
calculations. By projecting these operators to the different quantum numbers
of interest, we have matrices of correlation functions that are no larger 
than $O(100)$, which is manageable.

The operators we construct have shapes that lead to certain values of 
$J,P_t,P_l,$ and $p$. This is achieved by choosing a linear combination 
of generalised Polyakov loops whose paths consist of  various  transverse 
deformations of simple Polyakov loops, at various smearing and blocking 
levels. All the paths used for the construction of the operators are presented 
in Table~\ref{table_operators} and all together they lead to a basis of around 
700 operators. (The blocking levels for different parts of such an operator
need not be the same.) To construct, for example, an operator with a certain 
value of the spin $J$ we proceed as follows. Begin with the operator 
$\phi_{\alpha}(n_\mu)$ that has a deformation that we label as being at 
an angle $\alpha$ in the plane transverse to $x$ (assuming the flux tube 
winds around the $x$-torus) and that we label as being located at the 
site $x_\mu=an_\mu$. We can construct an operator $\phi_J(n_\mu)$ that 
belongs to a specific representation of $C_{4}$ by using the formula: 
\begin{equation}
\phi_J(n_\mu) = \sum_{n=1}^4 e^{iJ n \frac{\pi}{2}} 
\phi_{\alpha = n \frac{\pi}{2}}(n_\mu).
\label{eqn_opJ}
\end{equation}
This construction assumes that we are in the rest frame or in a frame 
where the momentum and spin are aligned. In our case, where the spin
is aligned along the flux tube, this means fixing the transverse
momentum to zero, which we can achieve by summing over all transverse
spatial translations: 
$\phi_J(n_t, p_{\shortparallel}, \vec{p}_{\perp}=0) \equiv 
\sum_{n_y,n_z}\phi_J(n_t, n_x, n_y, n_z)$. It is 
straight-forward to show that $\phi_{J=0}$ belongs to either $A_1$ or 
$A_2$ (depending on its value of $P_t$), that $\phi_{J=\pm 1}$ belongs 
to $E$, and that $\phi_{J=\pm 2}$ belongs to $B_1$ or $B_2$. The projection 
onto  certain values of $P_t$ and $P_l$ is demonstrated pictorially in 
eqns(\ref{eqn_A12},\ref{eqn_A13},\ref{eqn_A14}) for an operator of 
$J=0,2,1$ respectively. 
\begin{eqnarray}
\hspace{-0.25cm} \phi_{A}= {\rm Tr}\left[ \parbox{12.5cm}{\rotatebox{0}{\includegraphics[width=12.5cm]{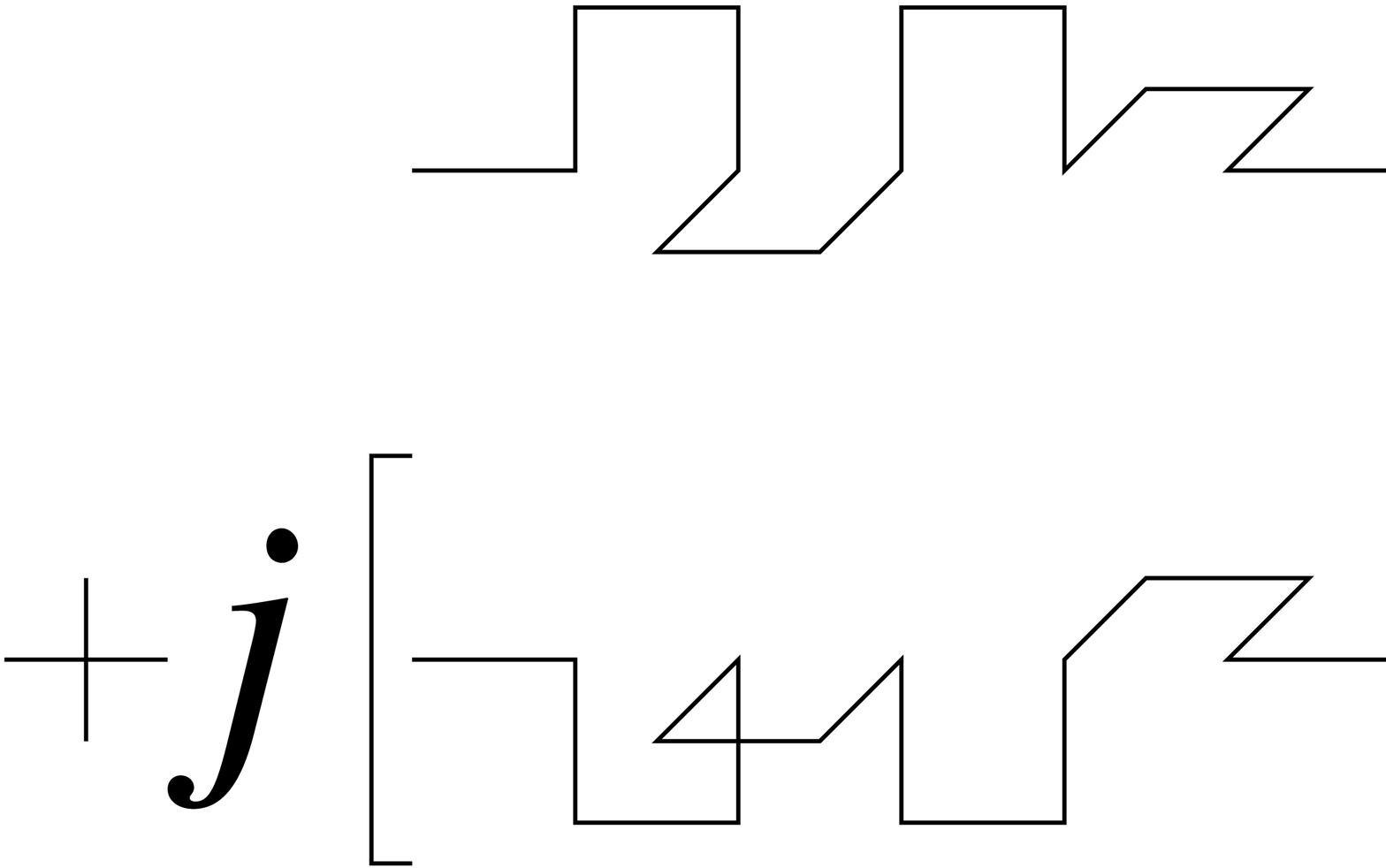}}} \ \right] 
\label{eqn_A12}
\end{eqnarray}
If $i=j=k=+1$ then the operator $\phi_{A}$ projects onto $\{ A_1, P_{l}=+ \}$,
if $i=+1,j=k=-1$ then it projects onto $\{ A_2, P_{l}=+ \}$, 
if $i=-1,j=+1,k=-1$ it projects onto $\{ A_1, P_{l}=- \}$ and finally, 
if $i=j=-1,k=+1$, it projects onto $\{ A_2, P_{l}=- \}$.
\begin{eqnarray}
\hspace{-0.25cm} \phi_{B}= {\rm Tr}\left[ \parbox{12.5cm}{\rotatebox{0}{\includegraphics[width=12.5cm]{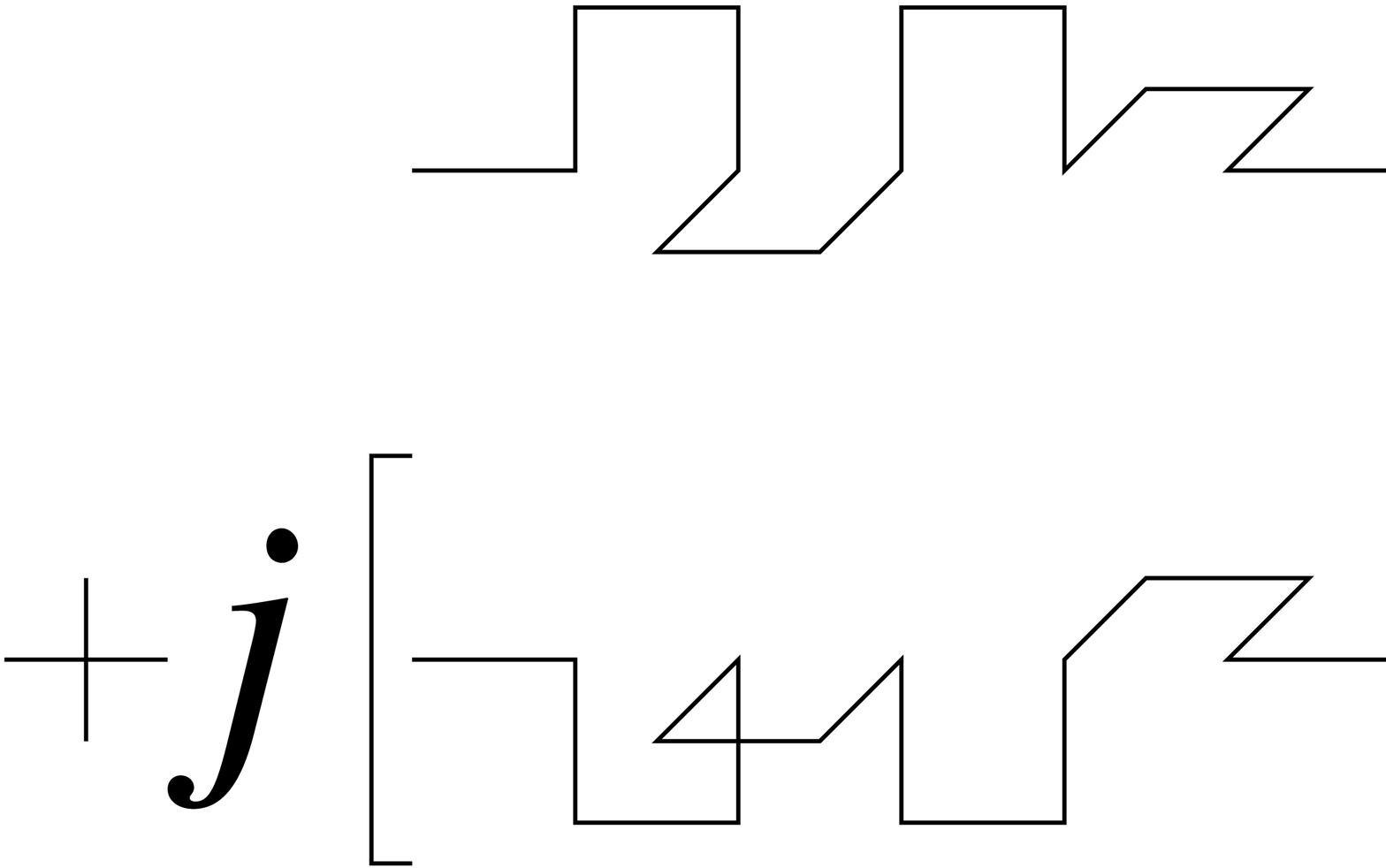}}} \ \right] 
\label{eqn_A13}
\end{eqnarray}
If $i=j=k=+1$ then the operator $\phi_{B}$ projects onto $\{ B_1, P_{l}=+ \}$,
if $i=+1,j=k=-1$ then it projects onto $\{ B_2, P_{l}=+ \}$, 
if $i=-1,j=+1,k=-1$ it projects onto $\{ B_1, P_{l}=- \}$ and finally, 
if $i=j=-1,k=+1$, it projects onto $\{ B_2, P_{l}=- \}$.
\begin{eqnarray}
\hspace{-0.25cm} \phi_{E}= {\rm Tr}\left[ \parbox{12.5cm}{\rotatebox{0}{\includegraphics[width=12.5cm]{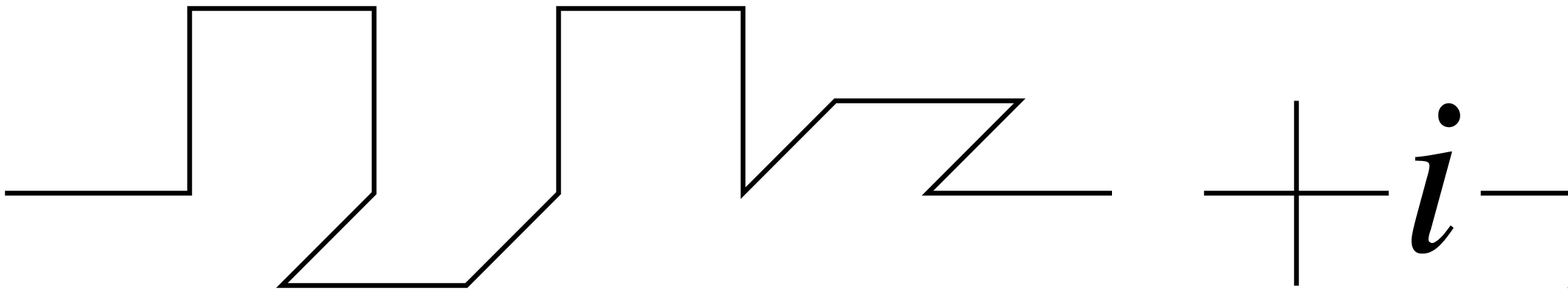}}} \ \right] 
\label{eqn_A14}
\end{eqnarray}
If $k=+1$ then the operator $\phi_{E}$ projects onto $\{ E, P_{l}=+ \}$ and 
if $k=-1$ it projects onto $\{ E, P_{l}=- \}$.
It is important to note that the above construction does not distinguish 
between $J$ and $J+4n$ where $n$ is an integer. Thus when we label a state 
as $J=0$, what we mean is $J=0,\pm 4, \pm 8, ...$, and similarly for our 
$J=\pm 1, \pm 2$ labels. Usually one can safely assume that the lightest
state will have the lowest value of $J$, but excited states may well
have higher spins. From now on we shall use the simple $J=0, \pm 1, 2$
label, but the reader should always be aware for what this is a
shorthand.
\begin{table}[htp]
\centering{\scalebox{1.0}{
\begin{tabular}[c]{|c||c||c||c||c||c||c||c||c||c||c||c||c||c|} \hline \hline
1 & 2 & 3 & 4 & 5 & 6 & 7 & 8 & 9 & 10 & 11 & 12 & 13 & 14 \\ \hline \hline
\rotatebox{90}{\includegraphics[height=0.022cm]{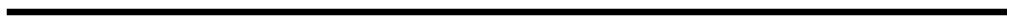}}
&
\rotatebox{90}{\includegraphics[height=0.32cm]{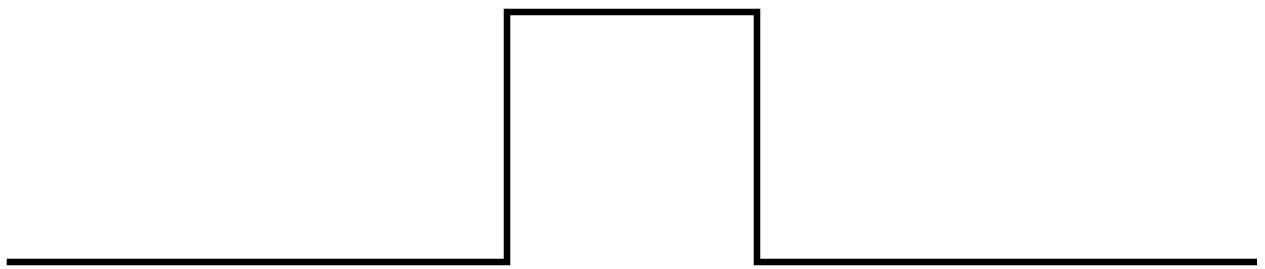}}
&
\rotatebox{90}{\includegraphics[height=0.4cm]{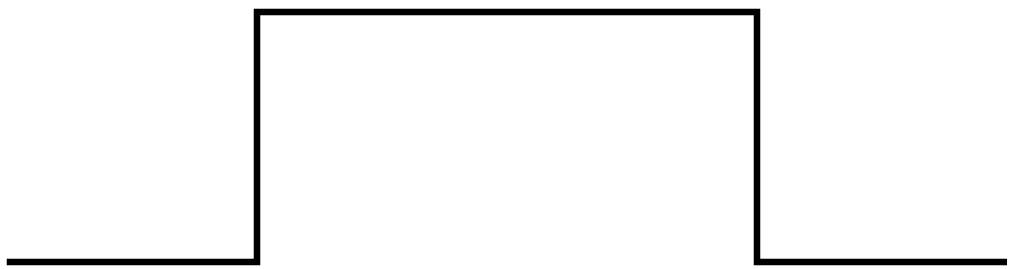}}
&
\rotatebox{90}{\includegraphics[height=0.8cm]{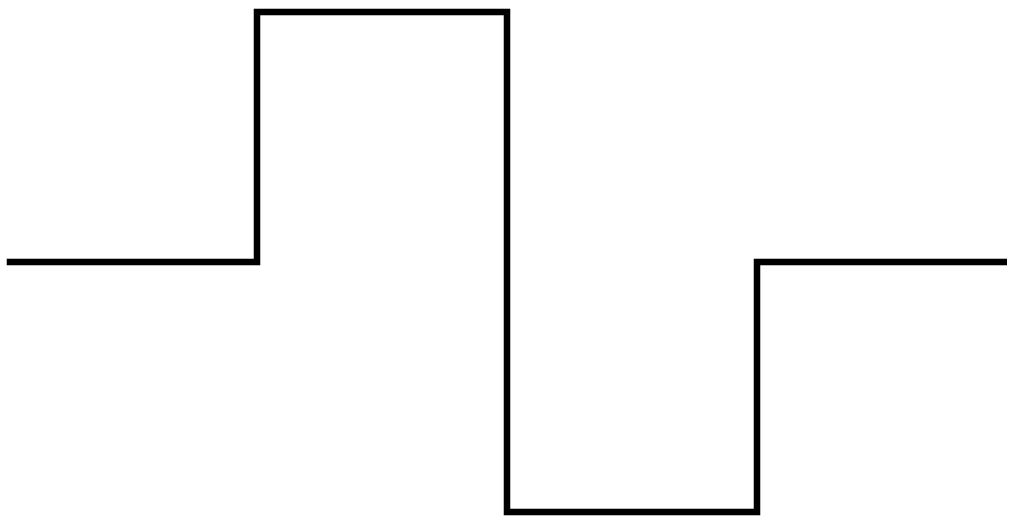}}
&
\rotatebox{90}{\includegraphics[height=0.4cm]{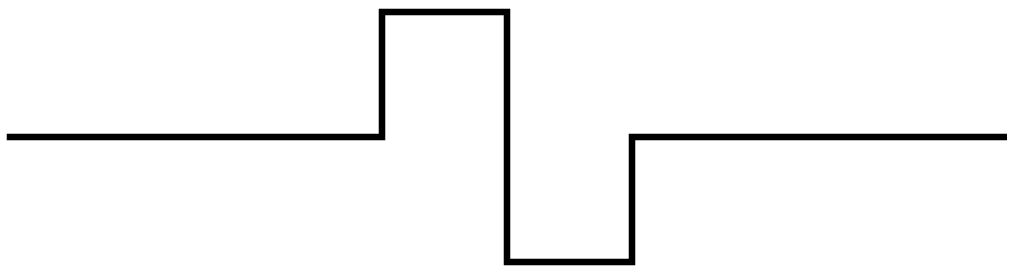}}
&
\rotatebox{90}{\includegraphics[height=0.4cm]{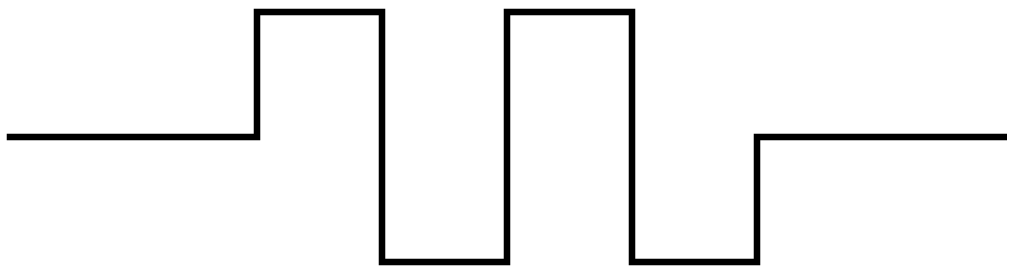}}
&
\rotatebox{90}{\includegraphics[height=0.4cm]{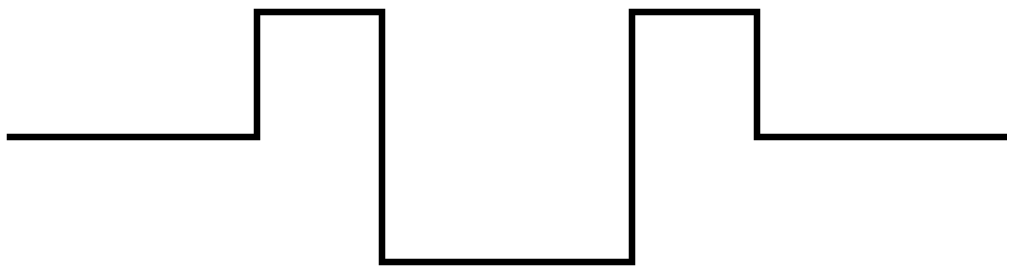}}
&
\rotatebox{90}{\includegraphics[height=0.21cm]{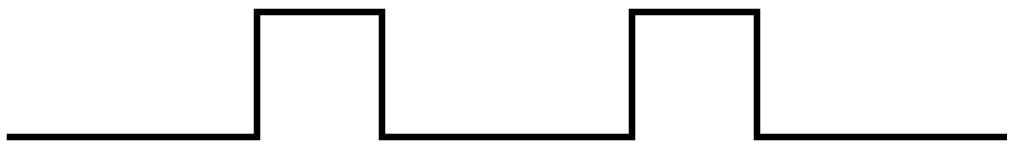}}
&
\rotatebox{90}{\includegraphics[height=0.4cm]{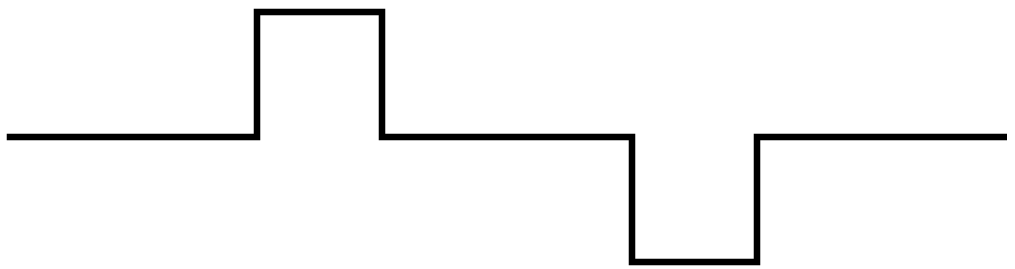}}
&
\rotatebox{90}{\includegraphics[height=0.21cm]{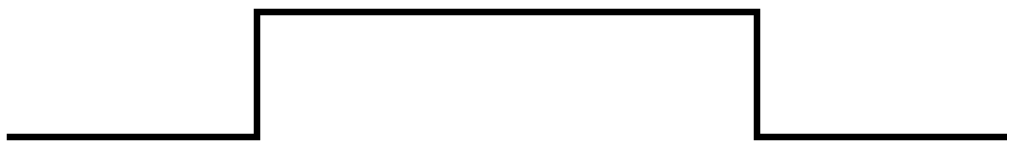}}
&
\rotatebox{90}{\includegraphics[height=0.4cm]{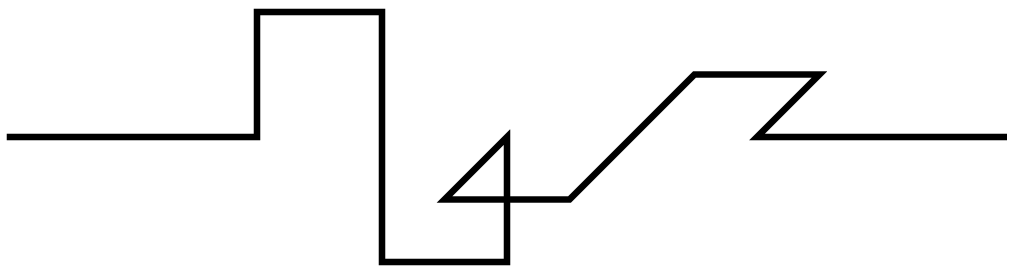}}
&
\rotatebox{90}{\includegraphics[height=0.32cm]{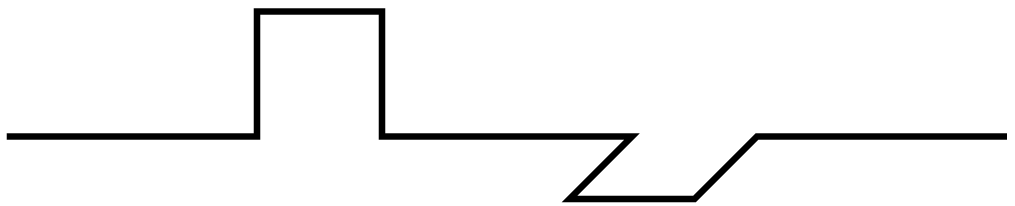}}
&
\rotatebox{90}{\includegraphics[height=0.32cm]{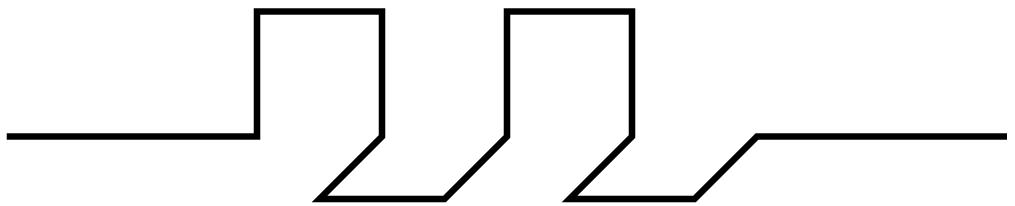}}
&
\rotatebox{90}{\includegraphics[height=0.4cm]{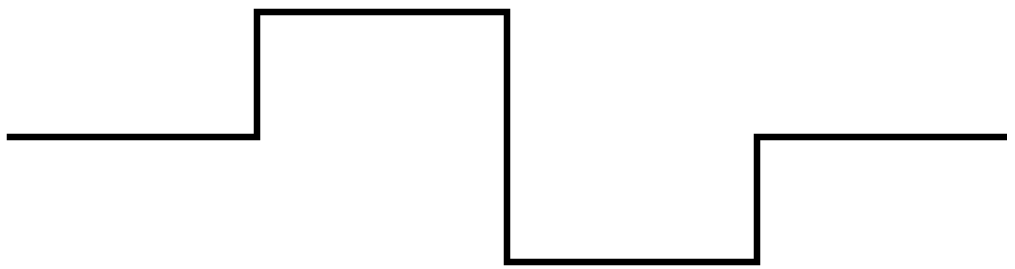}} \\ \hline \hline \hline
15&16&17&18&19&20&21&22&23&24&25&26&27&28 \\ \hline \hline
\rotatebox{90}{\includegraphics[height=0.32cm]{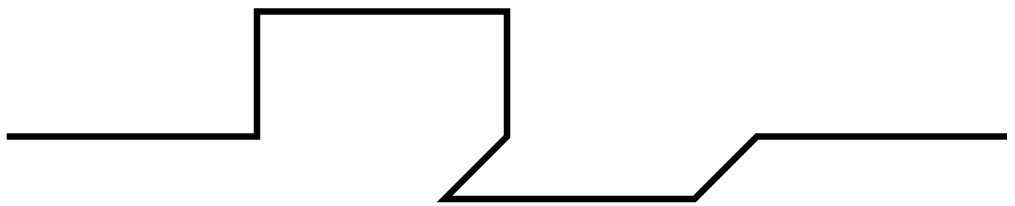}}
&
\rotatebox{90}{\includegraphics[height=0.32cm]{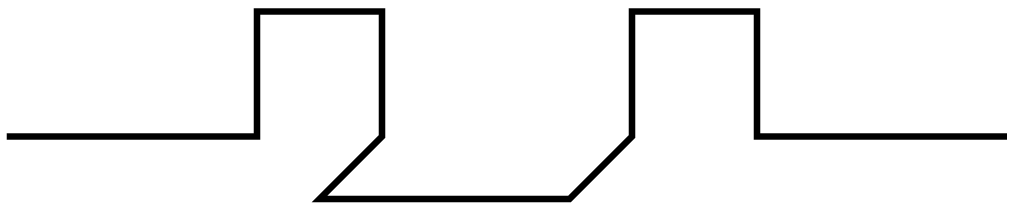}}
&
\rotatebox{90}{\includegraphics[height=0.42cm]{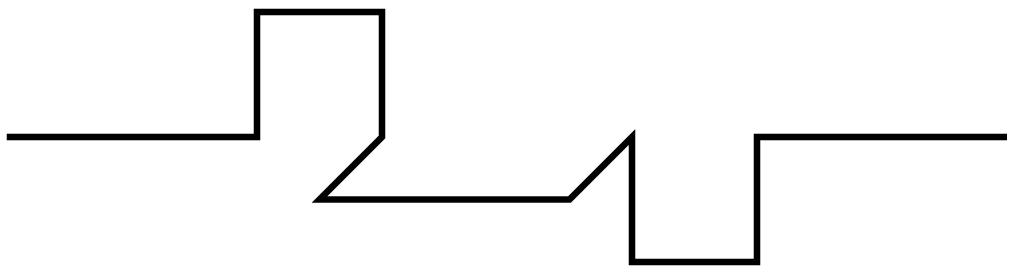}}
&
\rotatebox{90}{\includegraphics[height=0.32cm]{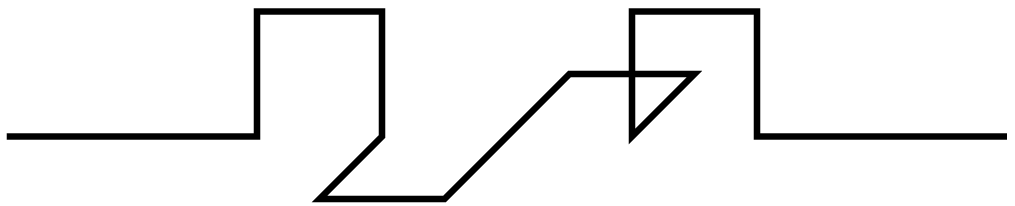}}
&
\rotatebox{90}{\includegraphics[height=0.42cm]{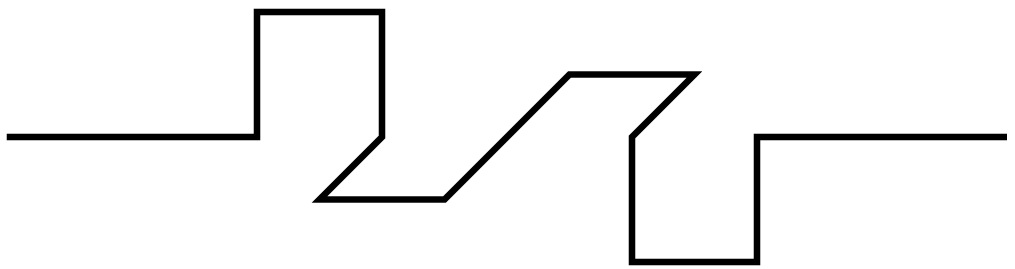}}
&
\rotatebox{90}{\includegraphics[height=0.32cm]{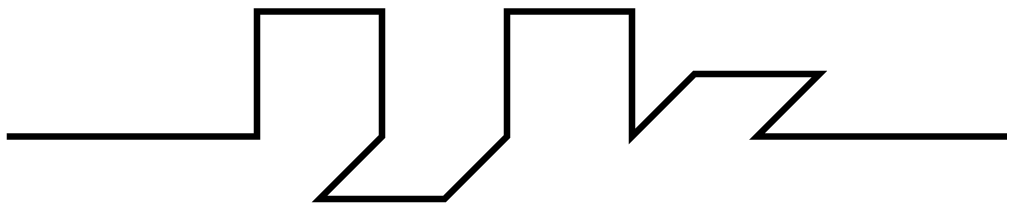}}
&
\rotatebox{90}{\includegraphics[height=0.4cm]{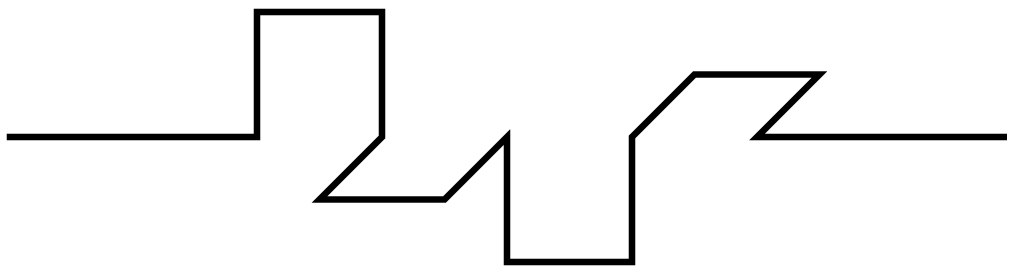}}
&
\rotatebox{90}{\includegraphics[height=0.32cm]{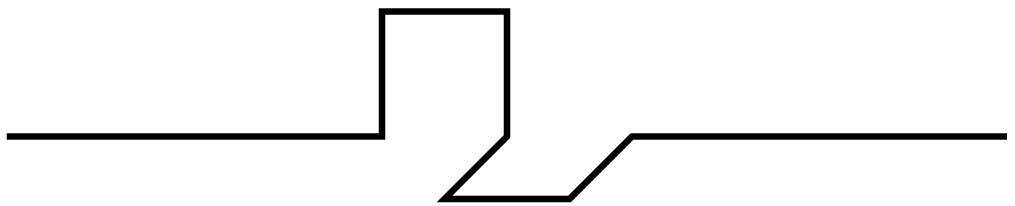}}
&
\rotatebox{90}{\includegraphics[height=0.64cm]{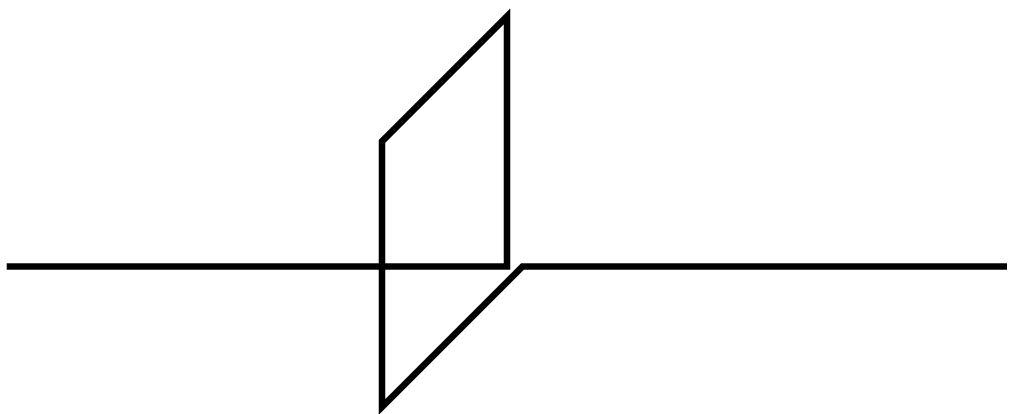}}
&
\rotatebox{90}{\includegraphics[height=0.64cm]{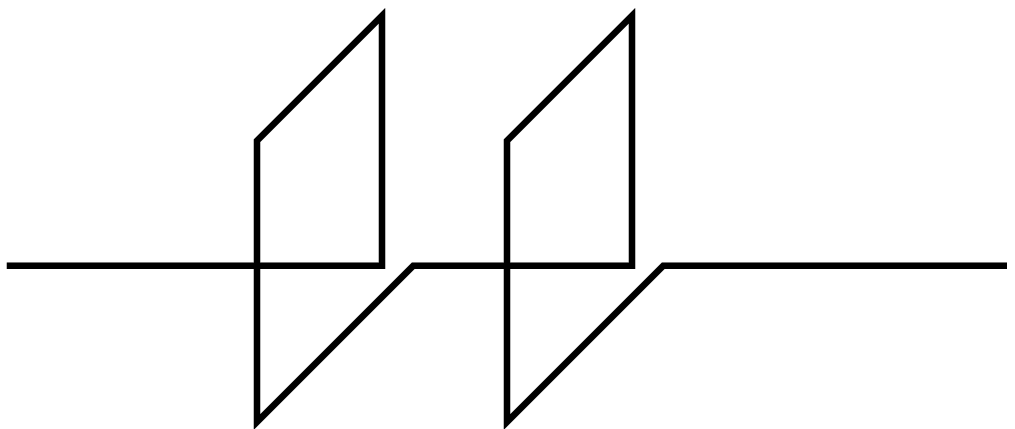}}
&
\rotatebox{90}{\includegraphics[height=0.8cm]{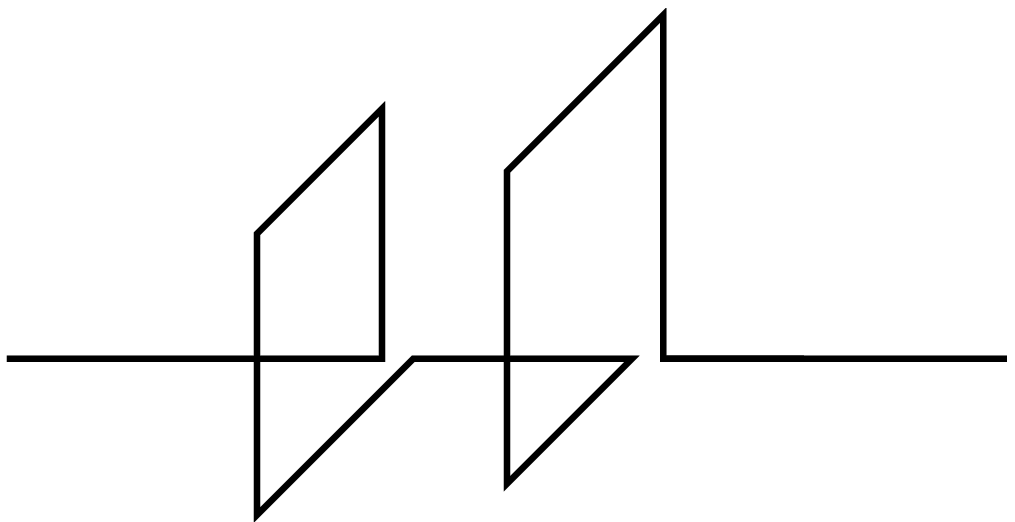}}
&
\rotatebox{90}{\includegraphics[height=0.62cm]{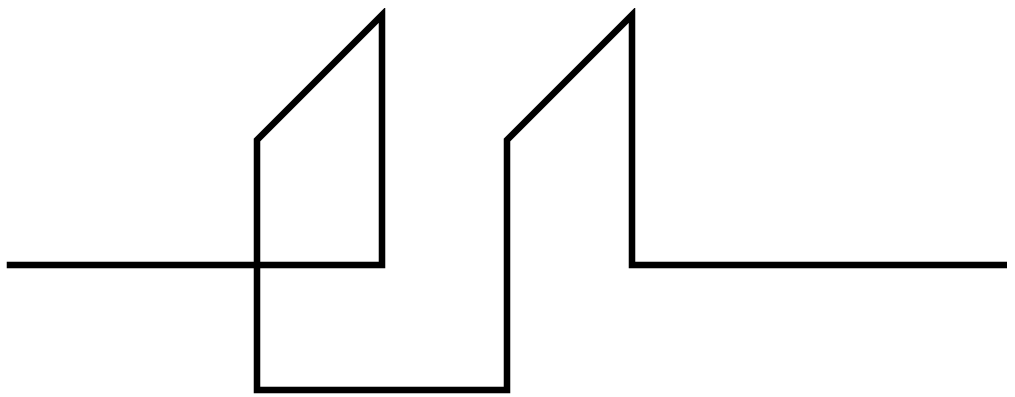}}
&
\rotatebox{90}{\includegraphics[height=0.74cm]{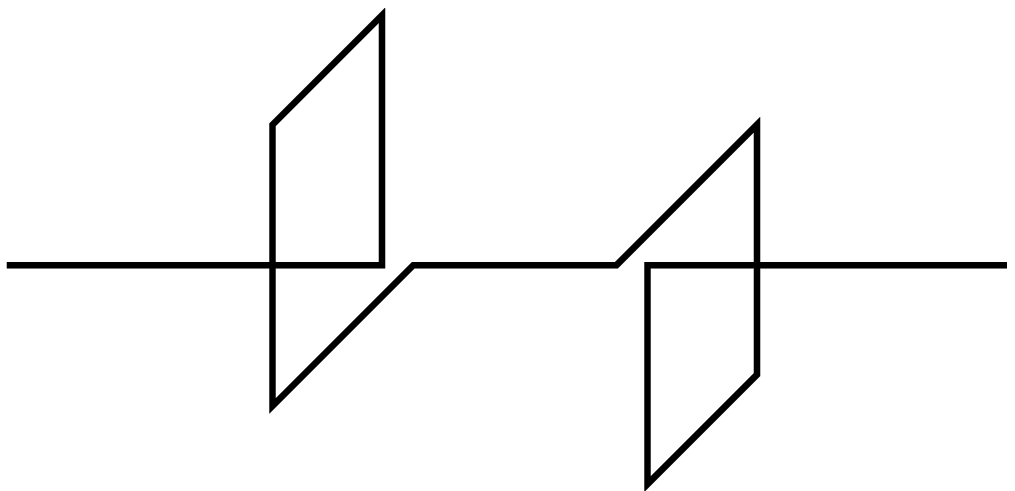}}
&
\rotatebox{90}{\includegraphics[height=0.76cm]{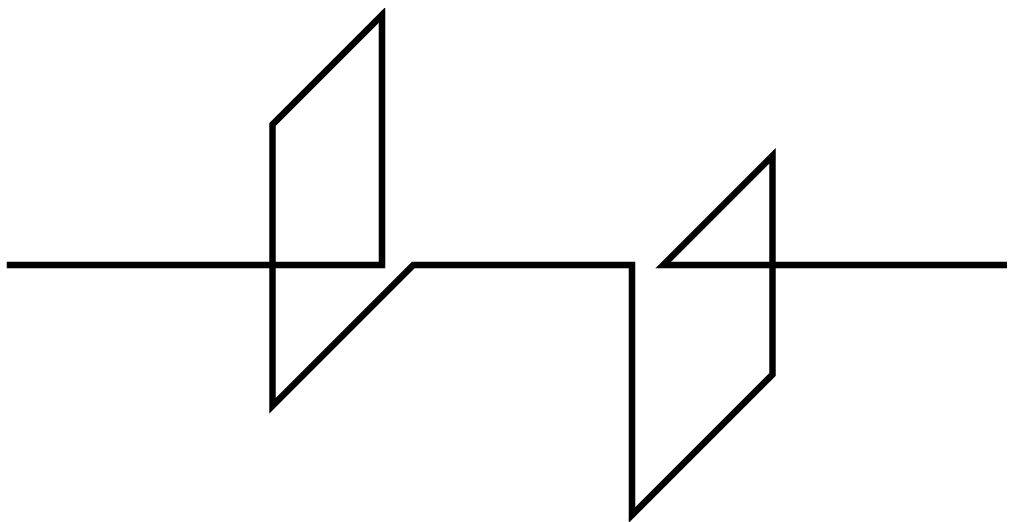}}\\ \hline
\end{tabular}}}
\caption{All the transverse deformations used for the construction of the operators.}
\label{table_operators}
\end{table}
%\vspace{-0.75cm}

So if we wish to project onto a flux tube that not only has spin $J$ but
in addition has  non-zero momentum $p_{\shortparallel} = 2\pi q/l$ along 
its axis, while still having 
zero transverse momentum $p_\perp = 0$, we can perform the sum
\begin{equation}
\phi_J(n_t, p_{\shortparallel}, p_\perp = 0) = \sum_{\vec{n}_\perp}
e^{ip_{\shortparallel} n_{\shortparallel}} \phi_J(n_t, n_{\shortparallel}, 
\vec{n}_\perp).
\label{eqn_opPJ}
\end{equation}
Since the spin component along the axis of a boost is invariant under 
boosts, such an operator produces states of the desired spin even though
it has not been defined in the rest frame. Note that since the longitudinal 
parity flips $p_{\shortparallel} \to -p_{\shortparallel}$, it is not a 
useful quantum number when $p_{\shortparallel}\neq 0$, and in that case 
we set $i=k=0$ in sums like those in 
eqn(\ref{eqn_A12}) and we simply label our states by $J$ or by $|J|,P_t$.

Since $J=2$ and $J=-2$ differ by $\Delta J =4$, the operators for them 
are identical. However we can construct lattice operators that have
transverse parity $P_t = \pm$ as described above. Hence for even $J$ we 
shall choose to label the states by $|J|$ and $P_t$. For odd $J$, on the
other hand, $\Delta J \neq 4n$ and we can use eqn(\ref{eqn_opJ}) to 
construct orthogonal $\pm |J|$ operators. Since these must be 
degenerate, we choose to calculate only one of these two states; 
but it should be understood that every such odd-$J$ state
is automatically accompanied by a degenerate $-J$ state that we
have chosen not to show explicitly.

In terms of our choice of operators, the present work should be regarded
as exploratory. Not surprisingly we will find that the overlaps are
not as good as in our earlier $D=2+1$ study. However for a number
of states we do have overlaps $|c|^2 \geq 0.9$, at which point we can
regard the difficult-to-estimate systematic errors as being under control.
Of course, for heavier states our identification of an `asymptotic'
exponential behaviour becomes less reliable, and therefore so does our
estimate of the overlap. On the other hand, in $D=3+1$ one has more 
channels with different quantum numbers than in $D=2+1$, and each of
these channels has its own ground state(s) which corresponds to some
excitation of the absolute ground state. Since these ground states are
afflicted by fewer systematic errors, this allows us to obtain
relatively reliable results for a number of string excitations even 
with our poorer overlaps.

\subsection{Earlier lattice calculations in $D=3+1$}
\label{subsection_previouslat}

Numerical exploration of the stringy nature of open and closed flux 
tubes, at the level of testing the L\"uscher correction to the ground
states, dates back to the mid-1980's. However the pioneering calculations 
for excited string states date to the early-90's, e.g.
\cite{CM}.
The interest here was both theoretical and phenomenological: the
excited string energy can be used as a potential in a Schr\"odinger equation
to get predictions for the masses of hybrid mesons where some
of the quantum numbers are carried by excited glue. More or less
simultaneously, there were calculations of Wilson loop expectation 
values investigating the match between string theory predictions 
and what one obtains in various gauge and spin models by,
in particular, the Torino group, e.g.
\cite{Caselle}.
Expectations of Wilson loops are transforms of eigenspectra - although 
care has to be taken with the self-energies associated with the 
boundaries - and provide an alternative way to test string models.
Indeed it is in this body of work that one first sees a prolonged 
and serious focus on matching the Nambu-Goto model to numerical results. 

Much of the work on flux tubes has focused on open flux tubes i.e. the
static potential and its excitations. In this case there is 
a smooth transition between short-distance perturbative physics (the 
Coulomb potential) and the long-distance confining physics (the 
flux tube). While this  transition is potentially
of great interest (especially at $N \to \infty$) it can introduce 
extra ambiguities in the extraction of the flux tube spectrum.
The spectrum of closed flux tubes, stabilised by closing them around 
a spatial torus, is a particularly clean way to investigate the 
properties of flux tubes, and that is why we have focused on that 
approach in both our $D=2+1$ 
\cite{AABBMT_k1d3}
and $D=3+1$ calculations.

As far as the eigenspectrum of closed flux tubes in $D=3+1$ gauge
theories is concerned, the most ambitious calculation we are aware of 
is the one in
\cite{kuti}
which was part of a larger calculation largely focused on the spectrum
of open flux tubes
\cite{kuti2}. 
This calculation was in SU(3) and was performed at one value of the 
lattice spacing and for three values of the flux tube length, $l$. 
The timelike and spacelike lattice spacings were chosen to be different, 
with the former quite small, while the  spatial lattice spacing
was chosen quite large, $a\surd\sigma \sim 0.5$.
The energies of 12 excited states, of various 
quantum numbers (some being ground states within their channels) were 
calculated. The three flux tube lengths chosen were all quite long, 
$l\surd\sigma \in [4,8]$, the intention being to compare with the leading 
$O(1/l)$ correction to the linear piece, $\sigma l$, at large $l$. 
The results show an initial tendency to approach the theoretical Nambu-Goto 
expectation but then, typically, to overshoot at the largest value of $l$. 
(See Fig.2 of
\cite{kuti}.)
At the larger two values of $l$ a rough agreement is observed with the 
theoretically expected level ordering and degeneracies,
although there is an additional fine structure that is still prominent 
even at the largest value of $l$. Shortly after this work a calculation 
for SU(4) and SU(6) appeared,
\cite{HMMT04}, 
that  focused on the ground state of the closed flux tube, but also 
pointed to the qualitative agreement between the first excited state and 
the full prediction of the Nambu-Goto model, even at modest values of $l$.

Our calculations attempt to improve upon these earlier calculations in\
several ways. First, our point of view is that the simplest picture
should emerge at $N=\infty$ and we therefore perform calculations
not only for SU(3), but also for SU(5) and SU(6). We find that
the $1/N^2$ corrections are negligible for $N\geq 3$, at our level
of statistical accuracy, which means we can treat our extensive
SU(3) calculations as being valid for larger $N$. Secondly,
what we are interested in is continuum rather than lattice physics.
Indeed, the (spatial) lattice spacing used in
\cite{kuti}
was large, and so carried the risk of significant finite-$a$ corrections. 
To control these in our calculations we mainly work at a much smaller 
value of $a$, and then explicitly check for $O(a^2)$ corrections by
performing a calculation at an even smaller value of $a$. We find 
that any dependence on $a$ is negligible and so we can treat our
results as being for the continuum limit. Thirdly, motivated by our 
$D=2+1$ calculations  
\cite{AABBMT_k1d3},
which show that the approximate agreement with the Nambu-Goto spectrum 
begins at remarkably small values of $l$, we focus within our calculations
mainly on small to medium values of $l$, i.e. $l\surd\sigma \in [1.8,4.8]$,
where we can be reasonably confident that we have the systematic errors 
under control. As we remarked in Section~\ref{subsection_energies_lat}, 
as one makes $l$ larger, the energies of interest become larger, and one 
runs a rapidly increasing risk of a systematic error that typically leads 
to an overestimate of the energy. It appears plausible to us 
that this is the source of the large-$l$ overshoot observed in
\cite{kuti}.
Since we find that we can obtain interesting results on the spectrum
without venturing to such large $l$, we choose not to do so in this
study, except for one calculation at $l\surd\sigma \sim 6.3$ which we
use both to check the qualitative behaviour of some states and to ensure
that we have some appreciable overlap with the range of $l$ where analytic
expansions in $1/l$ should be applicable (for the lightest excited states). 
In addition, and 
perhaps most importantly from the technical point of view, we incorporate a
a very large basis of operators in our calculations. Although the benefits 
are not as striking as in our earlier $D=2+1$ calculations, it is primarily 
this technical improvement that allows us to maintain usefully good
overlaps onto a large number of excited states at the small 
lattice spacings where we can confirm that we are close to the
continuum limit.

\section{Results}
\label{section_results}

Our calculations of the spectrum of closed flux tubes that wind 
around a spatial torus of length $l$ have been performed for
the SU($N$) groups and inverse bare couplings, $\beta=2N/g^2$,
listed in Table~\ref{table_physics}. For orientation we also
list some basic physical properties of the lattice gauge theories 
at those values of $\beta$. First there is the string tension 
$a^2\sigma$ which we extract from our calculation of the ground state 
string energy $E_0(l)$. (We fit $E_0(l)$ to the Nambu-Goto expression
plus a $O(1/l^7)$ correction. The value of $a^2\sigma$ extracted is not 
sensitive to any reasonable changes to the fitting function.) 
This tells us what $a$ is in physical 
units. To express $a$ in more intuitive `fermi' units one can set 
$\sigma$ to its real world value, $\surd\sigma \sim 440 {\mathrm{MeV}}$, 
although in all our field theories such units are, of course, fictitious. 
In these units we see that most of our calculations are for
$a\sim  0.09 {\mathrm{fm}}$ while the SU(3) calculation at $\beta = 6.338$ 
corresponds to $a\sim  0.06 {\mathrm{fm}}$. These values of $a$ 
are small enough that the lattice corrections are, in general, small.
We also show the critical value of the flux tube length, $l=l_c$,
below which the flux tube no longer exists because we are in a finite 
volume (anisotropic) deconfined phase that is a precise analogue of the 
finite temperature deconfined phase. (These values have been obtained
by extrapolating/interpolating in $\beta$, and for SU(5) in $N$, 
the values in 
\cite{BLMTUW_Tc}.) 
Our calculations will go down to values of $l$ that are very close to 
$l_c$. Here the flux tube should be about as wide as it is long and, naively, 
no longer looks anything like a thin string. Finally we list the value 
of the mass gap $am_G$, which here is the mass of the lightest scalar 
glueball. (These values are obtained from the values in
\cite{blmtuw04}
and, for $N=5$, in
\cite{blmt01b}.) 

\begin{table}[htb]
\begin{center}
\begin{tabular}{|c|c|c|c|c|c|}\hline
$N$ & $\beta$ & $l/a\in$  &  $a\surd\sigma$ & $l_c/a$ & $a m_G$ \\ \hline
3  & 6.338  & [16,24] &  0.12878(69) & 11.99(9) & 0.448(11) \\
3  & 6.0625 & [9,20]  &  0.19485(17) &  8.00(2) & 0.648(11) \\
   &        & [10,32] &  0.19526(38) &          &  \\
5  & 17.63  & [10,16] &  0.19664(81) &  8.32(5) & 0.630(22) \\
6  & 25.55  & [10,18] &  0.20187(27) &  8.30(4) & 0.588(13) \\  \hline
\end{tabular}
\caption{Parameters of our calculations with some corresponding 
properties of the gauge theories: the string tension, $\sigma$, the
deconfining length, $l_c$, and the mass gap, $m_G$.}
\label{table_physics}
\end{center}
\end{table}

We have performed high statistics calculations with a small number 
of operators in SU(6) at $\beta=25.55$ and in SU(3) at $\beta=6.0625$.
This allows us to obtain very accurate values for the ground
state energy, $E_0(l)$, and to analyse its $l$-dependence
in some detail. However to calculate the energies of excited
flux tubes we need much better overlaps on the excited states
than these calculations provide, and this we attempt to achieve with 
our much larger basis of operators, as discussed above, and with
correspondingly lower statistics (because of the
expense of such calculations). We perform the
latter calculations in SU(3) at both values of $\beta$, so as
to allow an explicit control of lattice spacing corrections.
These corrections turn out to be small, so we only need to perform a 
similar SU(5) calculation, which allows us to control finite-$N$
corrections, at the coarser value of $a$, as listed in
Table~\ref{table_physics}. We shall see that the finite-$N$ 
corrections are also small. This means we can focus our calculations
of the energy spectrum on one $N$ and on one lattice spacing, i.e. SU(3) 
at $\beta=6.0625$, without loss of generality. Here we cover a much 
larger range of $l$ and have much higher statistics, leading to
a more reliable extraction of flux tube energies. We
list separately the value of $a\surd\sigma$ extracted
from this large operator calculation in Table~\ref{table_physics},
since this is what we shall use in our later analysis of the excited 
states. We see that this value is considerably more accurate than 
the other calculations with a large number of operators.

We begin with a detailed analysis of our high statistics calculations 
of the ground state energy, $E_0(l)$. We then move on to the lightest
states with different non-zero values of longitudinal momentum, 
$p_{\shortparallel}=2\pi q/l$, for $q=1,2$. These, as we have seen, 
must possess non-trivial excitations, and we compare the excitation energy 
to the predictions of the free-string Nambu-Goto model. Next we consider 
those states which correspond to the first, and sometimes second, 
excited energy levels of the $p_{\shortparallel}=0$ and 
$p_{\shortparallel}=2\pi/l$ sectors, and whose energies we are able to 
calculate with some reliability. In one particular case we also
look at excited states in the $p_{\shortparallel}=4\pi/l$ sector.
We follow this with a brief aside on the ground state of the $k=2$
flux tube in SU(6). Finally we compare our results to the theoretical
predictions described in Section~\ref{section_EST}.

\subsection{Ground state}
\label{subsection_groundq0}

In this section we analyse the $l$-dependence of the ground state energy,
$E_0(l)$, of a flux tube that winds once around a spatial torus. Our aim 
is to see what we can say about the universal corrections to the leading 
linear piece, $\sigma l$. Previous lattice calculations, both for open 
and for closed flux tubes, e.g. 
\cite{kuti2,mlpw02,blmt01}, 
have provided good evidence that the coefficient of the $O(1/l)$ L\"uscher 
correction has the value appropriate to a simple bosonic string theory 
where the only massless modes are those of the transverse fluctuations, 
and, moreover, that this continues to be the case for larger $N$
\cite{HMMT04}. 
Motivated by the recent theoretical developments summarised above
\cite{LW04,JD,AHEK09},
we shall see if we can obtain any information on the coefficients of 
the higher order terms that we now believe to be universal.
In addition, motivated by what we have observed in $D=2+1$
\cite{AABBMT_k1d3},
we shall test how well the Nambu-Goto free string theory describes the
ground state energy. 

The calculations in this subsection differ from our main calculations
in that we use a small basis
of operators which we know, from previous experience, will provide
a very good overlap onto the ground state. Such calculations are much
faster and this enables us to achieve a greater statistical
accuracy for $E_0(l)$. We will present two calculations, one in
SU(3) and one in SU(6). The SU(3) calculation is at $\beta=6.0625$. 
Here one finds $a\surd\sigma \simeq 0.2$, which translates to 
$a \simeq 0.09 \mathrm{fm}$ if one wants to use such  semi-fictitious 
units. Thus this is a small value of $a$, where one expects lattice 
spacing corrections to be small. We perform calculations for
$l/a \in [9,20]$. The smallest length is only just above the first-order 
finite-volume deconfining transition which, at $\beta=6.0625$,
occurs at $l_c/a \simeq 8$
\cite{BLMTUW_Tc}.
The SU(6) calculation is at $\beta=25.55$ which, as we see in 
Table~\ref{table_physics}, corresponds to roughly the
same value of $a$ in units of $\surd\sigma$. The value of $l_c$ is slightly 
larger than in SU(3), and the range of flux tube lengths studied is 
$l/a \in [10,18]$. 

We shall begin by using the SU(3) calculations to quantify some of
the systematic errors, so as to ensure that they are small in our later
calculations of the flux tube spectrum. The first part of this
section is devoted to this issue, which can be skipped by the reader
who is primarily interested in our results.  

\subsubsection{systematic errors}
\label{subsubsection_systematics}

The systematic errors we focus upon are those associated with finite 
volume effects and with the extraction of an energy from a given correlation
function. 

The finite volume corrections are of two kinds: those that arise from
the finite spatial extent and those that arise from the finite 
temporal extent. An example of the former is the emission by a flux 
loop of a virtual glueball that propagates around one of the orthogonal 
spatial tori, of length $l_s$,  before being reabsorbed. This will alter 
the flux tube energy by $\Delta E \propto \tilde{g}^2 e^{-m_G l_s}$ where
$m_G$ is the mass of the glueball. If $l_s$ is small enough then the 
lightest state will be a combination of a flux loop and a conjugate
loop and in that case the leading large-$l_s$ 
contribution will come from its propagation around an orthogonal torus, 
giving $\Delta E \sim \tilde{g}^2 e^{-2E_0(l) l_s}$. So we expect that
as we reduce $l$ we will have to have a larger spatial volume to
minimise these corrections. Since we expect that $\tilde{g}^2 \propto 1/N^2$ 
all these finite volume corrections should disappear as $N\to\infty$
(as long as the volume is large enough that we remain in the usual 
confining phase) but will be present at finite $N$. 

The finite temporal extent $l_t$ means that the partition function $Z$ in 
the denominator of eqn(\ref{eqn_avPhi}) receives contributions not just 
from the vacuum but from other states propagating around the time torus. 
(It is a `finite temperature' partition function.) These same states 
will contribute to the path integral that appears in the numerator 
of eqn(\ref{eqn_avPhi}). At $N=\infty$ colour singlet states do not 
interact, and so this extra contribution will simply cancel between 
numerator and denominator. (In this sense there is no $T$ dependence 
at $N=\infty$ in the confining phase.) However at finite $N$ a state
propagating around the time torus will interact with the flux tube that 
propagates between the Wilson line operators in the numerator and this 
will imply an incomplete cancellation and a shift in the flux tube energy.
Typically such corrections will be $\Delta E \sim O(e^{-m l_t})$ where 
$m$ is the lightest state. For small $l$ the lightest state is the winding 
flux tube, so that $m = E_0(l)$, and we must make $l_t$ larger as $l$ 
decreases, so as to maintain  $E_0(l)l_t \gg 1$. 

To monitor these finite volume corrections one can perform calculations
for various transverse and temporal tori at each $l$. Such a detailed
study can be found for SU(6) in
\cite{HMMT04}.
Since it is not too expensive to make just one lattice torus very large,
we choose to do so for the temporal torus and then we vary the spatial
tori to monitor the corresponding finite spatial volume corrections.
The values of $l$ and lattice sizes of our SU(3) calculation are listed
in Table~\ref{table_gsesn3}. We calculate the ground and first excited state
energies as described in Section~\ref{subsection_energies_lat}. We show
the ranges of $t=n_ta$ which we fit with the single exponential in order to
obtain an energy. For the ground state we typically have a normalised overlap
$|c_0^2|\sim 0.98-0.99$ allowing an accurate and reliable extraction of its
energy. For the first excited state we have a much worse overlap, 
$|c_1^2|\sim 0.8$, and so we have to fit at larger $t$ and the statistical 
error is about ten times larger. Higher excited states are even worse: 
this is why we need to use a much larger basis of operators, as we shall 
do in later sections.

In Table~\ref{table_gsesn3} we can see that the finite volume corrections 
on $E_0(l)$ are not large: no more than $\sim 3-4\%$ even for the smallest 
volume. However since we are interested in identifying leading and 
subleading corrections to the linear dependence of $E_0(l)$, these 
are important. Since the leading
large volume correction decreases exponentially with the spatial size, we can 
assume that the corrections to our largest lattices are much smaller than any 
variation we observe between our second largest and largest spatial volumes. 
We therefore use the values of $E_0(l)$ calculated on the largest volumes
listed in Table~\ref{table_gsesn3}, in the expectation that any finite volume
effects are smaller than the quoted statistical errors. We also see from
Table~\ref{table_gsesn3} that the
finite volume corrections to the energy of the first excited state,
$E_1(l)$, are at most comparable to the statistical errors. We
therefore assume that even with our later more accurate calculations,
we are safe in using the largest, or even second largest, volumes in 
Table~\ref{table_gsesn3} for our eigenspectrum calculations.
 
\begin{table}[htb]
\begin{center}
\begin{tabular}{|c|c|c|cc|cc|}\hline
\multicolumn{7}{|c|}{SU(3) ; $\beta=6.0625$} \\ \hline
$l/a$ & $L_y\times L_z\times L_t$ & sweeps$\times 10^6$ & $n_t \in$ & $aE_0(l)$ & $n_t \in$ & $aE_1(l)$ \\ \hline
9  & $32\times 32\times 48$ & 0.6 & [2,7] & 0.1747(21) & [3,8] & 0.980(34) \\ 
   & $24\times 24\times 48$ & 0.5 & [2,7] & 0.1712(25) & [3,8] & 0.959(37) \\ \hline
10 & $20\times 20\times 36$ & 2.0 & [2,7] & 0.2425(16) & [3,8] & 0.967(18) \\ 
   & $16\times 16\times 36$ & 1.0 & [2,8] & 0.2391(26) & [3,8] & 0.952(26) \\ 
   & $10\times 16\times 36$ & 0.5 & [2,8] & 0.2307(26) & [3,8] & 0.970(27) \\ 
   & $10\times 10\times 36$ & 1.0 & [2,7] & 0.2346(13) & [3,7] & 0.967(18) \\ \hline
12 & $16\times 16\times 24$ & 5.0 & [2,7] & 0.3510(14) & [3,8] & 1.027(18) \\ 
   & $12\times 12\times 24$ & 0.8 & [2,7] & 0.3429(24) & [3,7] & 0.983(24) \\ \hline
16 & $16\times 16\times 16$ & 7.5 & [2,7] & 0.5361(14) & [3,8] & 1.074(9) \\
   & $12\times 16\times 24$ & 5.0 & [2,7] & 0.5335(15) & [3,8] & 1.039(13) \\ 
   & $10\times 10\times 36$ & 0.5 & [2,7] & 0.5183(51) & [3,8] & 1.031(51) \\ \hline
20 & $20\times 20\times 16$ & 6.0 & [2,7] & 0.7096(20) & [3,7] & 1.174(21) \\ \hline
\end{tabular}
\caption{\label{table_gsesn3} Energies of the ground and first excited states
of a closed flux tube of length $l$, on various spatial volumes, with
statistics and fitting ranges indicated. For SU(3) at $\beta=6.0625$.}
\end{center}
\end{table}

We now consider two of the main systematic errors in extracting a 
ground state energy from a correlation function $C(t)$. \\
1) We typically perform the fit using 
a single exponential to a range $t\in [t_1,t_2]$ as 
described in Section~\ref{subsection_energies_lat}. However this cannot
be entirely correct: the fact that there is a visible contribution from 
excited states for $t< t_1$ means that it also exists for $t\geq t_1$,
even if small. This is an error which leads to our estimates of $E_0(l)$ 
being too high, albeit by an amount which hopefully is no greater than
the statistical error. However since this error is systematic, its
contribution after a combined analysis of several calculations, may well
be significantly larger than the final statistical error. \\
2) Secondly, when performing the exponential fit we typically treat the 
errors on $C(t)$ at different values of $t$ as being independent, which 
is also incorrect. The calculations at different values of $t$ are based 
on the same lattice field configurations and so may be quite highly 
correlated (particularly if the energy in lattice units is small).  
The value of $\chi^2$ calculated with independent errors will be smaller 
than the true $\chi^2$ and we thus run the danger of extracting $E_0(l)$ 
from a range  $[t_1,t_2]$ where the fit appears acceptable but 
is in reality poor. (The error estimates on $E_0(l)$ are obtained by 
a jack-knife method and should  be less affected.)  To account for 
this, our heuristic procedure is typically to accept a fit only if the 
$\chi^2$ per degree of freedom is $\ll 1$. In practice we check that
the $\chi^2$ is not significantly decreased if we translate our fitting
interval to larger $t$. 

We will now explicitly estimate the effect of both the above errors 
on the SU(3) ground state calculations carried out on
the largest spatial volumes listed in Table~\ref{table_gsesn3}.

In the first column of Table~\ref{table_gsalln3} we display the ground
state energies that we obtain with our default method, which uses a 
single exponential fit to the correlator of the best (variational)
operator, and which assumes uncorrelated errors (`nocortt'), with a
heuristically reduced estimate of what constitutes an acceptable
$\chi^2$. We also show the range in $t$ used for each fit. 

In the second column of Table~\ref{table_gsalln3} we show the 
result of incorporating the error correlations in the statistical 
analysis (`cortt'). This is done in a standard way: one calculates
the correlation beween the fluctuations of the correlator at $t$ and
the fluctuations at $t^\prime$ for all $t,t^\prime$ that are of 
interest. This gives a correlation matrix whose inverse then appears
in the definition of $\chi^2$. What one is effectively doing is 
determining the linear combinations $\sum_t c_t \phi^\dagger(t)\phi(0)$ 
whose fluctuations are independent, and then using these to determine 
the $\chi^2$. In practice this procedure can produce anomalies if
all the elements of the correlation matrix, whose inverse we take,
are not determined very accurately. This is one reason we do not apply
it systematically to our later calculations but rather test what 
difference it makes in this high statistics calculation. Since this 
is the correct definition of $\chi^2$, the criterion for an acceptable 
fit is now simply the normal one: unity per degree of freedom. 
Comparing the values of $E_0(l)$ in the first and second columns of
Table~\ref{table_gsalln3}, we observe that for larger values of $l$ the 
results are, within statistical errors, the same. It is only for the 
smallest $l$ that there is any sign of what might be a small difference. 

We now try to estimate the maximum possible effect of an excited state 
contribution, by assuming that it is the first excited state that
is providing this. (Since it is plausible that in choosing the lightest
excited state we maximise the shift in $E_0$.) We take the energy of
the excited state, $E_1$, from Table~\ref{table_gsesn3}. We then
perform a 2-exponential fit, with $E_0$ and the two overlaps, $|c_0^2|$
and $|c_1^2|$, as free parameters. There is a partially subjective choice
to make about the start of the fitting range $t\in [t_1,t_2]$. (The
value of $t_2$ is usually chosen for convenience and plays a minor role.) 
In practice we use the lowest value of $t_1$ for which we can get an 
acceptable fit for some values of the parameters, to determine the 
possible range of these parameters. We show the results of this
in the third and fourth columns of Table~\ref{table_gsalln3}, using
uncorrelated and correlated errors respectively. Comparing these to the
values in the first two columns we see that the shift in $E_0$ is only 
noticeable at the largest values of $l$.
And, as we remarked, this estimate is intended to provide
something like an upper bound on the effect of excited state contributions.

\begin{table}[htb]
\begin{center}
\begin{tabular}{|c|cc|cc|cc|cc|}\hline
\multicolumn{9}{|c|}{ $aE_0(l)$ : SU(3) at $\beta=6.0625$} \\ \hline
$l/a$ & $n_t \in$ & nocortt & $n_t \in$ & cortt  & $n_t \in$ & nocortt + ex & $n_t \in$ & cortt + ex \\ \hline
9  & [2,8] & 0.1747(21) & [1,7] & 0.1777(16) & [0,9] & 0.1734(20) & [0,9] & 0.1747(25) \\
10 & [2,8] & 0.2424(15) & [4,7] & 0.2446(17) & [0,9] & 0.2402(14) & [1,7] & 0.2440(23) \\
12 & [2,7] & 0.3510(14) & [3,8] & 0.3516(11) & [1,9] & 0.3488(17) & [1,9] & 0.3475(15) \\
16 & [2,7] & 0.5361(14) & [3,8] & 0.5365(11) & [1,8] & 0.5333(16) & [1,8] & 0.5322(21) \\
20 & [2,7] & 0.7096(20) & [3,8] & 0.7071(23) & [1,5] & 0.7060(24) & [1,8] & 0.7032(31) \\ \hline
\end{tabular}
\caption{Ground state energies extracted in different
ways, from the fitting ranges shown, as described in the text.}
\label{table_gsalln3} 
\end{center}
\end{table}

The results in  Table~\ref{table_gsalln3} are reassuring. We will assume that 
in our excited state calculations, which are less accurate and where it 
would be 
difficult to perform explicit checks of this kind, the above systematic errors 
are small compared to the statistical errors. This is most plausible for
those `excited' states that are ground states in channels with non-trivial
quantum numbers. For other excitations, where secondary exponential may
have a lower energy, this must be only an assumption.   

\subsubsection{ground state analysis}
\label{subsubsection_gsanalysis}

In Tables~\ref{table_gsalln3} and ~\ref{table_gsn6} we present
our high statistics results for the ground state energy in
SU(3) and SU(6) respectively.

\begin{table}[htb]
\begin{center}
\begin{tabular}{|c|c|c|cc|}\hline
\multicolumn{5}{|c|}{SU(6) ; $\beta=25.55$} \\ \hline
$l/a$ & $L_y\times L_z\times L_t$ & sweeps$\times 10^6$ & $n_t \in$ & $aE^{k=1}_0(l)$  
\\ \hline
10  & $16\times 16\times 36$ & 1.5 & [2,5] & 0.2721(13) \\ 
12  & $16\times 16\times 36$ & 1.6 & [2,5] & 0.3858(23) \\ 
16  & $16\times 16\times 16$ & 2.0 & [2,5] & 0.5863(26) \\ 
18  & $18\times 18\times 18$ & 2.0 & [2,5] & 0.6719(25) \\ \hline
\end{tabular}
\caption{\label{table_gsn6} Energy of the ground state of a closed flux 
tube of length $l$, , with statistics and fitting ranges indicated. 
For SU(6) at $\beta=25.55$.}
\end{center}
\end{table}

What do these results for $E_0(l)$ tell us about the coefficient of 
the universal $1/l$ correction? In particular, do we find evidence for
\begin{equation}
E_0(l) \stackrel{l\to\infty}{=} \sigma l - \frac{\pi}{3l},
\label{eqn_bosonic}
\end{equation}
which would correspond to the simplest bosonic string universality 
class where the only massless modes on the flux tube are the
transverse oscillations?
There are many ways to address this question, and we proceed as follows.
We have calculated $E_0(l)$ for the lengths $l=l_1,l_2,...$
where $l_{i+1} \geq l_i$. We define effective L\"uscher coefficients 
and string tensions by
\begin{equation}
E_0(l) = \sigma_{eff} l - c_{eff}\frac{\pi(D-2)}{6l}
\label{eqn_MLceff}
\end{equation}
and determine the parameters $c_{eff}$ and $\sigma_{eff}$ for each 
pair of values $E_0(l_i)$ and $E_0(l_{i+1})$. We expect that at 
small $l$ the value of $c_{eff}$ may be very different from its 
large-$l$ limit, because of additional corrections that are higher 
powers in $1/l$. So the relevant question is whether $c_{eff}\to 1$ 
as $l\to\infty$. Motivated by what we have observed in $D=2+1$
\cite{AABBMT_k1d3}
we perform a similar exercise for the Nambu-Goto expression, writing
\begin{equation}
E_0(l) = \sigma_{eff} l 
\left( 1 - c_{eff}\frac{\pi(D-2)}{3\sigma_{eff} l^2} \right)^{\frac{1}{2}}.
\label{eqn_NGceff}
\end{equation}

We first consider SU(3). We take the values of $E_0(l)$ from the second
column of Table~\ref{table_gsalln3}. (The other sets of values give 
very similar results.) We obtain the values of $c_{eff}(l_i,l_{i+1})$
shown in Fig.~\ref{fig_ceffd4n3}. Note that while the vertical bar
on each point indicates the statistical error, the horizontal bar 
indicates the range, $l_i\surd\sigma$ to $l_{i+1}\surd\sigma$, from
which the value of $c_{eff}$ has been extracted. Given that additional 
massless modes typically contribute a multiple of $\pm 0.5$ to the L\"uscher 
term in eqn(\ref{eqn_MLceff}), what we see in Fig.~\ref{fig_ceffd4n3} 
provides quite convincing evidence for the validity of 
eqn(\ref{eqn_bosonic}). Equally interesting is the fact that when
we use  eqn(\ref{eqn_NGceff}) we obtain $c_{eff}$ close to
unity for all values of $l$. Thus any corrections to the  
Nambu-Goto expression for $E_0(l)$ must be small, even at our smallest
values of $l$ where the flux `tube' is presumably a fat blob, not much 
longer than it is wide. This recalls the situation in $D=2+1$, with
the difference that here the deviations from $c_{eff}=1$, although small, 
are large enough to be visible.

The corresponding results for SU(6) are shown in Fig.~\ref{fig_ceffd4n6}. 
This is very similar to what we have just seen in Fig.~\ref{fig_ceffd4n3}, 
so we may assume that for all $N\geq 3$ the central charge corresponds
to a bosonic string theory where the only massless modes on the
flux tube are the transverse oscillations. All this corroborates the 
claims from earlier calculations that the effective string theory of 
confining fux tubes in $D=3+1$ SU($N$) gauge theories is in the 
universality class of the simple bosonic string theory.

The theoretical analysis of
\cite{AHEK09}
goes much further than this and
tells us that when we expand the ground state energy $E_0(l)$
in powers of $1/l$ not only is the $O(1/l)$ L\"uscher correction
universal, but so also are the  $O(1/l^3)$ and $O(1/l^5)$ terms, 
and that their coefficients are precisely what one gets by expanding 
the Nambu-Goto free string expression in eqn(\ref{eqn_EnNG}) to 
that order. To test this prediction we fit 
\begin{equation}
E_0(l) = \sigma l - 
\sum^{n=2}_{n=0} \frac{c_n^{NG}}{\sigma^n l^{2n+1}}
+ \frac{\tilde{c}}{l^{\gamma}}
\label{eqn_E0fitAH}
\end{equation}
to our results for $E_0$, where the $c_n^{NG}$ are the appropriate 
Nambu-Goto coefficients, and on theoretical grounds we expect 
$\gamma\geq 7$. For a fixed $\gamma$ the fitted parameters are $\sigma$
and the coefficient $\tilde{c}$ of the leading unknown correction.
Beginning with SU(3) we take (as above) the values of $E_0(l)$
listed in the second column of Table~\ref{table_gsalln3}. We fit
these values with eqn(\ref{eqn_E0fitAH}) for various powers of 
$\gamma$ and show in Fig~\ref{fig_AHd4} the $\chi^2$ per degree of 
freedom of the best fit as a function of $\gamma$. We repeat
the exercise for SU(6) using the values in Table~\ref{table_gsn6}.
We see that the SU(3) calculation favours a value $\gamma < 7$. In fact 
the acceptable values are $\gamma=5$ and 3. Since our two lowest 
values of $l$ are very close to $l_c$ (see Table~\ref{table_physics})
and since the SU(3) transition at $l=l_c$ is weakly first order, it 
might be that these fits are influenced by this transition. In that case
one would expect a more reliable result for SU(6) where the transition
is robustly first order. As we see in Fig~\ref{fig_AHd4} a fit
to our SU(6) values with an $O(1/l^7)$ correction is indeed acceptable,
but it is clear that more accuracy is needed for a useful analysis.

\subsection{Ground states with $p_{\shortparallel}\neq 0$}
\label{subsection_groundq12}

We now consider the lightest states with non-zero momentum along 
the flux tube axis, $p_{\shortparallel}=2\pi q/l \neq 0$. To have
$p_{\shortparallel}\neq 0$ a flux tube cannot be invariant under
longitudinal translations: it must have transverse deformations,
corresponding to the excitation of some modes along the tube, and
it is the energies of these that interest us here. So to project
onto these we need to use our extended basis of operators. Of course, 
the energy not only increases from the inclusion of this excitation 
energy but also from the $p_{\shortparallel}^2$ term. So to avoid
the systematic errors associated with large energies, we only present
results for the lightest energy levels with $q=1$ and, with caveats,
for those with $q=2$. Of course we also simultaneously calculate the 
$q=0$ ground state energy, in the same basis of operators, 
to obtain a value for the string tension 
$a^2\sigma$, as described above. This value is then inserted into 
eqns(\ref{eqn_NLRmom},\ref{eqn_NLRspin},\ref{eqn_EnNG}) to provide
parameter free predictions for the energies of states with $q\neq 0$
in the Nambu-Goto model. The parameters of our calculations are
listed in Table~\ref{table_param} with some of the physical properties
listed in Table~\ref{table_physics}.

When presenting our results showing how the flux tube energies $E_n$ 
vary with the flux tube length $l$, it is clearly preferable
to do so using physical units rather than the lattice units,
($aE_n$ and $l/a$) in which these quantities are actually calculated.
We choose to use the calculated value of $a\surd\sigma$ as our unit
and typically plot $E_n/\surd\sigma$ versus $l\surd\sigma$. Since, for 
this purpose, any variation of the extracted value of $a\surd\sigma$  
with the choice of how one fits the $q=0$ ground state is completely 
negligible, the reader should not be concerned that there might
be some hidden circularity in our plots. Using physical
units makes it easier to assess the significance of the plots, 
and since we find that the $O(a^2)$ lattice spacing corrections to
our energies are not visible, there is nothing misleading about this 
use of physical units.

Before continuing, a technical caveat. Our $q=2$ energies 
are amongst the largest of any energies calculated in this 
paper. So the corresponding correlation functions decrease rapidly and it 
is frequently the case that one `identifies' an effective energy 
plateau starting at some $t = t_0$ where the error to signal ratio 
for $t>t_0$ is too large to be useful. This means that the plateau
identification is of low significance and the possibility of the
energy being a significant overestimate is serious. In this case
it makes sense to be guided, in identifying the energy plateaux,
by those calculations that have the greatest statistical accuracy.
As we see from  Table~\ref{table_param}, that is the SU(3) calculation
at $\beta=6.0625$. So in our SU(5) calculation, which is at roughly
the same value of $a\surd\sigma$, we use the same $t$-ranges as in
these SU(3) fits in order to extract the heavy $q=2$ energies. 

For orientation we begin by recalling the spectrum in the 
Nambu-Goto model. For $q=1$ the lightest state has one phonon with 
minimal non-zero momentum: $p=2\pi k/l$ with $k=1$. In the continuum 
limit this will produce two degenerate states, with $J=\pm 1$, 
or alternatively two $|J|=1$ states with $P_t = \pm$. The lightest 
$q=2$ states can either be formed from a single phonon with $k=2$
or from two phonons, each with $k=1$. The former gives two states
with $|J|=1$ and $P_t = \pm$, while the latter gives states with 
$J=0$ and $J=\pm2$ (the latter we organise into states with $|J|=2$ and 
$P_t = \pm$). There is only one $J=0$ state with $P_t = +$, because the 
phonon operators commute so that $a^+_1 a^-_1|0>$ and $a^-_1 a^+_1|0>$
are identical states. On a cubic lattice the two $|J|=1$ states belong 
to a two-dimensional representation and are exactly degenerate even
at finite $a$. So in this section we choose to show the energy of only 
one of these states (whether for $q=1$ or for $q=2$). 
With this convention, we expect to find one $J=1$ state 
occupying the lowest energy level of the $q=1$ sector, and 
four states ($J=0, P_t=+$; $|J|=2, P_t=+$; $|J|=2, P_t=-$; $J=1$)
in the lowest energy level of the $q=2$ sector. These low-lying states 
of the Nambu-Goto model are displayed in Table~\ref{table_NGstates}.

We begin with our high  statistics calculation of the lightest $q=1$ 
and $q=2$ energy levels, in SU(3) at $\beta=6.0625$. 
In Fig.~\ref{fig_Eqd4n3} we plot the energies against the flux tube
length, all in units of $\sigma$, and we also include
the $q=0$ ground state energies from which we extract the value of
$a^2\sigma$. We observe that the $q=1$ ground state energy agrees 
extremely well with the parameter-free Nambu-Goto prediction, as
do the nearly degenerate $q=2$ states, albeit within significantly 
larger statistical errors. The number and quantum numbers of these
$q=2$ states are also exactly as predicted by Nambu-Goto. It is
striking that the quantitative agreement persists down to the 
shortest flux tubes where $l\surd\sigma \sim 2$ and where the flux 
tube width, presumably $\sim 1/\surd\sigma$, is comparable to its length.  

In Fig.~\ref{fig_Eqd4n5} we display the analogous results for SU(5). 
Again the  $q=1$ ground state energy agrees very  well with 
the Nambu-Goto prediction. However for the  $q=2$ states the errors are 
now visibly larger, and while it is clear that they are roughly consistent
with Nambu-Goto, it is hard to make a stronger statement than that.

As we remarked above, a substantial part of the extra energy of the 
$q\neq 0$ states may arise from their non-zero momentum, i.e. from the 
$(2\pi q/l)^2$ term in $E^2$, and we are not interested in that. 
Rather we want to isolate the flux tube excitation energy and  see 
how that compares with, for example, the Nambu-Goto model prediction. 
To do so we follow
\cite{aabbmt_k2d3}
and define 
\begin{equation}
\Delta E^2(q,l)
=
E^2(q;l) - E_{0}^2(l) 
- \left ( \frac{2\pi q}{l}\right )^2
\stackrel{NG}{=} 4\pi\sigma (N_L+N_R) 
\label{eqn_exq} 
\end{equation} 
where the Nambu-Goto value follows from eqn(\ref{eqn_EnNG}), $E(q;l)$ 
is our result for the ground state energy with momentum 
$p_{\shortparallel}=2\pi q/l$, and $E_0(l)$ is the energy of the 
(absolute) ground state with $p_{\shortparallel}=0$.
We choose to subtract the calculated value 
of $E_{0}(l)$ rather than the Nambu-Goto prediction for it, since 
the former already has some small corrections which we would regard 
as belonging to the Casimir energy rather than to the excitation energy 
that we are trying to isolate here. Note also that we assume any lattice
corrections to the continuum energy-momentum dispersion relation
to be negligible at this $\beta$: an uncontroversial assumption, but
one which it would be useful to check explicitly. 

In Fig.~\ref{fig_DEqd4n3n5} we plot some of the 
$q=1$ and  $q=2$ excitation energies, obtained after processing
the energies plotted in Fig.~\ref{fig_Eqd4n3} and Fig.~\ref{fig_Eqd4n5}
through eqn(\ref{eqn_exq}). We omit the SU(5) $q=2$ values, because the 
errors are too large for them to add much here. By removing the momentum
contribution, which will presumably arise in any string model,
this analysis gives us a more precise appreciation of the significance 
of the apparent agreement in  Figs~\ref{fig_Eqd4n3} and \ref{fig_Eqd4n5} 
between our results and the predictions of the free string Nambu-Goto 
model. For $q=1$ the errors remain small and the agreement down
to the smallest values of $l\surd\sigma$ is as striking as before.
For  our lightest, nearly-degenerate, $q=2$ states, which we find to 
have precisely the quantum numbers predicted by the Nambu-Goto model,
the relative errors are now larger. Nonetheless, while one clearly 
sees significant deviations from the free string prediction
at the smaller values of $l$,  one also clearly sees that
these states converge rapidly to the Nambu-Goto value as $l$ increases,
and are in agreement with it within errors for $l\surd\sigma \sim 4$.
So, just as in $D=2+1$
\cite{AABBMT_k1d3},
the level of agreement with the free string model, for these modest flux 
tube lengths, is surprising.

It is clear from Fig.~\ref{fig_DEqd4n3n5} that the lightest
$q=1$ excitation energy shows no dependence on $N$ within the
small statistical errors. What can we say about the $q=2$ 
excitation energies? Since in this case the statistical errors are
too large for a direct comparison to be informative, what
we can do instead is to compare the effective energies
extracted from $t=a$ to $t=2a$, where our statistical errors
are small enough for a significant comparison. Of course these 
energy estimates will contain a significant admixture of higher 
excited states, since our SU(3) effective energy plateaux typically 
set in for $t\geq 2a$, so we are not looking for agreement with
the Nambu-Goto prediction. Rather we simply want to compare the 
SU(3) and SU(5) values as a measure of the $N$-dependence of the
$q=2$ spectrum. The result of processing these effective energies
through eqn(\ref{eqn_exq}) shows
that beyond the very shortest flux tube length, the SU(3)
and SU(5) excitation energies show no significant $N$-dependence
within reasonably small errors.

The above SU(3) and SU(5) results have been obtained at a single 
common value of the lattice spacing, $a\surd\sigma \sim 0.2 $. 
Do these results survive the continuum limit? We address this 
question with an SU(3) calculation at $\beta=6.338$ where the lattice 
spacing is significantly smaller, $a\surd\sigma \sim 0.13$. The
lightest $q=0,1,2$ energies are plotted in Fig.~\ref{fig_Eqd4n3f} 
and if we compare to Fig.~\ref{fig_Eqd4n3} we see that there
appears to be no significant $a$-dependence within the errors. 
Reassured by all this we choose not to attempt a comparable 
small-$a$ SU(5) calculation. 

We can assume from these checks that our calculations, although at 
fixed lattice spacings and fixed values of $N$, do in fact describe
the continuum limit of the gauge theory, at arbitrary $N\geq 3$,
within the errors of our calculations.

\subsection{Excited states}
\label{subsection_excited}

Having examined the ground state energy levels for $q=0,1,2$,
we now turn to the first, and sometimes second, excited energy levels. 
We shall mostly restrict 
ourselves to flux tubes with $q=0$ and $q=1$, since only here 
are the errors small enough for a precise analysis.

For each value of $p_{\shortparallel}=2\pi q/l$, we shall begin by 
recalling the excitation spectrum in the Nambu-Goto model. The string 
tension $\sigma$ is calculated from the ground state energy so there are 
no free parameters. We shall find that not only are the quantum numbers 
of the lightest states precisely as predicted by this model, but 
that the calculated energies of most of these states are close to the 
Nambu-Goto predictions down to remarkably small values of the flux tube 
length $l$. However we shall also find that, for each value of $q$,
the lightest excited state has an energy that is much lower than predicted,
and that this energy increases only weakly as $l$ grows. In each case
the anomalous state has $J=0, P_t=-$ quantum numbers and, partly
because it is light, it is accurately determined in our calculations.

\subsubsection{$q=0$ excited states}
\label{subsubsection_exq0}
    
In the Nambu-Goto model, the states in the first excited  $q=0$
energy level have two phonons of opposite momentum, with this momentum 
taking its minimal absolute value $|p|= 2\pi /l$. Each 
phonon carries unit spin, and these can be either parallel or 
anti-parallel, leading to two states with $J=0$ and two with $J=\pm 2$. 
(See Table~\ref{table_NGstates}.)
These can be re-organised into four states with quantum numbers
$(|J|,P_t,P_l) = (0,+,+),(0,-,-),(2,+,+),(2,-,+)$. In the 
Nambu-Goto model all these states are degenerate.

In our numerical calculations we are able to obtain, with reasonable 
accuracy, the first four excited states above the absolute ground state,
and we find that these have precisely the quantum numbers predicted 
by Nambu-Goto. We then process the calculated energies through 
eqn(\ref{eqn_exq}) so as to obtain the excitation energies, 
$\Delta E^2(q=0,l)$. We display the results in Fig~\ref{fig_DEq0exd4n3}, 
together with the Nambu-Goto prediction. We observe that 3 of the 4 states 
show the by now familiar rapid convergence to the Nambu-Goto prediction as 
$l\surd\sigma$ increases. However one state, the one with $0^{--}$ quantum 
numbers, displays a very different behaviour: it is much lower than the 
predicted value and approaches that value only very slowly as 
$l\surd\sigma$ increases. The fact that the energy is so much lower 
is robust against all our systematic errors: that is to say, this
anomalous behaviour is a reliable result.

Since it is only in the $N\to\infty$ limit that we should expect
a simple stringy description to be possible, it is natural to
ask if this anomalous behaviour is a finite $N$ effect. To 
investigate this we repeat the calculation in SU(5) at a value of $a$ 
that is similar to that of SU(3) at $\beta=6.0625$. Since the leading
large $N$ corrections are expected to be $O(1/N^2)$, if the anomaly
is a finite-$N$ correction it should be smaller in SU(5) by a factor 
of $\sim 3$. We plot the excitation energies obtained in SU(5)
in Fig~\ref{fig_DEq0exd4n5n3f}. We see that the anomalous behaviour
of the lightest $0^-$ state becomes more rather than less
pronounced in SU(5). It is clear that this anomaly is a prominent
feature of the gauge theory at $N=\infty$.

Does this anomalous behaviour survive the continuum limit? To address
this question we repeat the calculation at $\beta=6.338$ where $a^2\sigma$
is smaller by a factor of $\sim 2.3$. The latter is the relevant measure
for lattice spacing corrections, since these are expected to be $O(a^2)$ 
for our plaquette action. We plot the corresponding excitation energies 
of the $0^{--}$ state in Fig~\ref{fig_DEq0exd4n5n3f}. We see that it
continues to be anomalous, much as it was at the coarser value of $a$.
This confirms that what we are seeing is continuum physics.

\subsubsection{$q=1$ and $q=2$ excited states}
\label{subsubsection_exq1q2}

For $q=1$ the first excited energy level in the Nambu-Goto model is
composed of states that have either two or three phonons. 
(See Table~\ref{table_NGstates}.) In the states 
with two phonons, one has momentum $p = 2\pi k/l$ with $k=2$, and the other 
has $k=-1$. The two unit spins add to give 2 states with $J=0$ and 
two with $J=\pm 2$ and these can, as usual, be re-organised into states 
with $P_t=\pm$ for each of $|J|=0,2$. In the three phonon case, two of 
the phonons have $k=1$  and one has $k=-1$. There are now three unit spins 
to be added, giving three sets of states: two pairs with $J=\pm 1$ and 
another pair with $J=\pm 3$. On a cubic lattice the states that become 
$J=1$ and $J=3$ in the continuum limit reside in the same $E$ 
representation and cannot be distinguished on straightforward symmetry 
grounds. Moreover the $P_t=\pm$ partners are in the same multiplet,
and are exactly degenerate, and so it suffices to focus on the $P_t=+$ 
states. Thus we are looking for 4 states with even $J$ and 3 with odd $J$. 
We shall refer to the latter as $J=1$, although of course the state may 
in fact have any odd value of $J$ -- in particular $J=3$.

In our numerical calculations we are able to obtain, with reasonable 
accuracy, a number of  $q=1$  excited states above the  ground state. 
In the even-$J$ sector we find that the lightest four states have  
the quantum numbers $J^{P_t} = 0^\pm, 2^\pm$, which is 
precisely what one expects for the 2 phonon component
of the first excited energy level. Moreover, in the odd-$J$ 
sector, we find three states that are close in energy, 
which is, again, precisely what one expects for the 3 phonon component 
of the first excited energy level.

We begin by processing the even-$J$ energies through eqn(\ref{eqn_exq})
to obtain the excitation energies, $\Delta E^2(q=1,l)$. We first do
this for our most accurate and extensive calculation, SU(3) at 
$\beta=6.0625$, using the energies listed in Table~\ref{table_esq1}. 
The results, for lengths $l/a \leq 24$, are displayed in 
Fig~\ref{fig_DEq1exd4n3}, where they are compared to the Nambu-Goto 
prediction. For comparison we also show in this plot the $q=1$ ground 
state, which has $J=1$ and was shown earlier in Fig.~\ref{fig_DEqd4n3n5}. 
We see, just as we saw for the lowest 4 excited states with $q=0$, 
that 3 of the 4 states show a rapid convergence to the Nambu-Goto 
prediction as $l$ increases, but one state, again with $0^-$ quantum
numbers, displays a very different behaviour: it is much lower than the 
predicted value and only appears to approach that value very slowly as 
$l$ increases. As before, the fact that the energy is lower means that 
this anomalous result is robust against essentially all our systematic 
errors.

As in the case of $q=0$, we repeat the calculation at the smaller 
value of $a$. The results for the $0^-$ state are plotted in 
Fig~\ref{fig_DEq1j0-d4n5n3fn3}, and we see
that there is no significant dependence on $a$. So once again,
we are able to claim that this anomalous behaviour is clearly part 
of the physics of the continuum theory. Similarly a calculation in 
SU(5), whose results are also displayed in Fig~\ref{fig_DEq1j0-d4n5n3fn3}, 
shows that any large $N$ corrections to this anomalous behaviour 
are small. Once again we can claim that this represents the continuum
physics of all theories with $N\geq 3$.

We turn now to the $J=odd$ states with $q=1$. In Fig.~\ref{fig_Eq1j13exd4n3f}
we plot the energies of the ground state and the first four excited states.
We see that above the ground state there are three nearly degenerate 
excited states, consistent with the degeneracy predicted by Nambu-Goto 
for the first excited energy level. Moreover their energies are close 
to the Nambu-Goto prediction for that energy level. We may therefore 
assume that what we are seeing is in fact two $J=1$ states and one 
$J=3$ state. The fourth state is well separated from the other three 
and is very close in energy to what Nambu-Goto predicts for the second  
excited energy level. There should of course be several  $J=odd$
states in this energy level, but we do not attempt to identify these 
since their energies are quite large, so that the errors will be large 
as well. However we can say that there do not appear to be any anomalously 
light odd-$J$ states belonging to this energy level. So we can conclude 
that in the $q=1$ odd-$J$ sector we see no sign of any anomaly up to the 
second excited energy level.

We end this section by extending our analysis to the $q=2$ sector. In 
Fig.~\ref{fig_Eq2j0-d4n3n3f} we show the energy of the lightest 
$0^-$ state in our two SU(3) calculations. For comparison we
show the Nambu-Goto $q=2$ ground state energy, to which the lightest
$0^+,2^\pm$ states rapidly converge, and the energy of the first excited 
Nambu-Goto level, to which the lightest $0^-$ should converge
as $l\to\infty$. This plot clearly confirms that the lightest $0^-$ state 
is anomalous for $q=2$, just as it is for $q=0$ and $q=1$.

\subsubsection{The anomalous $0^-$ states}
\label{subsubsection_anomalous}

We have seen that nearly all the lightest states of a closed flux 
tube converge very rapidly to the free-string Nambu-Goto prediction
as $l$ increases. The striking and sole exceptions have been the 
lightest $J^{P_t}=0^-$ states, and this is so in every sector of 
longitudinal momentum that we have investigated. These $0^-$ states 
show a very large deviation from the free-string prediction, with 
what is at best a very slow approach to the latter. This deviation
persists, perhaps becoming even slightly larger, at larger $N$.
This makes these states very interesting: they are our first 
candidates, whether in 2+1 or 3+1 dimensions, for states that might 
be reflecting the non-stringy massive modes that one expects
to be present for a confining flux tube. 

One possibility is that this $0^-$ state is a mixture of the massive 
and massless modes, with this mixing becoming large at small $l$. 
The coupling will inevitably involve derivatives of the massless
Goldstone field and since the momenta of the phonons in the first 
excited Nambu-Goto state is $|p|=2\pi/l \, \to \, 0$ as $l\to\infty$,
the mixing will decrease as we increase $l$. Moreover as $l$ increases 
the gap between the massless modes decreases $\propto 1/l$ and so 
the number of these modes that are close enough in energy to the massive
mode to mix with it grows $\propto l$. So the mixing with any 
individual mode must eventually decrease as $l$ grows. In such a scenario
we expect this  $0^-$ state to approach the lightest $0^-$ Nambu-Goto 
energy level as  $l$ increases.

A different possibility is that the lightest $0^-$ state is simply a 
massive excitation of the ground state string. Because of the underlying
flux tube, the energy of this state should grow with $l$, and because 
its energy maintains a finite gap with respect to the ground state, 
it will initially appear to approach the first excited 
Nambu-Goto level, but at some higher $l$ it will cross that level,
and at even larger $l$ it will cross higher excited Nambu-Goto energy
levels. In such a scenario there should be an additional $0^-$ 
state that is the first massless excitation and which approaches 
the lightest $0^-$ Nambu-Goto energy level as $l$ grows. Although
such an excited  $0^-$ state will lie at a higher energy at small and
moderate values of $l$, and so will 
be harder to identify accurately, it is clear that its identification is
important if we are to make any progress with understanding the
nature of our anomalous states.

We therefore begin by trying to identify the next excited $0^-$ states,
to see if they include plausible candidates for stringy states.
In Fig.~\ref{fig_DEq0J0Pp-n3fn5n3d4} we plot the excitation energies,
obtained using eqn(\ref{eqn_exq}), of the
lightest and first excited   $p=0$ $0^-$ states, taken
from all our three calculations. We observe that in the range
of smaller $l$, $l\surd\sigma \leq 3.2$, where we have results 
from all three calculations, the excited $0^-$ state 
appears to be approaching the second excited Nambu-Goto energy 
level as we increase $l$, much as the lowest $0^-$ state appears 
to be approaching the first excited Nambu-Goto level, and with
a similar anomalously large deviation. So this would suggest
that there is no higher excited $0^-$ state to approach the first 
Nambu-Goto $0^-$ level, and therefore that the lightest $0^-$ will 
have to do so: and this would point to the first scenario discussed above.
However the larger $l$ calculations in 
Fig.~\ref{fig_DEq0J0Pp-n3fn5n3d4}, which we have performed only in 
SU(3) at $\beta=6.0625$, radically alter this picture: as $l$ grows the 
excited $0^-$ state turns over and begins to approach the first excited
energy level. We include here, for the first time, some results
from our $l=32$ ($l\surd\sigma \sim 6.3$) calculation. As warned, the 
errors are large, but the result is qualitatively striking: the lightest
$0^-$ appears to have crossed the first excited Nambu-Goto level
and is approximately degenerate with the first excited $0^-$, which
is consistent with approaching this Nambu-Goto level. This now
suggests that the physics is in fact closer to the second scenario described
above: the state that is the lightest $0^-$ at smaller $l$
is (mainly) a massive excitation that crosses the stringy energy
levels, while the first excited $0^-$ is (mainly) the stringy
state associated with the first excited Nambu-Goto energy level.
 
We now turn to the $q=1$ $0^-$ states where we are able to obtain
some results for the second as well as for the first excited states.
We plot the corresponding excitation energies in 
Fig.~\ref{fig_DEq1J0-exn3n5n3fd4}. For $l\surd\sigma \leq 3.2$, 
what we see is similar to what we saw for $q=0$ in this range of $l$:
the first excited $0^-$ state also appears to be anomalous, approaching 
only very slowly the corresponding free-string prediction 
(the upper horizontal line). However this behaviour
continues at our larger values of $l$, in contrast to what we have just 
seen for the $q=0$ states in Fig.~\ref{fig_DEq0J0Pp-n3fn5n3d4}: there 
appears to be no clear signal that the $0^-$ excited state turns over,
so as to approach the Nambu-Goto $0^-$ ground state (although the errors
are large enough to allow that at larger $l$), or that the $0^-$ 
ground state crosses this energy level. 

This difference between the $q=0$ and $q=1$ results is highlighted
if one considers the difference of the excitation energies of the lightest 
and first excited energy levels (labelled  $E_0$, $E_1$ respectively):
\begin{equation}
\delta \left\{\Delta E^2(q,l)\right\}
=
\Delta E_1^2(q,l) - \Delta E_0^2(q,l)
\stackrel{NG}{=} 8\pi\sigma .
\label{eqn_Dexq} 
\end{equation} 
Looking at the $q=1$ results in Fig.~\ref{fig_DDEq1J0-exn3d4}
we see that the difference remains consistent with the Nambu-Goto
expectation for all values of $l$. The large anomalous contribution
in the lightest and first excited $q=1$ $0^-$ states appears to be
a common additive contribution that simply cancels in the difference
leaving just the excitation energy of the massless stringy modes.
But with the important caveat that the errors are quite large at larger 
$l$. By contrast, the $q=0$ results in Fig.~\ref{fig_DDEq0J0-exn3d4}
show that after an initial behaviour similar to that seen for
$q=1$, the gap in the excitation energies unambiguously decreases
towards zero.

Since the $q=0$ results for the turnover appear unambiguous, and the 
$q=1$ results at larger $l$ have much larger errors, the simplest 
hypothesis consistent with these errors is that the $q=1$ excited 
$0^-$ has in fact begun to turn over at the largest values of $l$,
and so is paralleling the behaviour of the $q=0$ state, albeit at 
somewhat larger values of $l$. This latter behaviour is to be expected
because of the phonon contents of the $q=1$ and $q=0$ Nambu-Goto states, 
$ (a^+_{2} a^-_{-1} - a^-_{2} a^+_{-1})|0\rangle $ and
$ (a^+_{1} a^-_{-1} - a^-_{1} a^+_{-1})|0\rangle $ respectively.
As remarked earlier, one would expect higher momentum
phonons to interact more strongly with the massive modes,
because the interaction involves derivatives of the massless fields.
Since the phonon momentum is $2\pi k/l$, one expects the $k=2$
phonon for a length $l$ to have the same interaction as a
momentum $k=1$ phonon at length $l/2$. This heuristic argument
would roughly suggest that the interaction energy
of the $q=1$ state of length $l$ will have corrections that
are a mean of those of the  $q=0$ states of length $l$ and $l/2$,
so that the turn-over (indicating a decrease in this interaction energy)
will be delayed for the $q=1$ state as compared to the $q=0$ state.
All this would seem to be consistent with what we see for the
first excited $0^-$ states, although without  more accurate 
calculations one cannot say anything stronger than that. 

To decide whether the $q=0$ and $q=1$ $0^-$ ground states are in fact 
excitations of massive modes, a quantitative analysis is needed. 
Since we are not in a position to provide that at present, the 
following simple heuristic analysis may be useful in deciding whether
such a scenario is plausible\footnote{This arose in a discussion with 
Arttu Rajantie and David Weir}. 
Let us make the simple assumption that the lightest state with a massive 
mode excitation of the flux tube has an energy
\begin{equation}
E(p) = E_0(l) + \left(m^2+p^2\right)^{\frac{1}{2}}
\label{eqn_Eexmq} 
\end{equation} 
where $m$ is the mass scale of this mode, $p$ is the momentum it carries,
and $E_0(l)$ is the energy of the flux tube ground state. 
One then finds 
\begin{equation}
\frac{\Delta E^2(p)}{4\pi\sigma} 
\simeq 
\left(\frac{m}{\surd\sigma}\right)^2 \frac{1}{4\pi}
+
\frac{l\surd\sigma}{2\pi}
\left[\left(\frac{m}{\surd\sigma}\right)^2 +
\left(\frac{2\pi q}{l\surd\sigma}\right)^2 \right]^{\frac{1}{2}}.
\label{eqn_DEexmq} 
\end{equation} 
We plot this function in Figs~\ref{fig_DEq0J0Pp-n3fn5n3d4} and 
\ref{fig_DEq1J0-exn3n5n3fd4} with a value for the mass scale
\begin{equation}
\frac{m}{\surd\sigma} = 1.85.
\label{eqn_mfit} 
\end{equation} 
chosen to provide a qualitative fit. We note that this is an 
entirely reasonable value for the mass scale, being roughly half
the mass gap, $m \simeq M_G/2$, where $G$ is the lightest scalar
glueball. However, while this simple ansatz readily fits both
the $q=0$ and $q=1$ $0^-$ ground states and the differences 
between them, it does not have anything specific to say about the 
behaviour of the higher $0^-$ states. 

If on the other hand we wish to include both the ground and first excited
$0^-$ states in a single analysis, a simple way to do so is to assume 
a simple 2-body mixing between a Nambu-Goto stringy state and a new,
possibly massive state, with a mixing parameter that eventually decreases 
as $l$ increases (for the reasons given earlier). Such a mixing leads to an 
equal splitting in $\Delta E^2$ around the average of the un-mixed energies 
and since, as we see from  Fig.~\ref{fig_DEq0J0Pp-n3fn5n3d4}, this average
is close to the Nambu-Goto energy (taking seriously the large errors
on our largest value of $l$) and since the unmixed stringy state already
has this energy, this means that our putative massive mode
is also close in energy to that of the Nambu-Goto state and so
looks like a massless mode -- an unexpected conclusion.
Moreover the same analysis does not then describe the $q=1$ 
energies in Fig.~\ref{fig_DDEq1J0-exn3d4}. So such a simple
mixing picture does not appear to be tenable.

Turning now to the second excited state in the $q=1$ $0^-$ sector,
we see from Fig.~\ref{fig_DEq1J0-exn3n5n3fd4} that it does not 
seem to be anomalous, but is close to the 
appropriate Nambu-Goto prediction. Recall that while the lightest
$q=1$ $0^-$ Nambu-Goto state is constructed from two phonons, the
first excited $0^-$ energy level involves states with both 2 and 4 
phonons, with both sectors producing $0^-$ excited states. 
Thus there is room in the Nambu-Goto first excited energy level for 
this $0^-$ state, irrespective of whether the anomalous first excited 
state turns out to belong to the lightest or to the first excited string 
energy level.

\subsection{Closed loops of k=2 flux}
\label{subsection_k2}

In this paper our main focus is on confining flux tubes that carry
flux in the fundamental representation. However, one can also study
flux tubes carrying flux in higher representations. In particular
consider fluxes that emanate from sources that transform as the
product of $k$ fundamentals and any number of adjoints, where
$0 \leq k \leq N/2$.
The corresponding flux tubes are called $k$-strings. At finite $N$
such a flux tube will be unstable if it can decay via
gluon screening to a flux tube with lower energy per unit length.
However the latter will also be in the same $k$-sector since
gluons are adjoint particles. So the $k$-string with the smallest
string tension, $\sigma_k$, will be absolutely stable. It turns out 
\cite{d4kstring,blmt01}
that $\sigma_k < k \sigma$, so the lightest $k$-string is not
merely $k$ independent fundamental flux tubes, but may be thought
of as a bound state of these. Such a flux tube clearly has 
additional internal structure to that possessed by a fundamental
flux tube and it would be interesting to see how this affects
the energy spectrum of closed $k$-strings. This would provide
us with some insight into how such extra massive modes are encoded
in the effective string theory that describes the $k$-strings.
Precisely such a study has been carried out in $D=2+1$ gauge theories
\cite{aabbmt_k2d3}.

\begin{table}[htb]
\begin{center}
\begin{tabular}{|c|c|c|}\hline
\multicolumn{3}{|c|}{SU(6) ; $\beta=25.55$} \\ \hline
$l$ & $E^{k=2A}_0(l)$ & $E^{k=2S}_0(l)$ \\ \hline
10  & 0.5125(122)  & 0.694(38) \\ 
12  & 0.7030(190)  & 0.957(64) \\ 
16  & 1.0192(29)   & 1.291(7) \\ 
18  & 1.1647(40)   & 1.453(26) \\ \hline
\end{tabular}
\caption{\label{table_gsk2n6} Energies of the lightest closed 
$k=2$ flux tubes of length $l$, carrying flux in the totally 
symmetric, $k=2S$, and anti-symmetric, $k=2A$, representations
in  SU(6).}
\end{center}
\end{table}

In this paper we do not attempt such a study but initially limit 
ourselves to the $k=2$ ground state, as calculated in our high 
statistics SU(6) calculation (where our operator basis is too small 
to attempt a full spectrum calculation). Let $\psi_0$ be
the  best operator that our variational calculation produces
for the absolute ground state, using our full basis of operators.
We list in Table~\ref{table_gsk2n6} the energy of this ground state
as a function of its length $l$.
Now, the operator basis we use contains equal numbers of totally 
symmetric, $k=2S$, and totally anti-symmetric, $k=2A$, operators.
(We refer to
\cite{aabbmt_k2d3}
and
\cite{blmt01}
for a more detailed discussion of the appropriate operators.)
What we find is that the absolute ground state, $\psi_0$, is
almost entirely $k=2A$, corroborating the observation in
\cite{blmtuw04}. 
To be specific, if we denote by $\Phi_{2S}$ the space of $k=2S$ 
operators, $\phi_{2S}$, spanned by our basis, then the overlap
of $\psi_0$ on $\Phi_{2S}$ turns out to be as follows:
\begin{eqnarray}
\max_{\phi_{2S}\in\Phi_{2S}} 
| \langle \psi^\dagger_0 \phi_{2S} \rangle |^2
& = & 0.0079(6)  \qquad \ \,: l=10 \nonumber \\
& = & 0.00019(3) \qquad: l=12 \nonumber \\
& = & 0.00161(3) \qquad: l=16 \nonumber \\
& = & 0.00003(1) \qquad: l=18 
\end{eqnarray}
This is extremely small and tells us that to a very good approximation
the $k=2$ absolute ground state is in the totally antisymmetric
representation. (And indeed, if we restrict our operator basis to $k=2A$, 
the ground state  $\psi_0$ that we obtain differs insignificantly
from the one obtained using the full basis.) This strongly suggests
that, just as we showed in $D=2+1$
\cite{aabbmt_k2d3},
the low-lying eigenspectrum consists of states that are, to a good
approximation, either wholly $k=2S$ or $k=2A$. While we cannot attempt 
here an analysis as complete as the one we performed in $D=2+1$, we are 
able to calculate the ground state in the  $k=2S$ sector, whose energies
we list in  Table~\ref{table_gsk2n6}. This is again a state that changes 
negligibly if we use the full $k=2$ basis rather than the restriction
to $k=2S$.

In Fig.~\ref{fig_Ek2d4n6} we show how the energies of the $k=2A$ and 
$k=2S$ ground states vary with $l$. On the plot we also show the best
fits using the free string Nambu-Goto expression, where the parameters
being fitted are the respective string tensions, $\sigma_{k=2A}$ and  
$\sigma_{k=2S}$. On such a plot, where both $E$ and $l$ are rescaled
by the appropriate string tensions, the two Nambu-Goto curves will of
course coincide. What is interesting is that the two sets of energies
also appear to coincide: the (slightly) unstable $k=2S$ flux tube
behaves just like the stable $k=2A$ flux tube. And both show the
approximately linear rise with $l$ associated with the formation
of a confining flux tube.

It is clear from  Fig.~\ref{fig_Ek2d4n6} that at smaller $l$ the 
corrections to the Nambu-Goto fit become quite substantial. To resolve 
this more clearly we perform fits using eqns(\ref{eqn_MLceff})  
and (\ref{eqn_NGceff}) just as we did in the case of the fundamental
flux tube. The results are shown in Fig.~\ref{fig_ceffd4n6_k2A}
for the $k=2A$ flux tube. (The errors on our $k=2S$ energies
are too large to allow a useful analysis of this kind.)
Comparing to the comparable plots for fundamental flux tubes, 
in Figs~\ref{fig_ceffd4n3} and ~\ref{fig_ceffd4n6}, we see that
the corrections to Nambu-Goto are very much larger here. This is 
just what we observed in our $D=2+1$
\cite{aabbmt_k2d3}
calculations. (And has previously been observed for the effective 
L\"uscher correction in the $D=3+1$ SU(6) gauge theory in
\cite{HMMT04}.)
Nonetheless this plot does provide some additional significant 
evidence that the $k=2$ flux tube is also in the bosonic string 
universality class.

\subsection{Effective string theory: discussion}
\label{subsection_discussion}

Our above results show that most of the low-lying excited states of the
fundamental flux tube are remarkably close to the predictions of the 
free-string Nambu-Goto model, even down to values of $l$ where
the flux tube is not much longer than it is wide ($\sim 1/\surd\sigma$).
Now, we recall that the effective string analysis in
\cite{AHEK09}
has shown that if we expand the excited string state energies $E_n(l)$ 
in inverse powers of $l$ then they are universal to $O(1/l^5)$ in $D=2+1$ 
and to $O(1/l^3)$ in $D=3+1$, and that they must therefore 
agree with Nambu-Goto to that order. (If they are in the bosonic string 
universality class as is strongly implied by the numerical results for the 
ground state.) It is therefore natural to ask to what extent our numerical
results are merely reflecting this theoretical result.

To address this question we take, as an example, the $q=0$ excited states
in SU(3) at $\beta=6.0625$, whose excitation energies were plotted 
in Fig.~\ref{fig_DEq0exd4n3}, but omitting for the moment the `anomalous'
$J^{P_t} = 0^-$ state. We plot their energies as a function of length $l$
in Fig.~\ref{fig_Eq0exd4n3}. We also plot the theoretical prediction
to $O(1/l^3)$, 
\begin{equation}
E(l)
= \sigma l  
+ \frac{11\pi}{3}\frac{1}{l}
- \frac{121\pi^2}{18}\frac{1}{\sigma l^3}
\label{eqn_EnAH}
\end{equation}
which includes all the terms that are currently known to be universal.
The value of $\sigma$ comes from fitting the ground state energy
so this prediction is in fact parameter-free. (The variation of
$\sigma$ with the 
precise form of the ground state fit is completely negligible in the 
present context.) For comparison we also plot the prediction to just 
$O(1/l)$, i.e. just the `L\"uscher correction', and also the full 
Nambu-Goto prediction.

We see from Fig.~\ref{fig_Eq0exd4n3} that the agreement of our numerical 
results with Nambu-Goto is, in fact, far too good to be a mere corollary 
of the known theoretical universality results. Does this perhaps indicate 
that the next one or two terms, in powers of $1/\sigma l^2$, are also 
universal, and would this then suffice to explain our numerical results?
The answer to the latter question is no. As indicated by the way that 
the $O(1/l)$ and $O(1/l^3)$
curves in Fig.~\ref{fig_Eq0exd4n3} oscillate around the Nambu-Goto curve, 
the expansion of the Nambu-Goto expression in powers of $1/l$ is 
divergent in the range of $l$ where most of our numerical results are 
located. Indeed, as we see from eqn(\ref{eqn_EnNG}), the expansion of the
energy in powers of $1/\sigma l^2$, for these $N_R=N_L=1$ states, 
ceases to be convergent for
\begin{equation}
l\surd\sigma \leq \left(\frac{22\pi}{3}\right)^{\frac{1}{2}}
\simeq 4.8.
\label{eqn_NGdivg}
\end{equation}
So the analytical derivative expansion for this state only holds for 
$l\surd\sigma \gtrsim 4.8$. Although we have one or two values in this
range, and therefore make contact with the region to which the theoretical
analysis applies, the bulk of our calculations lie at smaller values of 
$l$. Thus in reality they complement the theoretical analysis, while
maintaining enough overlap to ensure some continuity between the
numerical and analytical results.

Our results for the range of smaller $l$ where the derivative expansion no 
longer converges, are telling us something simple about that expansion: 
it should, to all orders, remain close to that of Nambu-Goto.
In effect our numerical results are telling us that the appropriate 
starting point for an effective string picture of closed flux tubes is 
neither the classical background string nor the Gaussian action, but 
rather the full free string Nambu-Goto action. That is to say, if we 
write the energy of the the $n'th$ eigenstate as 
\begin{equation}
E_n(l) = E^{NG}_n(l) + \delta E_n(l),
\label{eqn_EcornNG}
\end{equation}
where $E^{NG}_n(l)$ is the Nambu-Goto energy, then Fig.~\ref{fig_Eq0exd4n3}
tells us that for the $0^+$ and $2^\pm$ states that are shown there this
correction term, $\delta E_n(l)$, is `small' over the whole range 
$l\surd\sigma \geq 2.5$: 
\begin{equation}
\delta E_n(l) \ll \sqrt{\sigma} \quad ; \quad 
l\surd\sigma \geq 2.5.
\label{eqn_delENG}
\end{equation}
Since the theoretical analysis in
\cite{AHEK09}
tells us that the expansion of  $\delta E_n(l)$ in powers of
$1/l$ cannot start with a power smaller than $1/l^5$, we know that
at large $l$ it must have the form: 
\begin{equation}
\frac{\delta E_n(l)}{\surd\sigma} 
\stackrel{l\to\infty}{=} 
\frac{c_n}{(\surd\sigma l)^5} 
\left( 1  + \frac{c_{n,1}}{\sigma l^2} + \frac{c_{n,2}}{(\sigma l^2)^2} 
+ ... \right).
\label{eqn_EcornNGa}
\end{equation}
The different possibilities can be usefully illustrated by the following
two scenarios. 

$\bullet$ The fact that the corrections are small down to small $l$
would follow most simply if the expansion coefficients in
eqn(\ref{eqn_EcornNGa}) were to be small so that the first term
dominated even at small $l$:
\begin{equation}
\frac{\delta E_n(l)}{\surd\sigma} 
\simeq
\frac{c_n}{(\surd\sigma l)^5} \quad ; \qquad l > l_c
\label{eqn_EcornNGb}
\end{equation}
Indeed if we fit the observed deviations of the states in 
Fig.~\ref{fig_Eq0exd4n3} from Nambu-Goto, we find that they can be roughly 
described by simple $O(1/l^5)$ corrections. This is demonstrated
in Fig.~\ref{fig_Eq0exd4n3_NGcorfit}, where the 
dotted curves incorporate such a correction to the Nambu-Goto
prediction. The magnitude of the coefficients in the two fits are
$c_n \simeq -16,\  +9$ respectively. These may appear quite large,
but they are in fact $\sim 50$ times smaller than the coefficient
of the $O(1/l^5)$ term in the expansion of $E_{NG}(l)$, and so should
be regarded as `very small' in natural units that include appropriate
powers of $4\pi$. Nonetheless, while this qualitative fit demonstrates the 
possibility of such a scenario, our calculations are not accurate enough
to encourage a more precise analysis.\\
$\bullet$ Alternatively, it may well be that as we decrease $l$ from large 
values, the series expansion for $\delta  E_n(l)$ diverges at some value 
of $l$ well above $l_c$, in the same way as the Nambu-Goto expansion does. 
For lower $l$ it should be resummable into some finite expression, since 
we know the flux tube energy does not diverge for any finite $l > l_c$. 
This means
that at small $l\surd\sigma$ the apparent power of $l$ may well be very 
different from the asymptotic $\propto 1/l^5$ behaviour. This can
be illustrated by the following simple example of a resummed series:
\begin{equation}
\frac{\delta E_n(l)}{\surd\sigma} 
=
\frac{c_n}{(\surd\sigma l)^5} 
\left( 1  + \frac{d_n}{\sigma l^2}\right)^{\gamma_n}
\stackrel{small \ l}{\propto} 
\frac{c_nd_n^{\gamma_n}}{(l\surd\sigma)^{5+2\gamma_n}}.
\label{eqn_EcornNGc}
\end{equation}
(We assume that $d_n/\sigma l^2 \gg 1$ over most of our range of $l$, 
as this is dictated by the radius of convergence we have assumed.)
At small $l$ this can decrease faster or slower than $1/l^5$
depending on the sign of $\gamma_n$ (which in turn depends on how 
the coefficients in eqn(\ref{eqn_EcornNGa}) oscillate in sign).
Let us assume, for example, that this series expansion diverges
around the same length as Nambu-Goto, i.e. at $l\surd\sigma \sim 5$
in the context of the states in Fig.~\ref{fig_Eq0exd4n3}. This implies 
that $d_n\sim 25$ in eqn(\ref{eqn_EcornNGc}). Given that the 
deviation from Nambu-Goto in Fig.~\ref{fig_Eq0exd4n3_NGcorfit} is
$\delta E/\surd\sigma \sim 0.5$ at $l\surd\sigma \sim 2$, we
see from eqn(\ref{eqn_EcornNGc}) that, very roughly, we need
$c_n \sim 16/6^{\gamma_n}$. If $\gamma_n > 0$ this leads to
a value for $c_n$ that seems unnaturally small when compared to
the corresponding Nambu-Goto coefficient. To allow a larger
value of $c_n$ one needs to have $\gamma_n < 0$. In that case,
as we see from eqn(\ref{eqn_EcornNGc}), the effective power of $1/l$ that
one sees at small $l$ will be smaller than 5. One sees from this example
the linkage between the coefficient $c_n$ of the leading 
asymptotic correction, the value of $l\surd\sigma$ at which the
expansion of $\delta E_n$ diverges, and the effective power of 
the correction at smaller $l\surd\sigma$.
We illustrate this in Fig.~\ref{fig_DEq0exd4n3_NGcorfit} with a fit 
to the excitation energies of these states, using a simple
$c/(l\surd\sigma)^4$ correction term in $\Delta E^2$. 
It is straightforward to see 
that this corresponds to a correction term in eqn(\ref{eqn_EcornNG})
of $\delta E = \left(E^2_{NG}+\Delta\right)^{1/2} - E_{NG}$
where $\Delta = 4\pi\sigma c/(l\surd\sigma)^4$, which is
a resummed expression that diverges at the same place as
Nambu-Goto and whose leading piece is of the form in 
eqn(\ref{eqn_EcornNGc}) with $\gamma_n=-1/2$. Clearly this illustrates
that such a fit is possible.

So far this is very similar to what we found in 2+1 dimensions
\cite{AABBMT_k1d3}. 
However, the important difference is that here we have also identified 
some additional states, the ground and first excited $0^-$ states with
longitudinal momenta $q=0$ and $q=1$ (as well as the ground state 
with $q=2$), which are far from the Nambu-Goto predictions over much
of our range of $l$. At moderate values of $l\surd\sigma$ the deviation 
from Nambu-Goto decreases much more slowly with increasing $l$
than $O(1/l^5)$: the dotted line in Fig.~\ref{fig_DEq0exd4n3_NGcorfit}
represents a very slowly varying $\propto 1/l^{1.7}$ correction 
to $\Delta E^2$. Such a behaviour could, in principle, be accommodated 
by the small-$l$ behaviour of a correction of the form shown in 
eqn(\ref{eqn_EcornNGc}). In addition, we have also seen that at larger 
$l$ the $p=0$ $0^-$ state does appear to break away from this slow 
variation, approaching the Nambu-Goto energy level more rapidly.
This can also be naturally accommodated by a correction as in 
eqn(\ref{eqn_EcornNGc}), since at larger $l$ the term $d_n/\sigma l^2$
becomes small enough that the power behaviour of the correction returns 
to its asymptotic value ($\propto 1/l^5$ in this case). Such an
interpretation is however less plausible if we take seriously the 
apparent crossing of the Nambu-Goto level at our largest value of $l$.
And it is not consistent with the apparent behaviour of the $p=0$ 
excited $0^-$ state which at larger $l$ appears to turn over and 
approach the lightest $0^-$ Nambu-Goto energy level. As discussed
in some detail in Section~\ref{subsubsection_anomalous}, a much more
natural explanation of these states is that they reflect massive
mode excitations, and these cannot be easily described in the framework
of a simple effective string action.

\section{Conclusions}
\label{section_conclusion}

In this paper we have calculated the energies of $\sim 20$ of the lightest 
excitations of a confining flux tube that is closed around a spatial torus.
We have varied the size $l$ of the torus from the very short, close to the 
critical length $l \surd\sigma \sim 1.6$ where the theory undergoes a 
finite-volume deconfining transition, up to moderately large lengths,
$l \surd\sigma \sim 6.3$. Although our most extensive calculations
have been in the SU(3) gauge theory at a fixed value of the lattice 
spacing $a$, we have also performed calculations in SU(5) and SU(6), 
at the same $a$, so as to have control over the large-$N$ limit, and 
in SU(3) at a smaller value of $a$, so as to have control over
the continuum limit. We find no significant $a$ or $N$ dependence
of our energy spectra, within our statistical errors, so that our results 
can be read as being for the continuum gauge theory, and for all $N\geq 3$.

Our goal has been to learn more about the effective string
action that describes confining flux tubes and which can hopefully teach
us something about the string theory that may describe the physics 
of a confining gauge theory, at least at large $N$. In the last few
years there has been startling analytic progress in this direction
\cite{LW04,JD,DM06,AHEK09},
and one has learned that if one expands the effective string action in
powers of $1/l$, then the first few terms are universal. This also
implies that, to this approximation, the effective action is the same
as Nambu-Goto in flat space-time, i.e. a free string theory.

Our calculations are largely complementary to this analytic work.
The latter focuses on long flux tubes, where $1/l\surd\sigma$
is small, and on excited states whose gap to the ground state
is small: $\Delta E(l) \ll m_G$ where $m_G$ is the mass gap of 
the theory. By contrast our calculations concentrate on small
to medium values of $l\surd\sigma$. This is largely dictated by the
fact that most of the systematic errors in a numerical estimate
of an energy increase as that energy increases (as discussed in
Section~\ref{subsection_energies_lat}). To control these
errors in our calculation, which makes important use of a variational 
approach, we use a very large basis of lattice operators, as described 
in Section~\ref{subsection_op_lat}. This is the main technical innovation 
of our calculations. This strategy works reasonably well, but is 
significantly less successful than in our similar calculations
in 2+1 dimensions
\cite{AABBMT_k1d3}.
For that reason the present calculations should be regarded as
exploratory. A better choice of operators could increase the
range and accuracy of such calculations quite dramatically. This
is an important direction for future work.

One of our main systematic errors does not affect the lightest state in
a sector with particular quantum numbers, and most of our calculated
states fall into this category. These include ground states with 
momenta along the flux tube $p=0,\, 2\pi/l$ and $4\pi/l$, and with spin 
around the flux tube axis of $|J|=0,1,2$ (a continuum and so 
not entirely reliable
labelling of the content of the relevant representations of
the cubic group), and with various parities, as described in 
Section~\ref{subsection_quanta}. Almost all these quantum numbers
entail non-trivial excitations of the flux tube and so are dynamically
interesting. (In contrast to transverse momentum which would teach us 
nothing new and which we therefore ignore.)

Our high statistics calculation of the absolute ground state,
described in Section~\ref{subsubsection_gsanalysis},
demonstrates quite accurately that the $O(1/l)$ universal L\"uscher 
correction is in the simple bosonic string universality class,
and so confirms earlier work. (In a brief aside in 
Section~\ref{subsection_k2}, we also confirmed that $k=2$ flux tubes 
appear to belong to the same universality class.) 
In the case of SU(6), but not SU(3), we obtained some weak numerical 
evidence that the leading non-universal $O(1/l^\gamma)$
term is consistent with $\gamma \in [3,7]$. This is to be compared 
to the recent analytic result that $\gamma \geq 7$
\cite{AHEK09}.
It will be hard to do much better than this because one is analysing 
very small `Casimir energy' corrections to the ground state  energy,
ideally on long flux tubes that have a large energy.
By contrast, the excited states receive a much larger contribution 
from the excitations of the string and thus provide a much better 
arena within which to test ideas about the effective string action. 
 
The most striking feature of our spectrum of excited states
is how closely most of these states track the predictions of the 
Nambu-Goto free string theory, and that they do so all the way down 
to lengths $l\surd\sigma \sim 2-3$ where the flux tube is presumably not 
much longer than it is wide. This is evident from the various plots in 
Section~\ref{subsection_groundq12} and Section~\ref{subsection_excited},
where we compare our calculated energies to the parameter-free
predictions of the free string theory. (The only parameter is $\sigma$
and that is determined by fitting the absolute ground state.)
It is of course natural to ask whether this might not simply be
a consequence of the fact that we now know
\cite{AHEK09}
on general grounds that an expansion of the energy $E(l)$ in powers
of $1/l$ must coincide with the corresponding expansion of the
Nambu-Goto energy  $E_{NG}(l)$ up to at least $O(1/l^3)$. The answer
to that question, 
as illustrated in Fig.~\ref{fig_Eq0exd4n3}, is `no'. Indeed, the
agreement with Nambu-Goto persists down to values of $l\surd\sigma$
where the expansion of $E_{NG}(l)$ in powers of $1/l$ is no longer
convergent and one has to use its full, resummed expression.
Note that our range of flux tube lengths, while focused on smaller
values, does in fact extend into the region where the Nambu-Goto
expansion converges (for these states), so that we overlap with
the range of $l$ to which the analytic calculations apply.
So, just as in our earlier 2+1 dimensional calculations
\cite{AABBMT_k1d3}, 
the implication of this result is very clear: the sensible starting 
point for analysing the effective string action is the Nambu-Goto 
free string theory and not the conventional Gaussian string 
action (in static gauge). If we do the latter then we will 
need to calculate corrections to all orders in $1/l$ in order to
access the shorter values of $l$ where we find that the spectrum 
still maintains a simple stringy character. If, on the other hand, we
do the former, then it appears that something like a modest $O(1/l^5)$ 
correction might be sufficient, as illustrated in 
Fig.~\ref{fig_Eq0exd4n3_NGcorfit}, or some resummed version, with
a modest leading coefficient, as illustrated in 
Fig.~\ref{fig_DEq0exd4n3_NGcorfit}.

This is, however, not the whole story. We have also found a few states 
for which the corrections to Nambu-Goto are very large. These
corrections reduce the energy, making the state relatively light --
incidentally ensuring that our calculations of these energies are 
accurate and reliable. These states are of particular interest 
because they potentially encode the excitation of the massive
modes that we expect to be associated with the non-trivial intrinsic
structure of the flux tubes. (Something we were unable to detect
in our more accurate analysis of fundamental flux tubes in $D=2+1$.)
All of these `anomalous' states, or at least
all those that we have identified so far, turn out to have 
$J^{P_t}=0^-$ quantum numbers. (Here $P_t$ is the two dimensional 
parity in the plane transverse to the flux tube axis.) The lightest
$0^-$ flux tube with the lowest non-zero longitudinal momentum, 
$p_{\shortparallel}=2\pi/l$, remains far from the Nambu-Goto
prediction over our whole range of lengths $l\surd\sigma \in [1.8,6.3]$
and, as we see in Fig.~\ref{fig_DEq1J0-exn3n5n3fd4}, it appears
to approach the stringy prediction only very slowly. By contrast,
the lightest $0^-$ flux tube in the $p_{\shortparallel}=0$ sector
shows a very similar behaviour till about $l\surd\sigma \sim 4$,
but then turns upwards and appears to have crossed the 
Nambu-Goto energy level in our largest-$l$ calculation at 
$l\surd\sigma \simeq 6.3$. At the same time, as we see in
Fig.~\ref{fig_DEq0J0Pp-n3fn5n3d4}, the first excited $0^-$, after
initially appearing to increase towards the first excited Nambu-Goto
level, turns downwards, appearing to approach the Nambu-Goto $0^-$ 
ground state as $l$ increases. A natural interpretation of this is
that the lightest state is (largely) a massive excitation, while
what we took to be the first excited state becomes the $0^-$ 
stringy ground state at large $l$. To go beyond such a 
qualitative observation requires a good understanding of how to 
simultaneously describe the massive and massless modes of a flux tube,
and this we do not have at present. However, as we saw in 
Section~\ref{subsubsection_anomalous}, a simple ansatz for the
massive modes naturally describes the behaviour of the lightest $0^-$ 
states for both $p_{\shortparallel}=0$ and $p_{\shortparallel}=2\pi/l$ with
a mass scale that is about one half of the lightest glueball mass.
While such a simple picture does not say anything specific about the
observed large corrections to the next, heavier $0^-$ states, it is 
natural, if these states are (approximately) the lightest massless
mode excitations, that they should show the somewhat different 
behaviour for  $p_{\shortparallel}=0$ and $p_{\shortparallel}=2\pi/l$ 
that is observed, as a results of the derivative couplings such
Goldstone fields must possess. 

In summary, our result is that the lightest closed flux tube states 
appear to fall into two distinct classes. The first class includes most 
of the states and here the spectrum is remarkably well described by 
the free bosonic string theory. This complements and extends to shorter 
lengths recent analytic results for very long flux tubes. 
The second class includes only a few states, all of which (so far) 
possess $0^-$ quantum numbers, and here the energies are remarkably 
far from the predictions of the simple Nambu-Goto model. 
The lightest of these states, for both zero and non-zero longitudinal
momenta, display a behaviour that is consistent with being primarily
a massive excitation that does not couple too strongly with the
massless modes. It is clear that a significant, but realistic, 
improvement to our calculations would have the potential of providing
convincing answers to some very interesting questions.

\vspace*{0.35in}

\leftline{\bf{Note added:}}
As the revised version of this paper was being prepared,
two further papers have appeared containing interesting 
new analytic results on the effective string action
\cite{aharony_new}. 
A summary of these and other results may be found in a recent talk
\cite{aharony_talk}. 
From the point of view of our work in this paper, 
a very interesting result is that the coefficients 
of $all$ those operators in the full effective action that 
appear in the expansion of the Nambu-Goto action (the `scaling 0'
operators) are in fact universal and equal to their Nambu-Goto
values
\cite{aharony_talk}. 
This result provides a conceptually clean separation of the full 
effective action into a piece that is Nambu-Goto and a qualitatively
different piece that is the correction to it. Another 
interesting result is that the leading correction to Nambu-Goto
is in fact universal
\cite{aharony_talk}; 
this arises not from the constraint of open-closed 
duality described earlier in this paper, but from the constraint that 
world-sheet conformal symmetry should be maintained at the quantum level
 -- a constraint that is not obvious in static-gauge, but can
be seen in the conformal gauge approach of Polchinski-Strominger, 
and which the Nambu-Goto model fails to satisfy.

\vspace*{0.7in}

\section*{Acknowledgements}

BB thanks O.~Aharony for many useful discussions and the Weizmann 
Institute for its hospitality during part of this work. As this paper 
was being completed, MT participated in the {\it Confining Flux 
Tubes and Strings} Workshop at the ECT*, Trento, where 
there were many talks relevant to the present work (available on 
the ECT web-site): MT is grateful to the participants for useful 
discussions and to the ECT for its support in part of this research
under the European Community - Research Infrastructure Action under 
the FP7 `Capacities' Specific Programme, project `HadronPhysics2'.
AA was supported by the EC 6th Framework Programme Research and Training 
Network MRTN-CT-2004-503369 and the Leventis Foundation during the earlier 
part of this work and by NIC, DESY during the final part of this work. 
BB has been supported by the U.S. Department of Energy under 
Grant No. DE-FG02-96ER40956. The computations were carried out on EPSRC 
and Oxford funded computers in Oxford Theoretical Physics. 

\vspace*{1.0in}
 
\clearpage

\begin{appendix}

\section*{Appendix:  $\quad$ Compilation of the energy spectra}
\label{section_appendix_results}

In this Appendix we list the numerical results which form the basis 
for the analysis in the body of this paper. We try to do so in sufficient
detail that they can be used by the interested reader to test
theoretical ideas about the spectrum.

Before doing so we list in Table~\ref{table_param} the parameters of the 
calculations that were used to calculate the closed flux tube eigenspectra.
Note that while the total number of
operators is $O(1000)$, after organising them into linear combinations
of particular quantum numbers, one typically has $\sim 50 - 200$ 
operators.

This Table complements the ealier Table~\ref{table_physics} where we listed
various physical properties in the lattice units appropriate to each of the
different calculations. Note that the string tension in lattice units, 
$a^2\sigma$, which is extracted
from the absolute ground state of the flux tube, will vary (very)
slightly with the functional form of the fit. Thus the reader may wish 
to perform his own fits to the $q=0$, $|J|=0$, $P_t=+$ energies
listed in Table~\ref{table_gsq012}, in order to obtain a value of
$a^2\sigma$ appropriate to the model they are testing.

In any given calculation all our dimensionful parameters and results are 
necessarily given in units of the lattice spacing, e.g. $aE$ or $L=l/a$.
Any expression for the spectrum should look exactly the same if all
dimensionful quantities are expressed in lattice units. For example,
to use the expression in eqn(\ref{eqn_EnNG}) with our tabulated values
simply replace $E,\sigma,l$ by $aE,a^2\sigma,l/a$ respectively.

Before listing our energies, some preliminary remarks. \\
1. The most accurate calculation, covering the largest range of flux tube
lengths, is the SU(3) one at $\beta=6.0625$. \\
2. The energies that are the largest in lattice units, are the least 
reliable. (Typically the larger the error, the greater the uncertainty on
the estimate of that error.) \\
3. Our use of the term `ground state' may
sometimes be ambiguous. Sometimes we mean the lightest state
with given values for all conserved quantum numbers, but sometimes
we mean the lightest state with a subset of those quantum numbers,
e.g. the ground state with $p=0$. The reader should not be perturbed
when this arises in the text or captions: the quantum numbers
are clearly indicated in the Tables and for each set of quantum
numbers all states are 
listed up to some energy, i.e. we do not miss any states of intermediate 
energy just because they may be difficult to estimate.\\
4. We remind the reader that since we only have the square group of
rotations in a plane as exact symmetries, what $|J|=0$, $|J|=1$, $|J|=2$ 
really mean is $|J|=0,4,...$, $|J|=1,3,...$, $|J|=2,6,...$.

In the following Tables we list the energy $E$ in lattice units as $aE$,
the length $l$ of the flux tube also in lattice units as $l/a$,
the longitudinal momentum $p=2\pi q/l$ (often labelled as
$p_{\shortparallel}$ in the text)  is labelled by the integer $q$, and
we give the absolute spin around the flux tube axis, $|J|$, and
the transverse parity $P_t$ (and the longitudinal parity $P_l$ for
$p=0$ states). The `phonon' content in the Nambu-Goto model of the lightest 
such states is listed in Table~\ref{table_NGstates}.

\begin{table}[h]
\begin{center}
\begin{tabular}{|cc|c|c|c|c|}\hline
$N$ & $\beta$ & $L_x=l/a$ & $L_y\times L_z\times L_t$ & sweeps$\times 10^6$ 
& num ops \\ \hline
3  & 6.0625 & 10  & $20 \times 20 \times 36$ & $1.5$ &  $580$ \\
   &        & 12  & $16 \times 16 \times 24$ & $1.5$ &  $580$ \\
   &        & 16  & $16 \times 16 \times 16$ & $1.5$ &  $725$ \\
   &        & 20  & $20 \times 20 \times 16$ & $1.4$ &  $725$ \\ 
   &        & 24  & $24 \times 24 \times 24$ & $4.0$ &  $725$ \\ 
   &        & 32  & $20 \times 20 \times 16$ & $4.5$ &  $725$ \\ \hline
3  & 6.3380 & 16  & $30 \times 30 \times 54$ & $0.5$ & $725$ \\
   &        & 18  & $24 \times 24 \times 36$ & $0.48$ &$725$  \\
   &        & 24  & $24 \times 24 \times 24$ & $0.5$ & $725$ \\ \hline
5  & 17.630 & 10  & $20 \times 20 \times 36$ & $0.4$ & $580$ \\
   &        & 12  & $16 \times 16 \times 24$ & $0.5$ & $580$ \\
   &        & 16  & $16 \times 16 \times 16$ & $0.5$ & $725$ \\ \hline
\end{tabular}
\caption{Parameters of our $SU(3)$ and $SU(5)$ calculations: 
lattice sizes, statistics, and number of operators. }
\label{table_param}
\end{center}
\end{table}

\begin{table}[h]
\begin{center}
\begin{tabular}{|ccc|ccc|cc|}\hline
\multicolumn{8}{|c|}{ $aE_{gs}(q,l)$ } \\ \hline
$q$ & $|J|$ & $P_t$ & $l/a$ & $N=3$ & $N=5$ & $l/a$ & $N=3f$ \\ \hline
 0 & 0 & + & 10 &  0.2448(27) & 0.2497(38) & 16 & 0.1841(27) \\
   &   &   & 12 &  0.3491(32) & 0.3591(54) & 18 & 0.2321(26) \\
   &   &   & 16 &  0.5393(38) & 0.5520(59) & 24 & 0.3499(55) \\
   &   &   & 20 &  0.7037(77) &  &   & \\ 
   &   &   & 24 &  0.8734(48) &  &   & \\ 
   &   &   & 32 &  1.165(19) &  &   & \\ \hline
 1 & 1 & $\pm$ & 10 & 0.9763(63)  & 0.971(15) & 16 & 0.6259(62)  \\
   &   &   & 12 & 0.9311(67) &  0.968(15) & 18 & 0.6299(60) \\
   &   &   & 16 & 0.970(10)  &  0.979(22) & 24 & 0.6545(57)\\
   &   &   & 20 & 1.057(17)  &   &   & \\ 
   &   &   & 24 & 1.152(12)  &   &   & \\ 
   &   &   & 32 & 1.408(35)  &   &   & \\ \hline
 2 & 0 & + & 10 & 1.686(10)  &  1.685(23) & 16 & 1.041(36) \\
   &   &   & 12 & 1.456(40)  &  1.468(81) & 18 & 1.021(13) \\
   &   &   & 16 & 1.338(37)  &  1.392(65) & 24 & 0.920(12) \\
   &   &   & 20 & 1.404(37)  &   &   & \\ 
   &   &   & 24 & 1.421(32)  &   &   & \\ \hline
 2 & 1 & $\pm$ & 10 & 1.540(37) &  1.508(85) & 16 & 1.040(43)  \\
   &   &   & 12 &  1.421(43) &  1.483(70) & 18 & 0.977(39) \\
   &   &   & 16 &  1.342(37) &  1.342(60) & 24 & 0.904(32)  \\
   &   &   & 20 &  1.389(35) &  &   & \\ 
   &   &   & 24 &  1.440(29) &  &   & \\ \hline
 2 & 2 & + & 10 &  1.612(49) &  1.56(11) & 16 & 1.077(50) \\
   &   &   & 12 &  1.525(44) &  1.63(10) & 18 & 0.970(46) \\
   &   &   & 16 &  1.428(37) &  1.38(8)  & 24 & 0.970(36) \\
   &   &   & 20 &  1.371(45) &   &   & \\ 
   &   &   & 24 &  1.382(24) &   &   & \\ \hline
 2 & 2 & - & 10 &  1.729(85) & 1.79(13) & 16 & 1.155(64) \\
   &   &   & 12 &  1.569(52) & 1.62(10) & 18 & 1.035(40) \\
   &   &   & 16 &  1.405(62) & 1.52(10) & 24 & 0.935(48)\\
   &   &   & 20 &  1.412(32) &  &   & \\ 
   &   &   & 24 &  1.442(33) &  &   & \\ \hline
\end{tabular}
\caption{The energy, $E_{gs}(q,l)$, of the lightest flux tube state with 
length $l$ and longitudinal momentum $p=2\pi q/l$ for $q=0,1,2$. 
(We include more than one state where these are approximately degenerate.) 
Quantum numbers are as shown.
Calculations are: SU(3) at $\beta=6.0625$ ($N=3$), SU(5) 
($N=5$) and SU(3) at $\beta=6.338$ ($N=3f$).}
\label{table_gsq012}
\end{center}
\end{table}

\begin{table}[htb]
\begin{center}
\begin{tabular}{|ccc|ccc|cc|}\hline
\multicolumn{8}{|c|}{$aE(q=0,l)$} \\ \hline
$|J|$ & $P_t$ &  $P_l$ & $l/a$ & $N=3$ & $N=5$ & $l/a$ & $N=3f$ \\ \hline
 0 & + & + & 10 & 0.905(12)  & 0.945(13) & 16 & 0.6211(97)\\
   &   &   & 12 & 0.9764(92) & 1.006(18) & 18 & 0.6419(77)\\
   &   &   & 16 & 1.0787(61) & 1.113(26) & 24 & 0.7248(80)\\
   &   &   & 20 & 1.1919(99) &  &   & \\ 
   &   &   & 24 & 1.278(26) &  &   & \\ \hline
 0 & - & - & 10 & 0.6606(68) & 0.614(12) & 16 & 0.4513(60)\\
   &   &   & 12 & 0.7472(59) & 0.733(13) & 18 & 0.4821(65)\\
   &   &   & 16 & 0.911(12) &  0.881(15) & 24 & 0.6278(70)\\
   &   &   & 20 & 1.054(20) &   &   & \\ 
   &   &   & 24 & 1.216(21) &   &   & \\ 
   &   &   & 32 & 1.646(54) &   &   & \\ \hline
 2 & + & + & 10 & 1.035(14) &  0.996(17) & 16 & 0.699(21)\\
   &   &   & 12 & 1.049(11) &  1.070(21) & 18 & 0.698(12)\\
   &   &   & 16 & 1.122(17) &  1.176(29) & 24 & 0.777(21)\\
   &   &   & 20 & 1.223(24) &   &   & \\ 
   &   &   & 24 & 1.327(23) &   &   & \\ 
   &   &   & 32 & 1.531(46) &   &   & \\ \hline
 2 & - & + & 10 & 1.083(11) &  1.077(20) & 16 & 0.738(11)\\
   &   &   & 12 & 1.094(16) &  1.113(23) & 18 & 0.740(10)\\
   &   &   & 16 & 1.158(18) &  1.129(32) & 24 & 0.792(15)\\
   &   &   & 20 & 1.201(24) &   &   & \\ 
   &   &   & 24 & 1.313(20) &   &   & \\ 
   &   &   & 32 & 1.503(61) &   &   & \\ \hline
\end{tabular}
\caption{Energies, $E(q,l)$, of the four lightest excited flux tube 
states with zero longitudinal momentum $p=2\pi q/l = 0$. Quantum 
numbers as shown. 
Calculations are labelled as in Table~\ref{table_gsq012}.}
\label{table_esq0}
\end{center}
\end{table}

\begin{table}[htb]
\begin{center}
\begin{tabular}{|cc|ccc|cc|}\hline
\multicolumn{7}{|c|}{$aE(q=1,l) \quad , \quad J={\mathrm{even}}$} \\ \hline
$|J|$ & $P_t$ & $l/a$ & $N=3$ & $N=5$ & $l/a$ & $N=3f$  \\ \hline
 0 & + & 10 & 1.341(26) & 1.330(47)  & 16 & 0.933(16) \\
   &   & 12 & 1.421(24) & 1.363(56)  & 18 & 0.883(25)\\
   &   & 16 & 1.378(36) & 1.422(75)  & 24 & 0.921(29)\\
   &   & 20 & 1.415(13) &   & & \\ 
   &   & 24 & 1.538(38) &   & & \\ 
   &   & 32 & 1.641(78) &   & & \\ \hline
 0 & - & 10 & 1.084(10) &  1.090(18)  & 16 & 0.724(11)\\
   &   & 12 & 1.089(11) &  1.064(18)  & 18 & 0.718(14)\\
   &   & 16 & 1.129(19) &  1.126(34)  & 24 & 0.746(16) \\
   &   & 20 & 1.229(24) &   & & \\ 
   &   & 24 & 1.358(25) &   & & \\ 
   &   & 32 & 1.575(71) &   & & \\ \hline
 2 & + & 10 & 1.457(37) &  1.384(72)  & 16 & 0.992(26)\\
   &   & 12 & 1.408(46) &  1.500(87)  & 18 & 0.954(39) \\
   &   & 16 & 1.422(50) &             & 24 & 0.926(38) \\
   &   & 20 & 1.501(63) &   & & \\ 
   &   & 24 & 1.480(43) &   & & \\ 
   &   & 32 & 1.582(71) &   & & \\ \hline
 2 & - & 10 & 1.682(65) &   & 16 & 1.098(74)\\
   &   & 12 & 1.474(47) &   & 18 & 1.056(52)\\
   &   & 16 & 1.526(57) &   & 24 & 0.969(43) \\
   &   & 20 & 1.421(76) &   & & \\ 
   &   & 24 & 1.567(40) &   & & \\ 
   &   & 32 & 1.778(110) &   & & \\ \hline
\end{tabular}
\caption{Energies, $E(q,l)$, of the lightest excited flux tube 
states with even spin $J$ and with the lowest non-zero longitudinal 
momentum, $p=2\pi q/l =2\pi/l$. Quantum numbers as shown.
Calculations are labelled as in Table~\ref{table_gsq012}.}
\label{table_esq1}
\end{center}
\end{table}

\begin{table}[htb]
\begin{center}
\begin{tabular}{|ccc|cc|}\hline
\multicolumn{5}{|c|}{$aE(q=1,l)$ $\quad , \quad$  $J=$odd} \\ \hline
$l/a$ & $N=3$ & $N=5$ & $l/a$ & $N=3f$  \\ \hline
10  & 1.417(24) & 1.359(44) & 16 &  0.8602(149)  \\  
12  & 1.337(27) & 1.377(58) & 18 &  0.8622(94) \\  
16  & 1.400(49) & 1.416(65) & 24 &  0.9261(123)  \\  
20  & 1.431(48) &  &    &    \\  \hline
10  &  &  & 16 &  0.9423(259) \\  
12  &  &  & 18 &  0.9366(116) \\  
16  &  &  & 24 &  0.9279(115) \\  
20  &  &  &    &    \\  \hline
10  &  &  & 16 &  0.9736(318) \\  
12  &  &  & 18 &  0.9864(139) \\  
16  &  &  & 24 &  0.9658(425) \\  
20  &  &  &    &    \\  \hline
10  &  &  & 16 &  1.1264(382) \\  
12  &  &  & 18 &  1.0804(165) \\  
16  &  &  & 24 &  1.1343(247) \\  
20  &  &  &    &    \\  \hline
 \end{tabular}
\caption{Energies of the four lightest excited flux tube states with 
longitudinal momentum $p=2\pi q/l = 2\pi/l$ and with $J=$odd. See
Table~\ref{table_gsq012} for the lightest odd-$J$ state.
Calculations are labelled as in Table~\ref{table_gsq012}.}
\label{table_esq1j1}
\end{center}
\end{table}

\begin{table}[htb]
\begin{center}
\begin{tabular}{|c|ccc|cc|}\hline
\multicolumn{6}{|c|}{$aE(q,l)$ $\quad , \quad$ $J^{P_t}=0^-$} \\ \hline
$q$ & $l/a$ & $N=3$ & $N=5$ & $l/a$ & $N=3f$ \\ \hline
 0  & 10  & 1.144(12)  & 1.179(32) & 16 &  0.7852(141)  \\  
    & 12  & 1.273(26)  & 1.237(31) & 18 &  0.7983(177) \\  
    & 16  & 1.340(31)  & 1.346(52) & 24 &  0.8555(278)  \\  
    & 20  & 1.366(38)  &  &    &    \\  
    & 24  & 1.432(31)  &  &    &    \\  
    & 32  & 1.633(79)  &  &    &    \\  \hline
 1  & 10  & 1.456(34)  & 1.372(46) & 16 &  0.983(13)  \\  
    & 12  & 1.388(33)  & 1.469(73) & 18 &  0.984(13) \\  
    & 16  & 1.512(50)  & 1.576(19) & 24 &  0.991(21)  \\  
    & 20  & 1.562(86)  &  &    &    \\  
    & 24  & 1.689(78)  &  &    &    \\  \hline
 1  & 10  & 1.607(48)  & 1.622(18) & 16 &  1.107(16)  \\  
    & 12  &    --      & 1.630(22) & 18 &  -- \\  
    & 16  & 1.674(18)  & 1.662(25) & 24 &  1.140(21)  \\  
    & 20  & 1.705(19)  &           &    &    \\   \hline
 2  & 10  & 1.771(12)  &           & 16 &  1.150(21)   \\  
    & 12  & 1.622(15)  &           & 18 &  1.060(18)  \\
    & 16  & 1.522(10)  &           & 24 &  0.999(14)  \\  
    & 20  & 1.554(15)  &           &    &    \\   \hline
\end{tabular}
\caption{Energies of the lightest excited  $0^-$ flux tube states with 
longitudinal momentum, $p=2\pi q/l$, apart from those states in
Table~\ref{table_esq0} and Table~\ref{table_esq1}
Calculations are labelled as in Table~\ref{table_gsq012}.}
\label{table_esqj0-}
\end{center}
\end{table}

\end{appendix}

\clearpage

\clearpage

\begin{figure}[hp!]
\begin	{center}
\leavevmode
\input	{plot_ceffd4n3b}
\end	{center}
\caption{Effective central charge in SU(3): from L\"uscher ($\bullet$) and 
Nambu-Goto ($\circ$) using eqns(\ref{eqn_MLceff},\ref{eqn_NGceff}).}
\label{fig_ceffd4n3}
\end{figure}

\begin{figure}[hp!]
\begin	{center}
\leavevmode
\input	{plot_ceffd4n6}
\end	{center}
\caption{Effective central charge in SU(6): from L\"uscher ($\bullet$) and 
Nambu-Goto ($\circ$) using eqns(\ref{eqn_MLceff},\ref{eqn_NGceff}).}
\label{fig_ceffd4n6}
\end{figure}

\begin{figure}[htb!]
\begin	{center}
\leavevmode
\input	{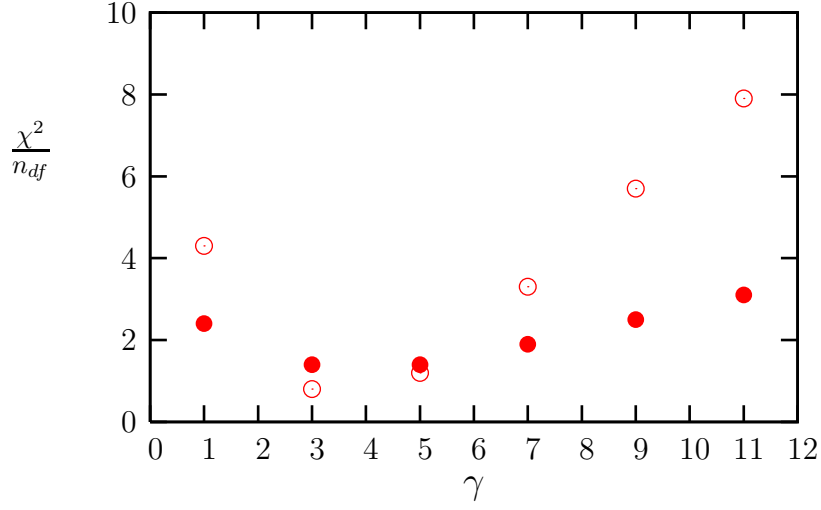}
\end	{center}
\caption{$\chi^2$ per degree of freedom for the best fit to $E_0(l)$
using eqn(\ref{eqn_E0fitAH}), for both SU(3), $\circ$, and SU(6), $\bullet$.} 
\label{fig_AHd4}
\end{figure}

\begin{figure}[hp]
\begin	{center}
\leavevmode
\input	{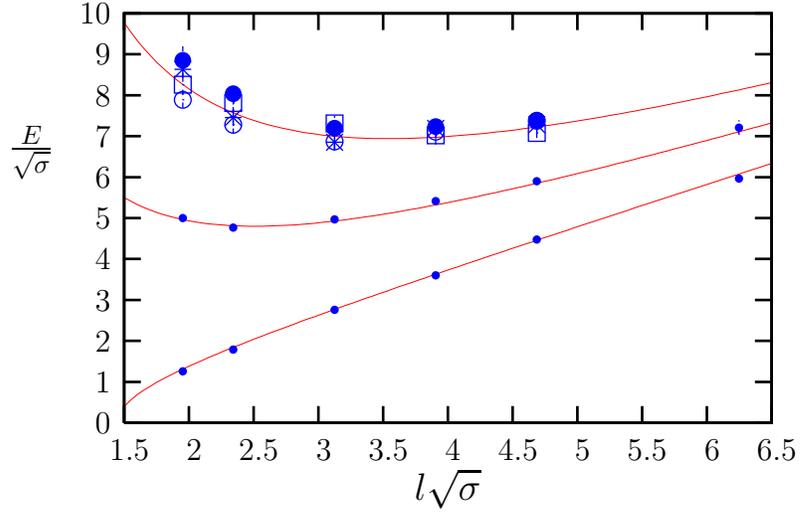}
\end	{center}
\caption{Lightest flux tube energies for longitudinal momenta $q=0$,
$\bullet$, $q=1$, $\bullet$, and  $q=2$  in SU(3) at $\beta=6.0625$. 
The four $q=2$ states are  $J^{P_t}= 0^+ (\star ), \ 1^\pm (\circ ),  
\ 2^+ (\Box ), \ 2^- (\bullet )$. Lines are Nambu-Goto predictions.}
\label{fig_Eqd4n3}
\end{figure}

\begin{figure}[hp]
\begin	{center}
\leavevmode
\input	{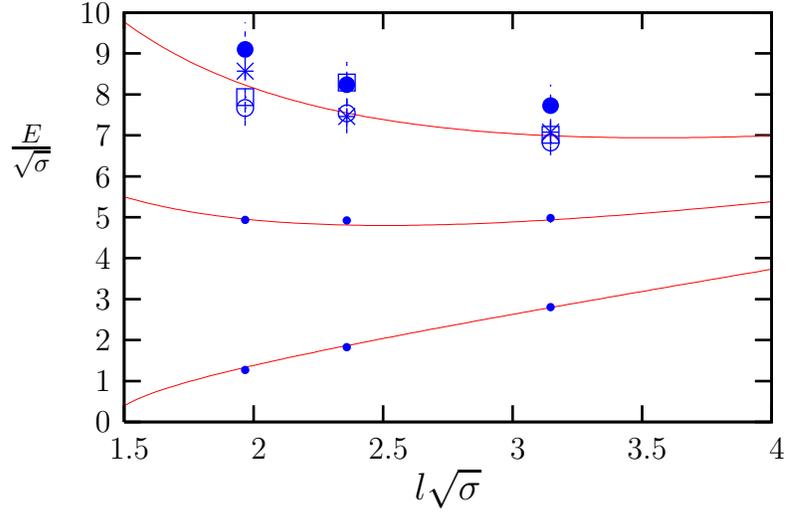}
\end	{center}
\caption{As in Fig.~\ref{fig_Eqd4n3} but for SU(5) at $\beta=17.63$.}
\label{fig_Eqd4n5}
\end{figure}

\begin{figure}[htb]
\begin	{center}
\leavevmode
\input	{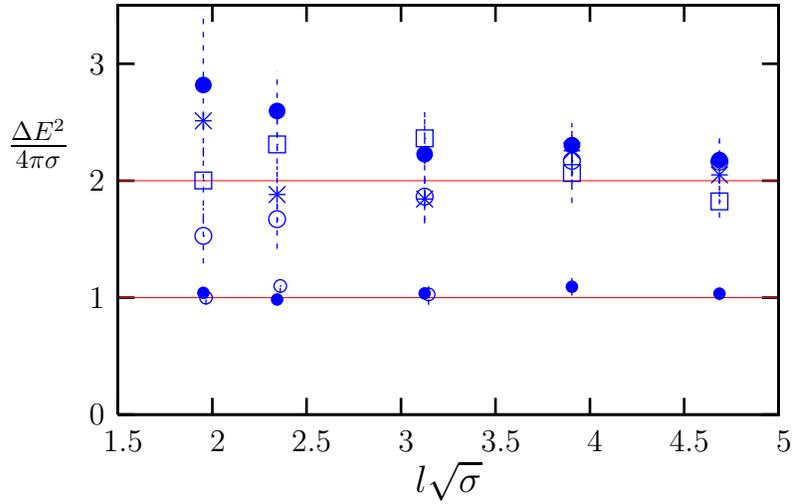}
\end	{center}
\caption{Excitation energies, as defined in eqn(\ref{eqn_exq}), 
of the lightest $q=1$ states in SU(3)
at $\beta=6.0625$, $\bullet$, and in SU(5), $\circ$, and also of the 
lightest $q=2$ states in SU(3).  The four $q=2$ states have quantum
numbers $J^{ P_t}=0^+$ ($\star$), $1^{\pm}$ ($\circ$),  $J=2^+$
($\Box$),  $J=2^-$ ($\bullet$). Lines are Nambu-Goto predictions.} 
\label{fig_DEqd4n3n5}
\end{figure}

\begin{figure}[hp]
\begin	{center}
\leavevmode
\input	{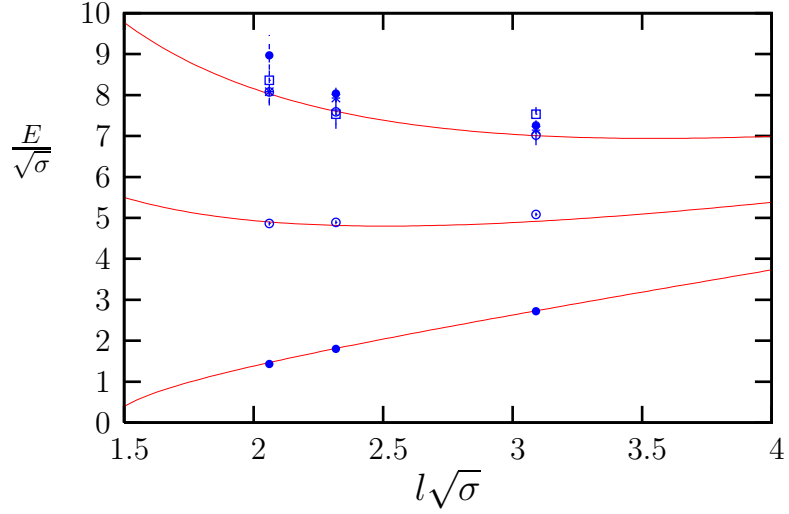}
\end	{center}
\caption{Lightest flux tube energies for $q=0$, $q=1$ and $q=2$, 
in SU(3) at $\beta=6.338$. States labelled as in Fig.~\ref{fig_Eqd4n3}.
Lines are Nambu-Goto predictions.}
\label{fig_Eqd4n3f}
\end{figure}

\begin{figure}[p]
\begin	{center}
\leavevmode
\input	{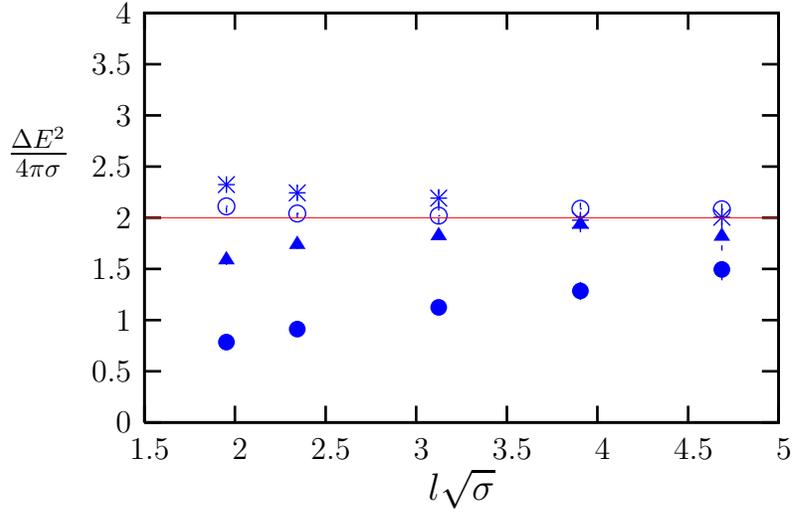}
\end	{center}
\caption{Excitation energies, as defined in eqn(\ref{eqn_exq}), 
of the lightest few $q=0$ states in SU(3) at $\beta=6.0625$.
The states are $J^{P_tP_l}=$ $0^{++}$ ($\blacktriangle$),
$0^{--}$ ($\bullet$), $2^{++}$ ($\circ$), 
$2^{-+}$ ($\star$). Line is Nambu-Goto prediction.}
\label{fig_DEq0exd4n3}
\end{figure}

\begin{figure}[h]
\begin	{center}
\leavevmode
\input	{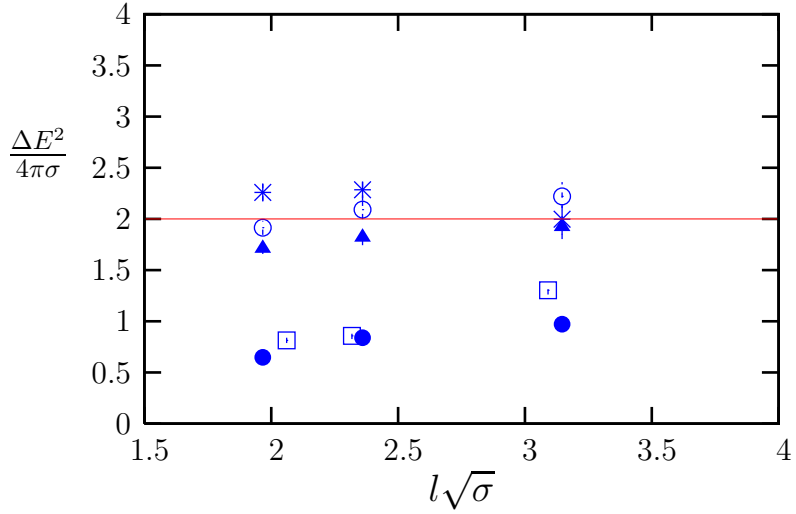}
\end	{center}
\caption{Excitation energies,  as in Fig.~\ref{fig_DEq0exd4n3},
but for SU(5), together with the $0^{--}$ in SU(3) at $\beta=6.338$
($\square$). Line is Nambu-Goto prediction.}
\label{fig_DEq0exd4n5n3f}
\end{figure}

\begin{figure}[h]
\begin	{center}
\leavevmode
\input	{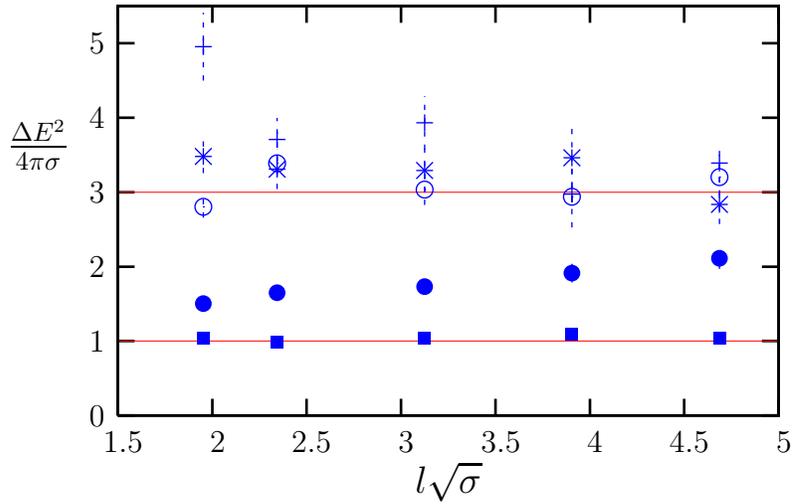}
\end	{center}
\caption{Excitation energies, as defined in eqn(\ref{eqn_exq}), 
of the lightest few excited $q=1$ states with even $J$. In SU(3) at 
$\beta=6.0625$. These are $J^{P_t}= 0^+ (\circ), 0^- (\bullet),
2^+ (\ast), 2^- (+)$. Also shown is the $q=1$ ground state
($\blacksquare$) which has $J=1^+$. Lines are Nambu-Goto 
predictions.}
\label{fig_DEq1exd4n3}
\end{figure}

\begin{figure}[p]
\begin	{center}
\leavevmode
\input	{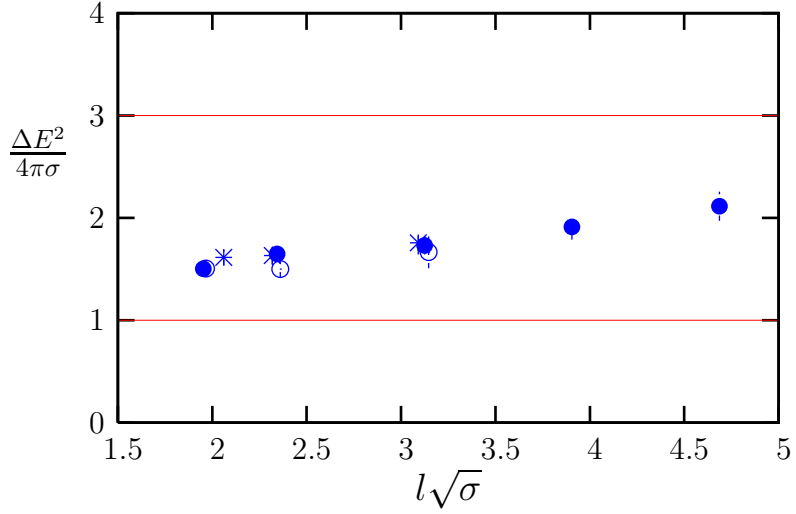}
\end	{center}
\caption{Excitation energy, as defined in eqn(\ref{eqn_exq}), 
of the lightest $0^-$ flux tube with
$q=1$: for SU(5), $\circ$, and  SU(3) at $\beta=6.338$, $\ast$,
compared to the values in  SU(3) at $\beta=6.0625$, $\bullet$.
Lines are Nambu-Goto predictions.}
\label{fig_DEq1j0-d4n5n3fn3}
\end{figure}

\begin{figure}[h]
\begin	{center}
\leavevmode
\input	{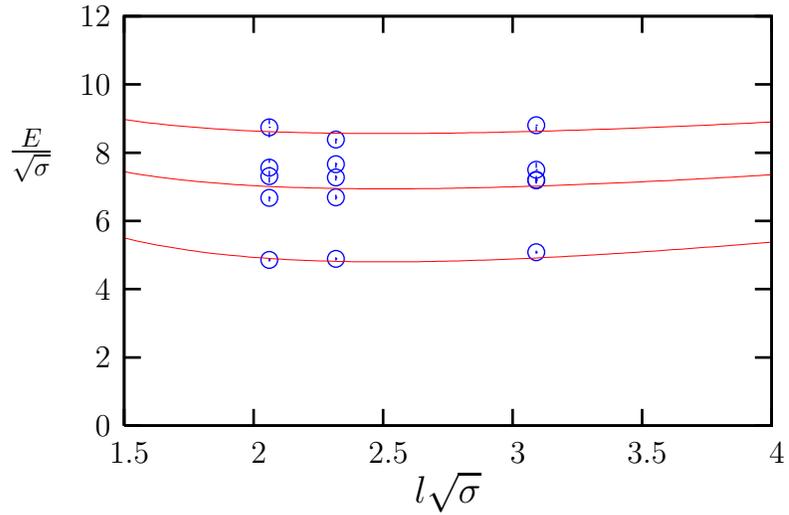}
\end	{center}
\caption{Energies  of the lightest five $q=1$ states 
with $J=$odd, in SU(3) at $\beta=6.338$.
Lines are Nambu-Goto predictions.}
\label{fig_Eq1j13exd4n3f}
\end{figure}

\begin{figure}[h]
\begin	{center}
\leavevmode
\input	{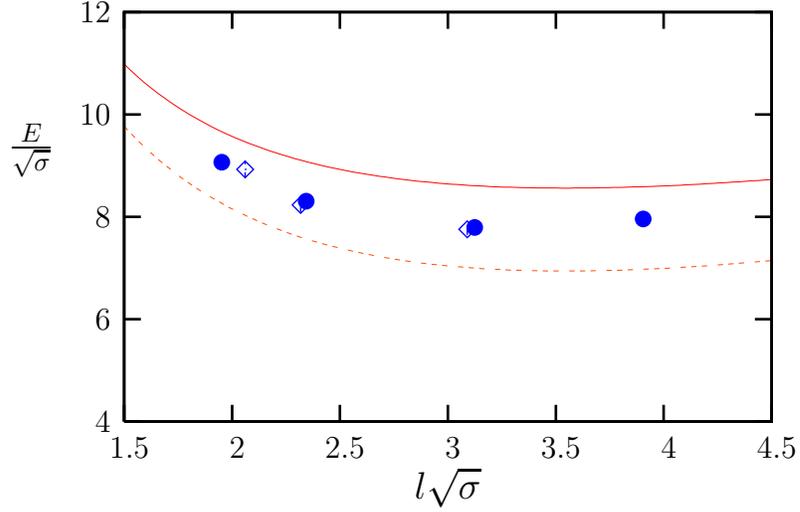}
\end	{center}
\caption{Energy of the lightest $q=2$ state with $J^{P_t}=0^-$
in SU(3) at  $\beta=6.0625$, $\bullet$, and at  $\beta=6.338$,
$\lozenge$. Solid line is the Nambu-Goto prediction;
dotted line is the  prediction for the $q=2$ ($0^+,2^\pm$) ground energy
level.}
\label{fig_Eq2j0-d4n3n3f}
\end{figure}

\begin{figure}[h]
\begin	{center}
\leavevmode
\input	{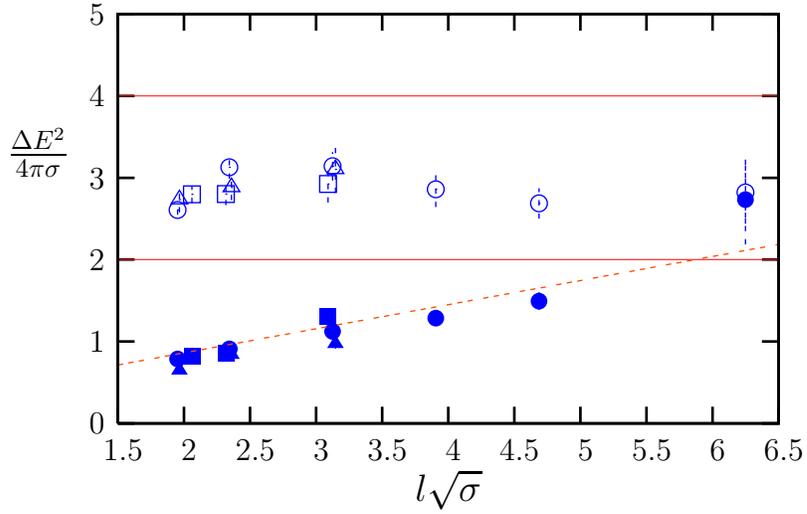}
\end	{center}
\caption{Excitation energies, as defined in eqn(\ref{eqn_exq}),
of $q=0$, $0^-$ states: the ground state in SU(3) 
at $\beta=6.0625$, $\bullet$, in SU(3) at $\beta=6.338$, $\blacksquare$, 
and in SU(5), $\blacktriangle$; and the first excited states,
$\circ$,  $\square$ and $\vartriangle$ 
respectively. Solid lines are the Nambu-Goto predictions; dashed line
is the massive mode ansatz in eqn(\ref{eqn_DEexmq}).}
\label{fig_DEq0J0Pp-n3fn5n3d4}
\end{figure}

\begin{figure}[h]
\begin	{center}
\leavevmode
\input	{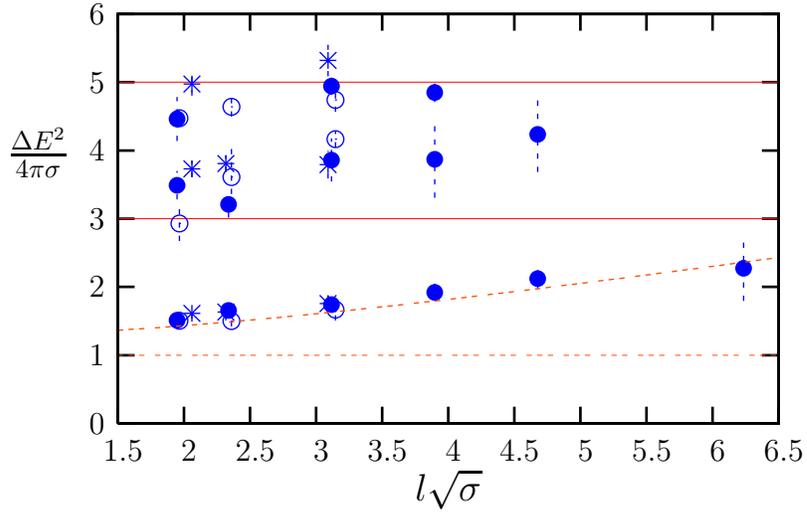}
\end	{center}
\caption{Excitation energies, as defined in eqn(\ref{eqn_exq}),
of the lightest three $q=1$, $0^-$ states
in SU(3) at $\beta=6.0625$, $\bullet$, in SU(3) at $\beta=6.338$, 
$\ast$, and in SU(5), $\circ$. Solid lines are Nambu-Goto predictions for
the $0^-$; dashed line is the massive mode ansatz in eqn(\ref{eqn_DEexmq}). 
(Dotted horizontal line is for the $J=1^{\pm}$ ground state).}
\label{fig_DEq1J0-exn3n5n3fd4}
\end{figure}

\begin{figure}[h]
\begin	{center}
\leavevmode
\input	{plot_DDEq1J0-exn3d4.tex}
\end	{center}
\caption{Difference of the excitation energies of the lightest 
two $q=1$, $0^-$ states in SU(3) at $\beta=6.0625$, $\bullet$. 
Line is Nambu-Goto prediction.}
\label{fig_DDEq1J0-exn3d4}
\end{figure}

\begin{figure}[h]
\begin	{center}
\leavevmode
\input	{plot_DDEq0J0-exn3d4.tex}
\end	{center}
\caption{Difference of the excitation energies of the lightest 
two $q=0$, $0^-$ states in SU(3) at $\beta=6.0625$, $\bullet$. 
Line is Nambu-Goto prediction.}
\label{fig_DDEq0J0-exn3d4}
\end{figure}

\begin{figure}[h]
\begin	{center}
\leavevmode
\input	{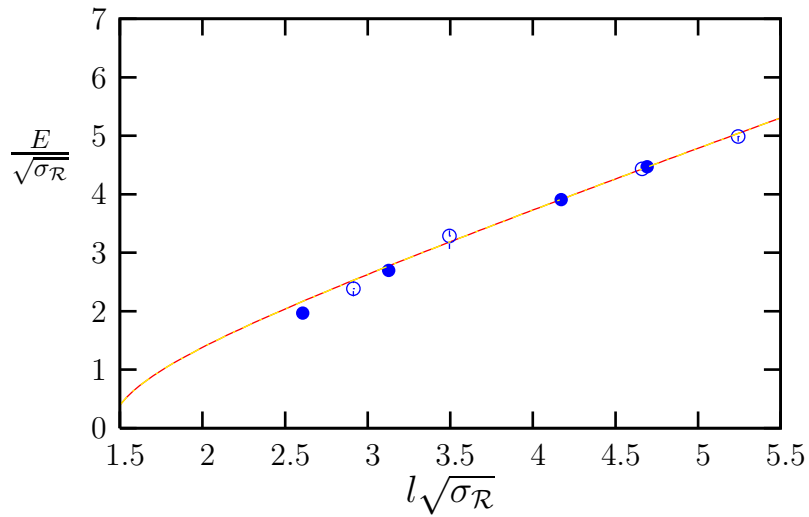}
\end	{center}
\caption{Energies of the lightest closed $k=2$ 
flux tubes of length $l$, carrying flux in the totally antisymmetric
representation ${\cal{R}}=2A$, $\bullet$, and the  totally symmetric
one ${\cal{R}}=2S$, $\circ$. In SU(6). Line is Nambu-Goto fit.}
\label{fig_Ek2d4n6}
\end{figure}

\clearpage

\begin{figure}[htb]
\begin	{center}
\leavevmode
\input	{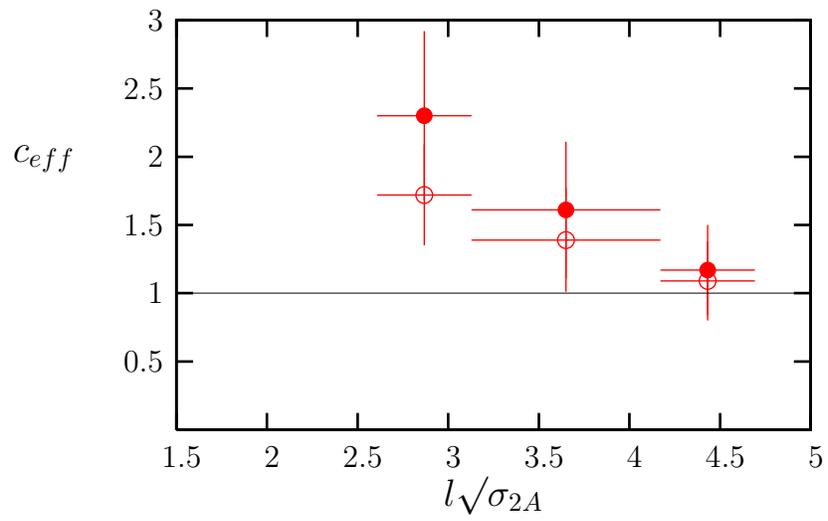}
\end	{center}
\caption{Effective central charge for the lightest totally antisymmetric
$k=2$ flux tube in SU(6): from L\"uscher ($\bullet$) and 
Nambu-Goto ($\circ$) using eqns(\ref{eqn_MLceff},\ref{eqn_NGceff}).}
\label{fig_ceffd4n6_k2A}
\end{figure}

\begin{figure}[htb]
\begin	{center}
\leavevmode
\input	{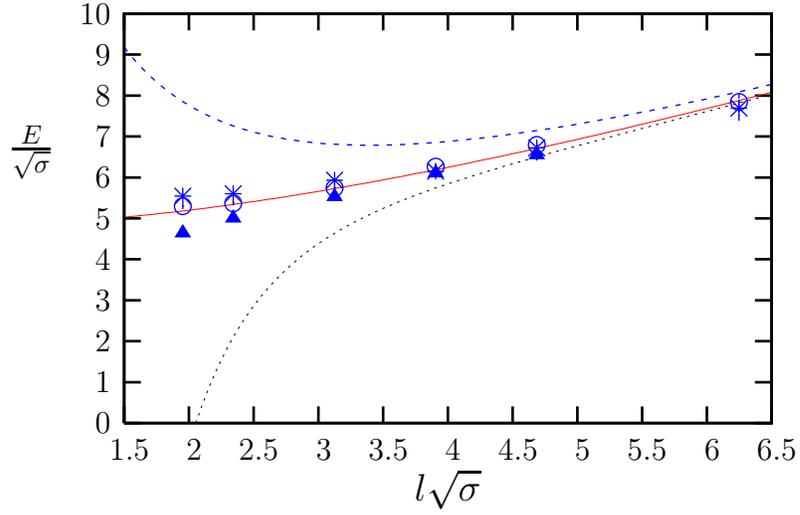}
\end	{center}
\caption{Energies of the lightest $q=0$ excited states in SU(3) at 
$\beta=6.0625$. States are $J^{P_t P_l}= 0^{++}$, $\blacktriangle$,
$2^{++}$, $\circ$, $2^{-+}$, $\star$.
Solid line is Nambu-Goto, dotted is universal expansion to $O(1/l^3)$, 
and dashed is expansion to $O(1/l)$ (L\"uscher correction). }
\label{fig_Eq0exd4n3}
\end{figure}

\begin{figure}[htb]
\begin	{center}
\leavevmode
\input	{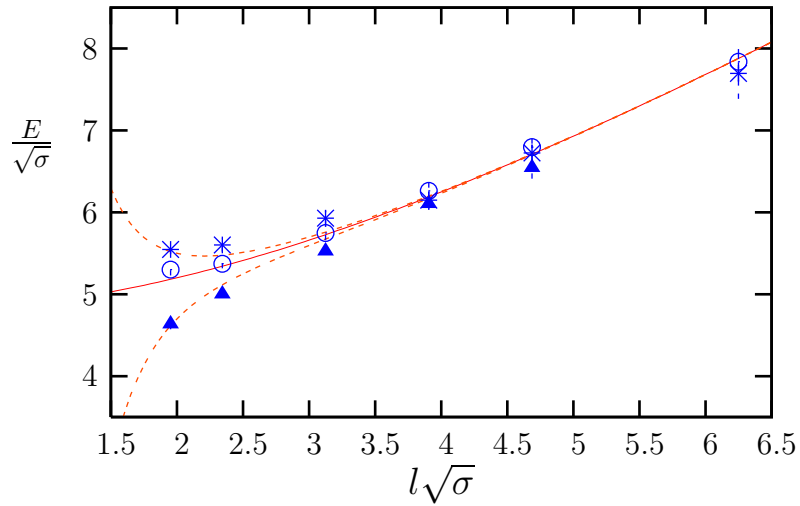}
\end	{center}
\caption{Energies of the lightest $q=0$ excited states in SU(3) at 
$\beta=6.0625$. States are $J^{P_t P_l}= 0^{++}$, $\blacktriangle$,
$2^{++}$, $\circ$, $2^{-+}$, $\star$. Solid line is Nambu-Goto, dotted 
line is with an additional $O(1/l^5)$ correction.}
\label{fig_Eq0exd4n3_NGcorfit}
\end{figure}

\begin{figure}[htb]
\begin	{center}
\leavevmode
\input	{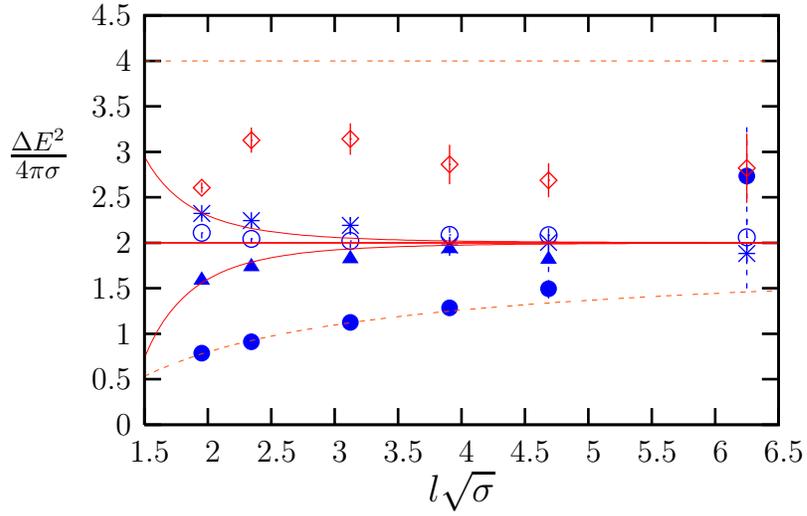}
\end	{center}
\caption{Excitation energies of the lightest $q=0$ excited states in 
SU(3) at $\beta=6.0625$. The states are $J^{P_t P_l}=$
$0^{--}$ ($\bullet$), $0^{++}$ ($\blacktriangle$), $2^{++}$ ($\circ$)
and  $2^{-+}$ ($\star$), the solid horizontal line is the Nambu-Goto 
prediction, the curved lines are fits (see text) with a resummed 
$O(1/l^5)$ correction, and the curved 
dotted line is a fit with a resummed $O(1/l^{1.7})$ correction.
Also shown is the next excited $0^{--}$ ($\Diamond$) and the horizontal
dotted line is the NG prediction for that state.}
\label{fig_DEq0exd4n3_NGcorfit}
\end{figure}

\end{document}